%% file: EXO-19-002_temp.tex
\pdfoutput=1

\documentclass[11pt,twoside,a4paper,cmspaper,final,collab]{cms-tdr}
\def\svnVersion{f17174c-D}\def\svnDate{2020/01/16}\def\cmsCernNoTag{CERN-EP-2019-237}\def\cmsCernDate{\today}\def\cmsMessage{"Published in the Journal of High Energy Physics as \href{http://dx.doi.org/10.1007/JHEP03(2020)051}{\doi{10.1007/JHEP03(2020)051}.}"}

\begin{document}\cmsNoteHeader{EXO-19-002}

\hyphenation{had-ron-i-za-tion}
\hyphenation{cal-or-i-me-ter}
\hyphenation{de-vices}
\newcommand{\MTthree}{\ensuremath{M_{\mathrm{T}}}}
\newcommand{\LT}{\ensuremath{L_{\mathrm{T}}}}
\newcommand{\ST}{\ensuremath{S_{\mathrm{T}}}}
\newcommand{\nossf}{\ensuremath{N_{\mathrm{OSSF}}}}
\newcommand{\mossf}{\ensuremath{M_{\mathrm{OSSF}}}}
\newcommand{\nell}{\ensuremath{N_{\mathrm{leptons}}}}
\newcommand{\lightlepton}{\ell}
\newcommand{\nbj}{\ensuremath{N_{\cPqb}}}
\newcommand{\ttphi}{\ensuremath{\ttbar\phi}}
\newcommand{\JHUGen}{{\textsc{JHUGen}}\xspace}
\newlength\cmsTabSkip\setlength{\cmsTabSkip}{1ex}
\providecommand{\NA}{\ensuremath{\text{---}}}

\cmsNoteHeader{EXO-19-002}
\title{Search for physics beyond the standard model in multilepton final states in proton-proton collisions at $\sqrt{s} = 13\TeV$}

\author*[inst1]{Maximilian Dieter Heindl}

\date{\today}

\abstract{A search for physics beyond the standard model in events with at least three charged leptons (electrons or muons) is presented. The data sample corresponds to an integrated luminosity of 137\fbinv of proton-proton collisions at $\sqrt{s} = 13\TeV$, collected with the CMS detector at the LHC in 2016--2018. The two targeted signal processes are pair production of type-III seesaw heavy fermions and production of a light scalar or pseudoscalar boson in association with a pair of top quarks. The heavy fermions may be manifested as an excess of events with large values of leptonic transverse momenta or missing  transverse momentum. The light scalars or pseudoscalars may create a localized excess in the dilepton mass spectra. The results exclude heavy fermions of the type-III seesaw model for masses below 880\GeV at 95\% confidence level in the scenario of equal branching fractions to each lepton flavor. This is the most restrictive limit on the flavor-democratic scenario of the type-III seesaw model to date. Assuming a Yukawa coupling of unit strength to top quarks, branching fractions of new scalar (pseudoscalar) bosons to dielectrons or dimuons above 0.004 (0.03) and 0.04 (0.03) are excluded at 95\% confidence level for masses in the range 15--75 and 108--340\GeV, respectively. These are the first limits in these channels on an extension of the standard model with scalar or pseudoscalar particles.}

\hypersetup{
pdfauthor={CMS Collaboration},
pdftitle={Search for physics beyond the standard model in multilepton final states in proton-proton collisions at 13 TeV},
pdfsubject={CMS},
pdfkeywords={CMS, physics, seesaw, scalar, multileptons}}

\maketitle

\section{Introduction}

A search for new phenomena in final states with at least three charged leptons (electrons or muons) is presented, using 137\fbinv of proton-proton ($\Pp\Pp$) collision data at $\sqrt{s} = 13\TeV$ collected by the CMS experiment at the CERN LHC from 2016 to 2018.
The results are interpreted in the context of two beyond the standard model (SM) theories, namely the type-III seesaw and light scalar or pseudoscalar sector extensions to the SM. The event selection and signal region definitions are chosen in a way that allows other models to be tested.
Phenomenologically, these models show complementary signatures of resonant and nonresonant multilepton final states, as described below.

The seesaw mechanism introduces new heavy particles coupled to leptons and to the Higgs boson, in order to explain the light masses of the neutrinos~\cite{Minkowski:1977sc,Mohapatra:1979ia,Magg:1980ut,Mohapatra:1980yp,Schechter:1980gr,Schechter:1981cv,Foot:1988aq,Mohapatra:1986aw,Mohapatra:1986bd}.
Within the type-III seesaw model, the neutrino is assumed to be a Majorana particle whose mass arises via the mediation of new massive fermions.
These massive fermions are an SU(2) triplet of heavy Dirac charged leptons ($\Sigma^\pm$) and a heavy Majorana neutral lepton ($\Sigma^0$).
In $\Pp\Pp$ collisions, these massive fermions may be pair-produced through electroweak interactions in both charged-charged and charged-neutral pairs.
Multilepton final states arise from the decays of each of the $\Sigma^+\Sigma^-$, $\Sigma^+\Sigma^0$, and $\Sigma^-\Sigma^0$ pairs to the nine different pairs of $\PW$, $\PZ$, and Higgs bosons with SM leptons and the subsequent leptonic decays of the SM bosons.
A complete decay chain example would be $\Sigma^\pm \Sigma^0\to (W^\pm \nu) (W^\pm \ell^\mp) \to (\ell^{\pm} \nu \nu) (\ell^{\pm} \nu \ell^\mp)$, where $\ell$ and $\nu$ are the three flavors of charged and neutral SM leptons, respectively.
All 27 distinct signal production and decay combinations of the seesaw signal are simulated~\cite{Biggio:2011ja}.
The $\Sigma^{\pm,0}$ are degenerate in mass, their decays are prompt, and the $\Sigma$ decay branching fractions are identical across all lepton flavors (flavor-democratic scenario).
This is achieved by taking the mixing angles to be $V_{\Pe}=V_{\mu}=V_{\tau}=10^{-4}$, values that are compatible with the existing constraints~\cite{Abada:2008ea,Abada:2007ux,delAguila:2008pw,Biggio:2011ja,Goswami:2018}.

New light scalars or pseudoscalars are a ubiquitous feature of many theories of physics beyond the SM, including, but not limited to, extended Higgs sectors, supersymmetric theories, and dark sector extensions~\cite{Cacciapaglia:2019bqz,Ellwanger:2009dp,Maniatis:2009re,Buckley:2014fba}.
We consider a generalization of a simple model \cite{Casolino:2015cza,Chang:2017ynj}, where a new light CP-even scalar or CP-odd pseudoscalar boson ($\phi$) is produced in $\Pp\Pp$ collisions via a Yukawa coupling of the $\phi$ to top quarks, $g_{\PQt}$, either in three-body associated production with top quark pairs, or in top quark pair production with three-body top quark decays, ${\PQt}\to\cPqb\PW\phi$.
The signal is collectively labeled as $\ttphi$.
In this paper, we search for decays of the $\phi$ boson via a Yukawa coupling to the charged leptons, $g_{\ell}$, into dielectron or dimuon pairs within multilepton events.
The decays of the $\phi$ boson into tau-tau lepton pairs are not considered.
It is assumed that $g_{\ell}\ll g_{\PQt}$ and that all other couplings of the $\phi$ boson are negligible.
Furthermore, the $\phi$ boson decays are taken to be prompt, and the $\phi$ branching fractions into different flavors of charged lepton pairs, $\mathcal{B}(\phi\to\ell\ell)$, as well as $g_{\PQt}$, are left as free parameters.

\begin{figure}[!htp]
\centering
\includegraphics[width=.50\textwidth]{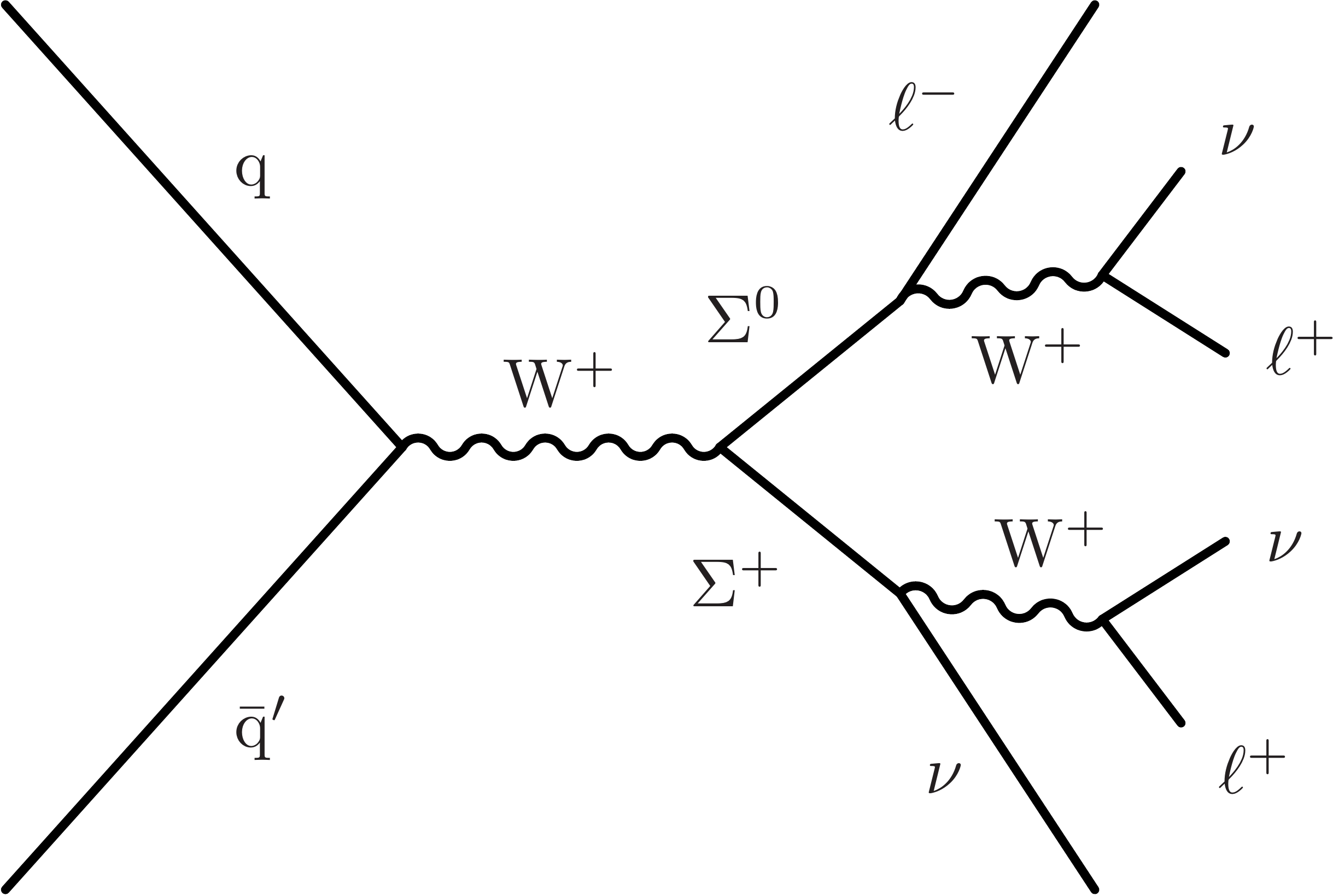} \hspace{.025\textwidth}
\includegraphics[width=.43\textwidth]{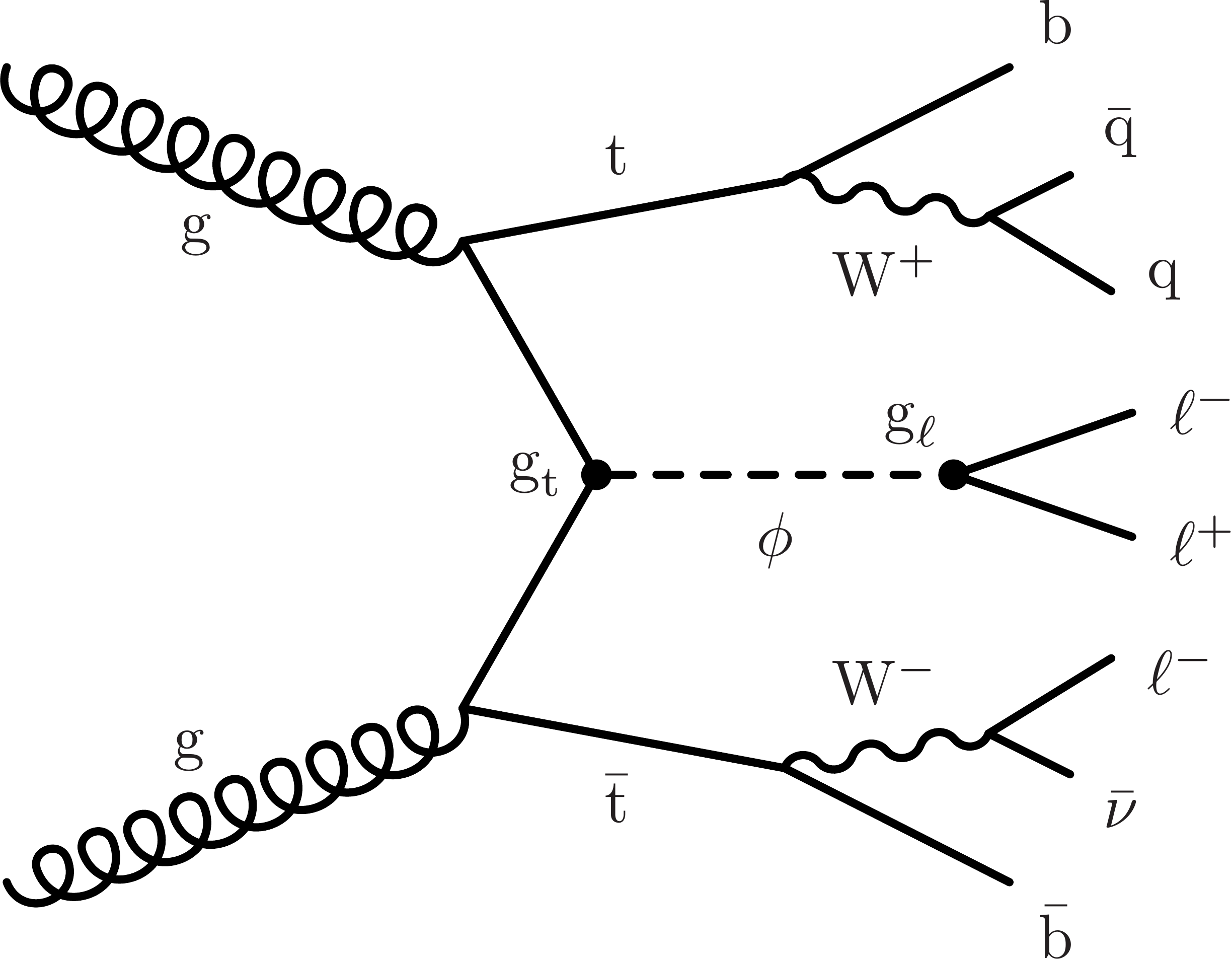}
\caption{
Leading order Feynman diagrams for the type-III seesaw (left) and $\ttphi$ (right) signal models, depicting example production and decay modes in $\Pp\Pp$ collisions.
\label{fig:FeynmanDiagrams}}
\end{figure}

Figure~\ref{fig:FeynmanDiagrams} illustrates example diagrams for the production and decay of heavy fermions in the type-III seesaw model (left) and a light scalar or pseudoscalar boson in the $\ttphi$ model (right).

Prior searches for the manifestation of the type-III seesaw model have been conducted by the ATLAS and CMS Collaborations using data recorded at $\sqrt{s} = 7$, 8, and 13\TeV~\cite{CMS:2012ra,Aad:2015cxa,Aad:2015dha,Sirunyan:2017qkz}.
The most stringent constraints in the flavor-democratic scenario are from a CMS search using 13\TeV data collected in 2016, which excluded $\Sigma$ masses below 850\GeV~\cite{Sirunyan:2017qkz}.
The present study of the $\ttphi$ model is the first direct search for a light scalar or pseudoscalar boson in leptonic decays produced in association with a top quark pair.

\section{The CMS detector}
The central feature of the CMS apparatus is a superconducting solenoid of 6\unit{m} internal diameter, providing a magnetic field of 3.8\unit{T}.
Within the solenoid volume are a silicon pixel and strip tracker, a lead tungstate crystal electromagnetic calorimeter (ECAL), and a brass and scintillator hadron calorimeter, each composed of a barrel and two endcap sections.
Forward calorimeters extend the pseudorapidity ($\eta$) coverage provided by the barrel and endcap detectors.
Muons are detected in gas-ionization chambers embedded in the steel flux-return yoke outside the solenoid.
A more detailed description of the CMS detector, together with a definition of the coordinate system used and the relevant kinematic variables, can be found in Ref.~\cite{Chatrchyan:2008zzk}.
The CMS detector uses a two-tiered trigger system \cite{Khachatryan:2016bia}.
The first level, composed of custom hardware processors, uses information from the calorimeters and muon detectors to select the most relevant pp collision events at rates up to 100\unit{kHz}.
These are further processed by a second level consisting of a farm of processors, known as the high level trigger, that combines information from all CMS subdetectors to yield a final event rate of less than 1\unit{kHz} for data storage.

\section{Data samples and event simulation}

The data samples analyzed in this search correspond to a total integrated luminosity of 137\fbinv (35.9, 41.5, and 59.7\fbinv in years 2016, 2017 and 2018, respectively), recorded in $\Pp\Pp$ collisions at $\sqrt{s} = 13\TeV$.
A combination of isolated single-electron and single-muon triggers was used with corresponding transverse momentum (\pt) thresholds of 24 and 27\GeV in 2016,
27 and 32\GeV in 2017, and 24 and 32\GeV in 2018.
Event samples from Monte Carlo (MC) simulations are used to estimate the rates of signal and relevant SM background processes.
The $\PW\PZ$, $\PZ\gamma$, $\ttbar\PZ$, $\ttbar\PW$, and triboson backgrounds are generated using \MGvATNLO (2.2.2 in 2016, 2.4.2 in 2017 and 2018 data analyses)~\cite{Alwall:2014hca} at next-to-leading order (NLO) precision. The top quark mass used in all simulations is 172.5\GeV.
The $\PZ\PZ$ background contribution from quark-antiquark annihilation is generated using \POWHEG 2.0~\cite{Nason:2004rx,Frixione:2007vw,Alioli:2010xd} at NLO, whereas the contribution from gluon-gluon fusion is generated at leading order (LO) using \MCFM 7.0.1~\cite{Campbell:2010ff}.
Backgrounds from Higgs boson production for a Higgs boson mass of 125\GeV are generated at NLO using \POWHEG and \JHUGen 7.0.11~\cite{Gao:2010qx,Bolognesi:2012mm,Anderson:2013afp,Gritsan:2016hjl}.
Simulated event samples for Drell--Yan (DY) and $\ttbar$ processes, generated at NLO with \MGvATNLO and \POWHEG, respectively, are used for systematic uncertainty studies.

All signal samples are simulated using \MGvATNLO 2.6.1 at LO precision.
The production cross section for the type-III seesaw signal model $\sigma(\Sigma\Sigma)$ is calculated at NLO plus next-to-leading logarithmic precision, assuming that the heavy leptons are SU(2) triplet fermions \cite{Fuks:2012qx,Fuks:2013vua}, while the $\ttphi$ production cross section $\sigma(\ttphi)$ comes directly from the \MGvATNLO 2.6.1 generator at LO precision.

All background and signal samples in 2016 are generated with the NNPDF3.0 NLO or LO parton distribution functions (PDFs), with the order matching that in the matrix element calculations. In 2017 and 2018, the NNPDF3.1 next-to-next-to-leading order PDFs~\cite{Ball:2014uwa,Ball:2017nwa} are used.
Parton showering, fragmentation, and hadronization for all samples are performed using \PYTHIA 8.230~\cite{Sjostrand:2014zea} with the underlying event tune CUETP8M1~\cite{Khachatryan:2015pea} for the 2016 analysis, and CP5~\cite{Sirunyan:2019dfx} for the 2017 and 2018 analyses.
Double counted partons generated with \PYTHIA and \MGvATNLO are removed using the FxFx~\cite{Frederix:2012ps} matching schemes.
The response of the CMS detector is simulated using dedicated software based on the \GEANTfour toolkit~\cite{Agostinelli:2002hh},
and the presence of multiple $\Pp\Pp$ interactions in the same or adjacent bunch crossing (pileup) is incorporated by simulating additional interactions, that are both in-time and out-of-time with the hard collision according to the pileup in the data samples.

\section{Event reconstruction}

A particle-flow (PF) algorithm~\cite{Sirunyan:2017ulk} aims to reconstruct and identify each individual particle in an event, with an optimized combination of information from the various elements of the CMS detector.
In each event, the candidate vertex with the largest value of summed physics-object $\pt^2$ is taken to be the primary $\Pp\Pp$ interaction vertex (PV). Here the physics objects are the jets, clustered using the jet finding algorithm ~\cite{Cacciari:2008gp,Cacciari:2011ma} with the tracks assigned to candidate vertices as inputs, and the associated missing transverse momentum, taken as the negative vector sum of the \pt of those jets.
The energy of photons is obtained from the ECAL measurement.
The energy of electrons is determined from a combination of the electron momentum at the PV as determined by the tracker, the energy of the corresponding ECAL cluster, and the energy sum of all bremsstrahlung photons spatially compatible with originating from the electron track.
The energy of muons is obtained from the curvature of the corresponding track.
The energy of charged hadrons is determined from a combination of their momentum measured in the tracker and the matching ECAL and HCAL energy deposits, corrected for zero-suppression effects and for the response function of the calorimeters to hadronic showers.
Finally, the energy of neutral hadrons is obtained from the corresponding corrected ECAL and HCAL energies.

Jets used in this analysis are reconstructed using the anti-\kt algorithm~\cite{Cacciari:2008gp} with a distance parameter of 0.4, as implemented in the \textsc{FastJet} package~\cite{Cacciari:2011ma}. Jets are required to have $\pt>30\GeV$ and, to be fully in the tracking system volume, $\abs{\eta}<2.1$.
Jet momentum is determined as the vectorial sum of all particle momenta in the jet, and is found from simulation to be, on average, within 5--10\% of the true momentum over the whole \pt spectrum and detector acceptance.
The effect of the pileup on reconstructed jets is mitigated through a charged hadron subtraction technique, which removes the energy of charged hadrons not originating from the PV~\cite{Sirunyan:2017ulk}.
The impact of neutral pileup particles in jets is mitigated by an event-by-event jet-area-based correction of the jet four-momenta~\cite{Cacciari:2008ca,Cacciari:2008ps,Sirunyan:2017jes}.

Jet energy corrections are derived from simulation studies so that the average measured response of jets becomes identical to that of particle level jets.
In situ measurements of the momentum balance in dijet, photon+jet, leptonically decaying $\PZ$+jet, and multijet events are used to determine any residual differences between the jet energy scale in data and in simulation, and appropriate corrections are made to the jet \pt~\cite{Sirunyan:2017jes}.
Additional quality criteria are applied to each jet to remove those potentially dominated by instrumental effects or reconstruction failures~\cite{CMS:2017jme}.
Finally, all selected jets are required to be outside a cone of $\Delta R \equiv \sqrt{\smash[b]{(\Delta\eta)^2+(\Delta\phi)^2}}=0.4$ around a selected electron or muon as defined below, where $\Delta \phi$ is the azimuthal distance.

A subset of these reconstructed jets originating from $\Pb$ hadrons is identified using the DeepCSV {\cPqb} tagging algorithm~\cite{Sirunyan:2017ezt}.
This algorithm has an efficiency of 60--75\%  to identify {\cPqb} quark jets, depending on jet \pt and $\eta$, and a misidentification rate of about 10\% for {\cPqc} quark jets as well as 1\% for light quark and gluon jets.

The missing transverse momentum vector \ptvecmiss is computed as the negative vector sum of the transverse momenta of all the PF candidates in an event, and its magnitude is denoted as \ptmiss~\cite{Sirunyan:2019kia}.
The \ptvecmiss is modified to account for corrections to the energy scale of the reconstructed jets in the event.

Electrons and muons are reconstructed by geometrically matching tracks reconstructed in the tracking system with energy clusters in the ECAL~\cite{Khachatryan:2015hwa} and with the tracks in the muon detectors~\cite{Sirunyan:2018fpa}, respectively.
Electrons are required to be within the tracking system acceptance, $\abs{\eta}<2.5$, and muons are required to be within the muon system acceptance, $\abs{\eta}<2.4$. Both electrons and muons must have $\pt>10\GeV$.
Furthermore, electrons must satisfy shower shape and track quality requirements to suppress those originating from photon conversions in detector material as well as hadronic activity misidentified as electrons.
Similarly, muons must satisfy track fit and matching quality requirements to suppress muon misidentification due to hadron shower remnants that reach the muon system.

Prompt isolated leptons produced by SM boson decays (either directly, or via an intermediate tau lepton) are indistinguishable from those produced in signal events.
Thus, SM processes that can produce three or more isolated leptons, such as $\PW\PZ$, $\PZ\PZ$, $\ttbar\PZ$, $\ttbar\PW$, triboson, and Higgs boson production, constitute the irreducible backgrounds.
Reducible backgrounds arise from SM processes, such as $\PZ$+jets or $\ttbar$+jets production, accompanied by additional leptons originating from heavy quark decays or from misidentification of jets. Such leptons arising not from boson decays, but from leptons inside or near jets, hadrons that reach the muon detectors, or hadronic showers with large electromagnetic energy fractions, are referred to as misidentified leptons.

The reducible backgrounds are significantly suppressed by applying a set of lepton isolation and displacement requirements in addition to the quality criteria in the lepton identification~\cite{Khachatryan:2015hwa,Sirunyan:2018fpa}.
The relative isolation is defined as the scalar \pt sum, normalized to the lepton \pt, of photon and hadron PF objects within a cone of $\Delta R $ around the lepton.
This relative isolation is required to be in the range of 5--15\% for $\Delta R=0.3$ for electrons, scaling inversely with the electron \pt, and to be less than 15\% for $\Delta R= 0.4$ for muons.
The isolation quantities are corrected for contributions from particles originating from pileup vertices.
In addition to the isolation requirement, electrons must satisfy $\abs{d_{z}}<0.1$\cm and $\abs{d_{xy}}<0.05$\cm in the ECAL barrel ($\abs{\eta}<1.479$), and $\abs{d_{z}}<0.2$\cm and $\abs{d_{xy}}<0.1$\cm in the ECAL endcap ($\abs{\eta}>1.479$), where $d_{z}$ and $d_{xy}$ are the longitudinal and transverse impact parameters of electrons with respect to the primary vertex, respectively.
Similarly, muons must satisfy $\abs{d_{z}}<0.1$\cm and $\abs{d_{xy}}<0.05$\cm.
All selected electrons within a cone of $\Delta R<0.05$ of a selected muon are discarded, since these are possibly due to bremsstrahlung from the muons.

In trilepton events, where misidentified-background contributions are dominant, additional 3-dimensional impact parameter significance and {\cPqb} tag veto requirements are imposed on the leptons, removing those with significant displacement with respect to the PV or whose matching jet is {\cPqb} tagged.
A PF jet with $\pt>10\GeV$ and $\abs{\eta}<2.5$ is considered to be matched if it is located within a cone of $\Delta R<0.4$ around the lepton without any further quality criteria on the jet.
These electron and muon reconstruction and selection requirements result in typical efficiencies of 40--90 and 75--95\%, respectively, depending on the lepton \pt and $\eta$~\cite{Khachatryan:2015hwa,Sirunyan:2018fpa}.

\section{Event selection}

In both data and simulated event samples, events satisfying the trigger criteria are required to pass additional offline selections.
Each event is required to have at least one electron with $\pt>35\GeV$ (30\GeV in 2016) or at least one muon with $\pt>26\GeV$ (29\GeV in 2017) to be consistent with the trigger thresholds, depending on the trigger used to collect the event.
Throughout this analysis, we consider events with exactly 3 leptons (3L) in one category and four or more leptons (4L) in another category.
In the 4L event category, only the 4 leading-\pt leptons are considered.
All events containing a lepton pair with $\Delta R<0.4$ or a same-flavor lepton pair with dilepton invariant mass below 12\GeV are removed to reduce background contributions from low-mass resonances as well as final-state radiation.
The 3L events containing an opposite-sign same-flavor (OSSF) lepton pair with the dilepton invariant mass below 76\GeV, when the trilepton invariant mass is within a $\PZ$ boson mass window ($91\pm15\GeV$), are also rejected.
This suppresses events from the $\PZ\to\ell\ell^{*}\to\ell\ell\gamma$ background process, where the photon converts into two additional leptons, one of which is lost.
The event selection criteria for both the type-III seesaw and $\ttphi$ signal models are orthogonal to those used in the estimation of SM backgrounds.

In the context of the type-III seesaw extension of the SM, pair production of heavy fermions gives rise to events with multiple energetic charged leptons or neutrinos in the final state.
Given the relatively high momenta of bosons and leptons originating from the decays of these heavy particles, kinematic quantities, such as the scalar \pt sum of all leptons, are instrumental in suppressing  SM  contributions.
This  is  especially  valid  for  decay  modes  such  as $\Sigma^{\pm}\to\ell^{\pm}{\PZ}\to\ell^{\pm}\ell^{\pm}\ell^{\mp}$, where all of the daughter particles of the heavy fermion can be reconstructed in the detector.
However, \ptmiss can be used as a complementary kinematic quantity in other decay modes, such as $\Sigma^0\to\nu \PH\to\nu{\PW}^{\pm}{\PW}^{\mp}$
or $\Sigma^\pm\to\nu \PW^{\pm} \to \nu \ell^{\pm} \nu$, where neutrinos can carry a significant fraction of the outgoing momentum.
We define $\LT$ as the scalar \pt sum of all charged leptons, and the quantity $\LT$+\ptmiss is chosen as the primary kinematic discriminant to select this variety of decay modes.

We classify the selected multilepton events into statistically independent search channels using the multiplicity of leptons, \nell, as well as the multiplicity and mass of distinct OSSF pairs, $\nossf$ and $\mossf$, respectively.
In cases of ambiguity, $\mossf$ is calculated using the OSSF pair with the mass closest to that of the $\PZ$ boson, considering both electrons and muons.
The 3L events with an OSSF lepton pair are labeled as OSSF1, whereas those without are labeled as OSSF0.
The OSSF1 events are further classified as on-$\PZ$, below-$\PZ$, and above-$\PZ$, based on the $\mossf$ relative to the $\pm15\GeV$ window around the $\PZ$ boson mass,
where the latter two categories are also collectively labeled as off-$\PZ$.
Similarly, the 4L events are classified as those with zero, one, and two distinct OSSF lepton pairs, OSSF0, OSSF1, and OSSF2, respectively.

In the 3L on-$\PZ$ search region, the sensitivity is increased by considering the transverse mass discriminant $\MTthree=\sqrt{\smash[b]{2\ptmiss\pt^{\lightlepton}[1-\cos(\Delta\phi_{\ptvecmiss,\ptvec^{\lightlepton}})]}}$ instead of $\LT$+\ptmiss, where $\lightlepton$ refers to the lepton that is not part of the on-$\PZ$ pair.
We reject 3L on-$\PZ$ events with $\ptmiss<100\GeV$, and 4L OSSF2 events with $\ptmiss<100\GeV$ and two distinct OSSF lepton pairs on-$\PZ$, as these are used in the estimation of SM backgrounds.

This event selection and binning scheme yields a total of 40 statistically independent search bins for the type-III seesaw model, as summarized in Table~\ref{tab:seesawSRs}.

\begin{table}
\centering
\topcaption{Multilepton signal region definitions for the type-III seesaw signal model.
All events containing a same-flavor lepton pair with invariant mass below 12\GeV are removed in the 3L and 4L event categories.
Furthermore, 3L events containing an OSSF lepton pair with mass below 76\GeV when the trilepton mass is within a $\PZ$ boson mass window ($91\pm15\GeV$) are also rejected.
The last $\LT$+\ptmiss or $\MTthree$ bin in each signal region contains the overflow events.}
\label{tab:seesawSRs}
\resizebox{\textwidth}{!}{
\begin{tabular}{l c r c c r l c }
\hline
Label                & \nell & $\nossf$ & $\mossf$ ({\GeVns}) & \ptmiss ({\GeVns})            & \multicolumn{2}{c}{Variable and range ({\GeVns})} &  Number of bins \\[0.5ex] \hline
3L below-$\PZ$         & ~~3       &  1~~      & $<$76         & \NA                    & ~~~~$\LT$+\ptmiss & $[0,1200]$ & \multicolumn{1}{c}{6} \\[0.5ex]
3L on-$\PZ$            & ~~3       &  1~~      & 76--106         & $>$100              & ~~~~$\MTthree$ & $[0,700]$ & \multicolumn{1}{c}{7}  \\[0.5ex]
3L above-$\PZ$         & ~~3       &  1~~      & $>$106          & \NA                  & ~~~~$\LT$+\ptmiss & $[0,1600]$ & \multicolumn{1}{c}{8} \\[0.5ex]
3L OSSF0           & ~~3       &  0~~      & \NA    & \NA                                & ~~~~$\LT$+\ptmiss & $[0,1200]$ & \multicolumn{1}{c}{6} \\[0.5ex]
4L OSSF0           & $\geq$4 &  0~~      & \NA      & \NA                                & ~~~~$\LT$+\ptmiss & $[0,600]$ & \multicolumn{1}{c}{2} \\[0.5ex]
4L OSSF1           & $\geq$4 &  1~~      & \NA      & \NA                                & ~~~~$\LT$+\ptmiss & $[0,1000]$ & \multicolumn{1}{c}{5} \\[0.5ex]
\multirow{2}{*}{4L OSSF2}           & \multirow{2}{*}{$\geq$4}      & \multirow{2}{*}{2~~}     & \multirow{2}{*}{\NA}        & $>$100 if both   & \multirow{2}{*}{~~~~$\LT$+\ptmiss} & \multirow{2}{*}{$[0,1200]$}  & \multicolumn{1}{c}{\multirow{2}{*}{6}} \\
 & & & &  pairs are on-$\PZ$ & & &  \\ \hline
\end{tabular}}
\end{table}

In contrast, the $\ttphi$ model yields events with a resonant OSSF lepton pair originating from the $\phi$ decays produced in association with a $\ttbar$ pair.
We consider only 3L or 4L events with at least one OSSF lepton pair and exclude those with $\mossf$ on-$\PZ$.
This event selection requires semileptonic or dileptonic $\ttbar$ decays in the $\ttphi$ signal.
Unlike the type-III seesaw heavy fermions, relatively light scalar or pseudoscalar decays do not necessarily produce energetic charged leptons,
but can yield striking resonant dilepton signatures in events with high hadronic activity and {\cPqb} tagged jets.
Therefore, we seek events with resonances in the OSSF dilepton mass spectra in various $\ST$ bins, where $\ST$ is defined as the scalar \pt sum of all jets, all charged leptons ($\LT$) and \ptmiss.

We probe the $\ttphi$ signal in light and heavy $\phi$ mass ranges, namely 15--75 and 108--340\GeV.
Signal masses below 15\GeV and in the range of 75--108\GeV are not considered because of background from low-mass quarkonia and $\PZ$ boson resonances, respectively. Masses above 340\GeV are not considered as the $\phi\to\ttbar$ decay channel becomes kinematically accessible here.

To account for the effects of radiation and resolution on the invariant mass reconstruction, we consider the 12--77\GeV (low) and 106--356\GeV (high) reconstructed dilepton mass ranges for the light and heavy signal mass scenarios, respectively, in both 3L and 4L channels.
Because there can be an ambiguity caused by additional leptons originating from the $\ttbar$ system, the reconstruction of the correct $\phi$ mass is not always possible.
Therefore, we define the $\mossf^{20}$ and the $\mossf^{300}$ variables as the OSSF lepton pair masses of a given lepton flavor closest to the targeted mass of 20 and 300\GeV, respectively. The $\mossf^{20}$ variable is used for the low dilepton mass range, while the $\mossf^{300}$ variable is used for the high dilepton mass range.
Events with a value of $\mossf^{20}$ ($\mossf^{300}$) outside the low (high) dilepton mass ranges are not considered.
The analysis is insensitive to the choice of the targeted mass value, and this simplified scheme allows multiple $\ttphi$ signal scenarios to be probed with a single mass spectrum.

The $\mossf^{20}$ and $\mossf^{300}$ masses are calculated separately for each lepton flavor scenario, yielding two nonorthogonal categories labeled as 3/4L($\Pe\Pe$) and 3/4L($\mu\mu$).
Hence, a given event can qualify for both the low and high dilepton mass regions, as well as for both lepton flavor channels.
For example, a $\mu^\pm\mu^\pm\mu^\mp$ event could be present in both low and high dilepton mass regions in the 3L($\mu\mu$) category, and similarly, an $\Pe^\pm\Pe^\mp$$\mu^\pm\mu^\mp$ event could qualify for both the 4L($\Pe\Pe$) and 4L($\mu\mu$) categories.
However, for any one given $\ttphi$ signal mass and flavor scenario, only one of the dilepton mass ranges of a single flavor category is considered.

Events that satisfy the low or high dilepton mass ranges are considered in orthogonal $\nbj=0$ (0B) and $\nbj\geq1$ (1B) selections, where $\nbj$ is the multiplicity of {\cPqb} tagged jets in an event.
Events in the 3L signal channels are further split into 3 $\ST$ bins (0--400\GeV, 400--800\GeV, and $\geq$800\GeV) for both $\nbj$ selections, those in the 4L signal channels are split into 2 $\ST$ (0--400\GeV and $\geq$400\GeV) bins for the 0B selection, and only one inclusive bin in $\ST$ is used for the 1B selection.

This event selection and binning scheme results in a total of 70 (68) statistically independent low (high) dilepton mass search bins in each of the 3/4L($\Pe\Pe$) and 3/4L($\mu\mu$) channels for the $\ttphi$ signal model, as summarized in Table~\ref{tab:ttphiSRs}.
The signal mass hypotheses that are closer to the mass bin boundaries than to the bin centers are probed with a modified binning scheme, where the mass bin boundaries are shifted by half the value of the bin widths.

\begin{table}
\centering
\topcaption{Multilepton signal region definitions for the $\ttphi$ signal model.
All events containing a same-flavor lepton pair with invariant mass below 12\GeV are removed in the 3L and 4L event categories.
Furthermore, 3L events containing an OSSF lepton pair with mass below 76\GeV when the trilepton mass is within a $\PZ$ boson mass window ($91\pm15\GeV$) are also rejected.}
\label{tab:ttphiSRs}
\resizebox{\textwidth}{!}{
\begin{tabular}{l r r r r r l l c c c}
\hline
Label                & \nell & $\nossf$ & $\mossf$ & $\nbj$        & \multicolumn{2}{c}{Variable and range ({\GeVns})}   & \multicolumn{4}{c}{Number of bins} \\[0.5ex] \hline
 & & & & & & & $\ST$ ({\GeVns}) &  0--400 & 400--800 &  $>$800~ \\ [\cmsTabSkip]
\multirow{2}{*}{3L(${\Pe\Pe}/\mu\mu$) 0B} & \multirow{2}{*}{3} & \multirow{2}{*}{1} & \multirow{2}{*}{off-$\PZ$} & \multirow{2}{*}{0}  & ~~~~~~~$\mossf^{20}$ & $[12,77]$ & & 13  & 13  & 5 \\[0.5ex]
  &       &        &          &   & ~~~~~~~$\mossf^{300}$ & $[106,356]$ & & 10  & 10  & 10 \\[0.5ex]
\multirow{2}{*}{3L(${\Pe\Pe}/\mu\mu$) 1B} & \multirow{2}{*}{3} & \multirow{2}{*}{1} & \multirow{2}{*}{off-$\PZ$} & \multirow{2}{*}{$\geq$1} & ~~~~~~~$\mossf^{20}$ & $[12,77]$ & & 13  & 13  & 5 \\[0.5ex]
&       &        &        &   & ~~~~~~~$\mossf^{300}$ & $[106,356]$ &  & 10  & 10  &10  \\  [\cmsTabSkip]
 & & & & & & & $\ST$ ({\GeVns}) &  0--400 & $>$400  \\ [\cmsTabSkip]
\multirow{2}{*}{4L(${\Pe\Pe}/\mu\mu$) 0B} & \multirow{2}{*}{$\geq$4} & \multirow{2}{*}{$\geq$1} & \multirow{2}{*}{off-$\PZ$} & \multirow{2}{*}{0} & ~~~~~~~$\mossf^{20}$ & $[12,77]$ & & 3  & 2   & \\[0.5ex]
  &       &        &         &   & ~~~~~~~$\mossf^{300}$ & $[106,356]$ & & 3  &  2  & \\  [\cmsTabSkip]
 & & & & & & & \multicolumn{2}{l}{$\ST$ inclusive}  &  \\ [\cmsTabSkip]
\multirow{2}{*}{4L(${\Pe\Pe}/\mu\mu$) 1B} & \multirow{2}{*}{$\geq$4} & \multirow{2}{*}{$\geq$1} & \multirow{2}{*}{off-$\PZ$} & \multirow{2}{*}{$\geq$1} & ~~~~~~~$\mossf^{20}$ & $[12,77]$ & & 3 & & \\[0.5ex]
  &       &        &         &   & ~~~~~~~$\mossf^{300}$ & $[106,356]$ &  & 3 & & \\[0.5ex] \hline
\end{tabular}}
\end{table}

\section{Background estimation and systematic uncertainties}\label{sec:backgroundEstimation}

The irreducible backgrounds are estimated using simulated event samples and are dominated by the $\PW\PZ$, $\PZ\PZ$, $\ttbar\PZ$, and $\PZ\gamma$ processes.
The event yields of these processes are obtained from theoretical predictions, with normalization corrections derived in dedicated control regions as described below. These estimates for the $\PW\PZ$, $\PZ\PZ$ and $\PZ\gamma$ processes are largely independent of each other. Since these backgrounds make significant contributions to the $\ttbar\PZ$-enriched control region, the normalization correction for this process is measured after the corresponding corrections have been obtained for the other backgrounds.
The normalization correction factors and their associated uncertainties, which include both statistical and systematic contributions, take the contamination of events from other processes into account and are applied to the corresponding background estimates in the signal regions.

\begin{figure}[!htp]
\centering
\includegraphics[width=.4\textwidth]{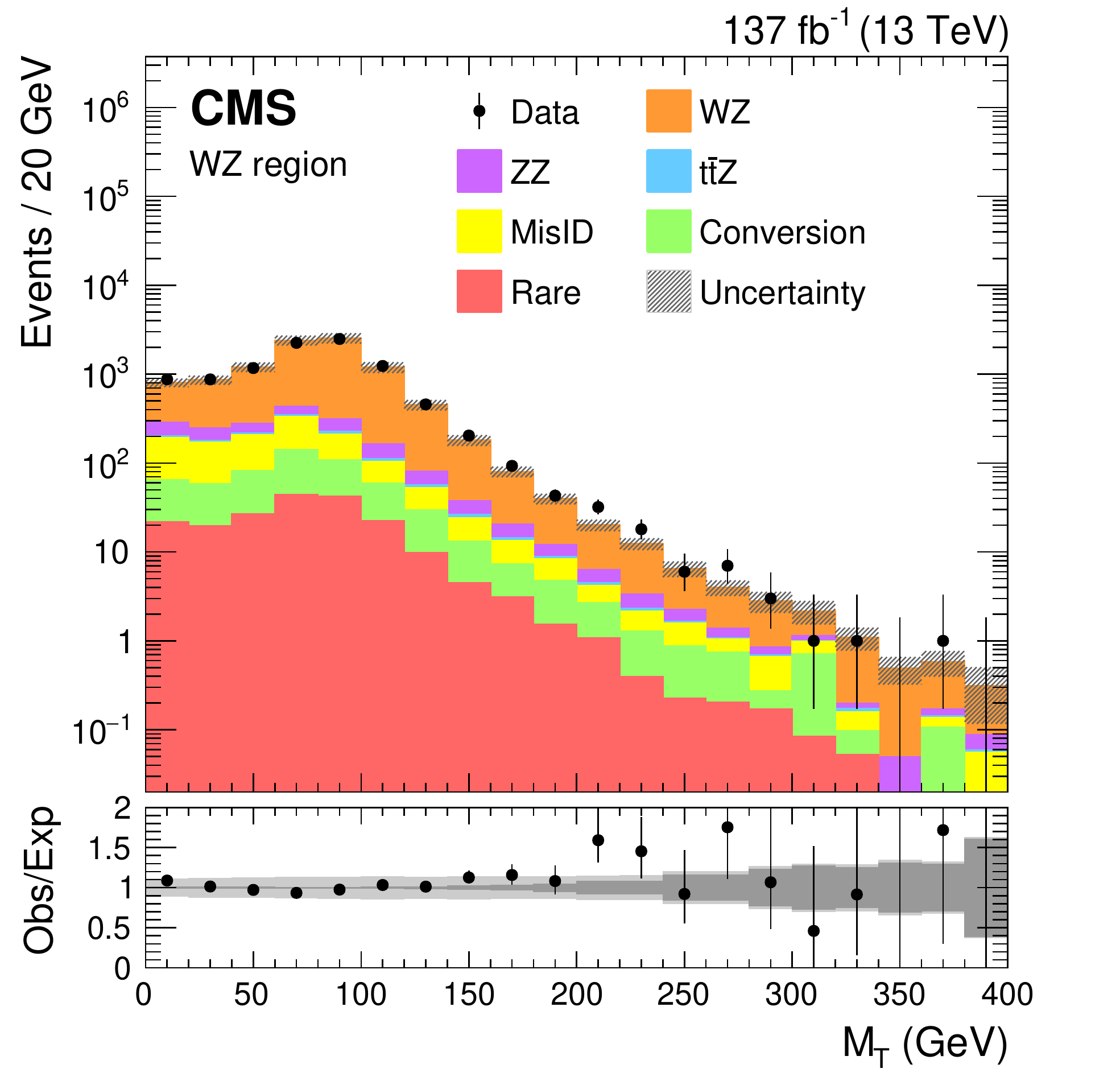}\hspace{.05\textwidth}
\includegraphics[width=.4\textwidth]{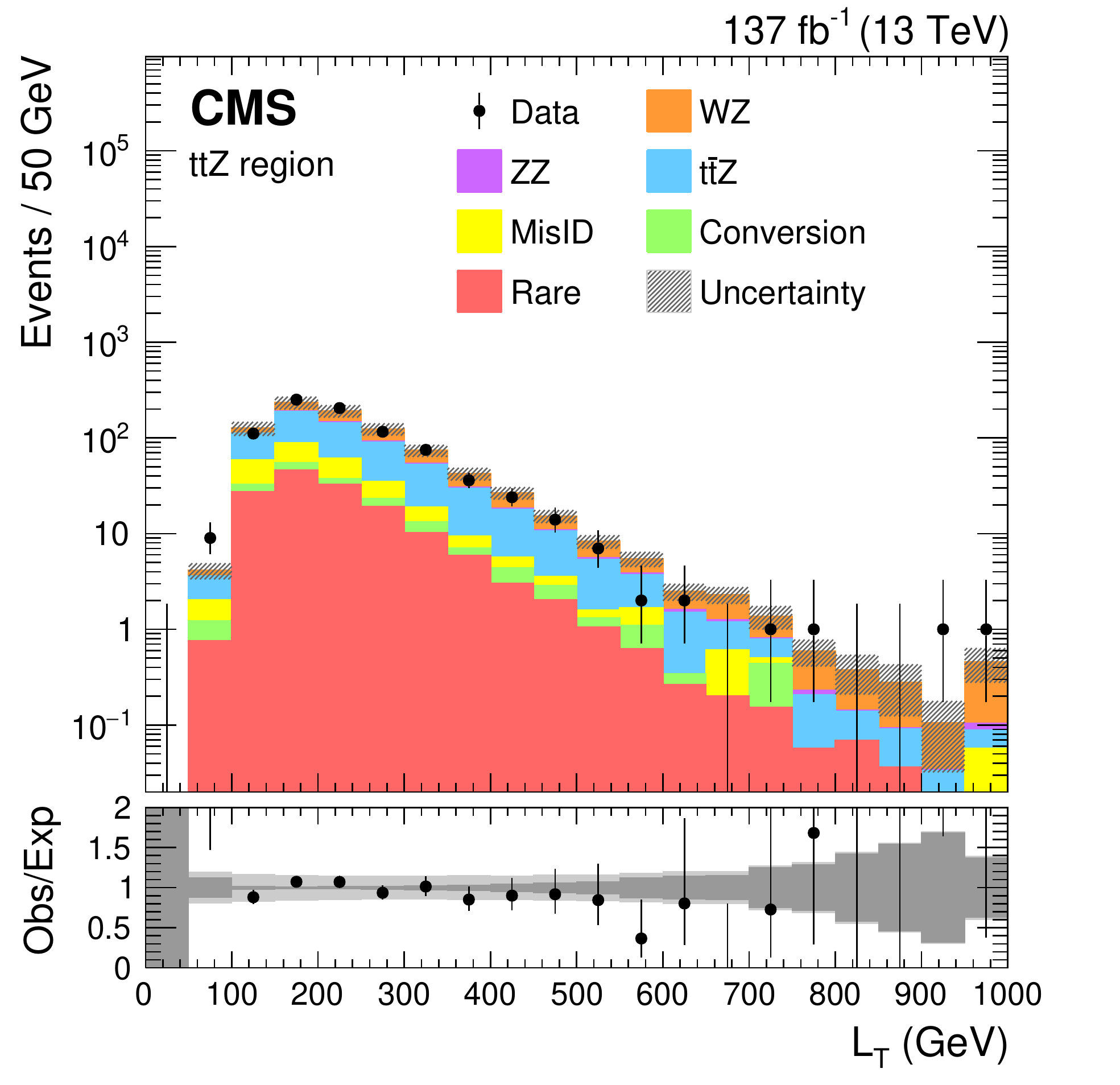}
\includegraphics[width=.4\textwidth]{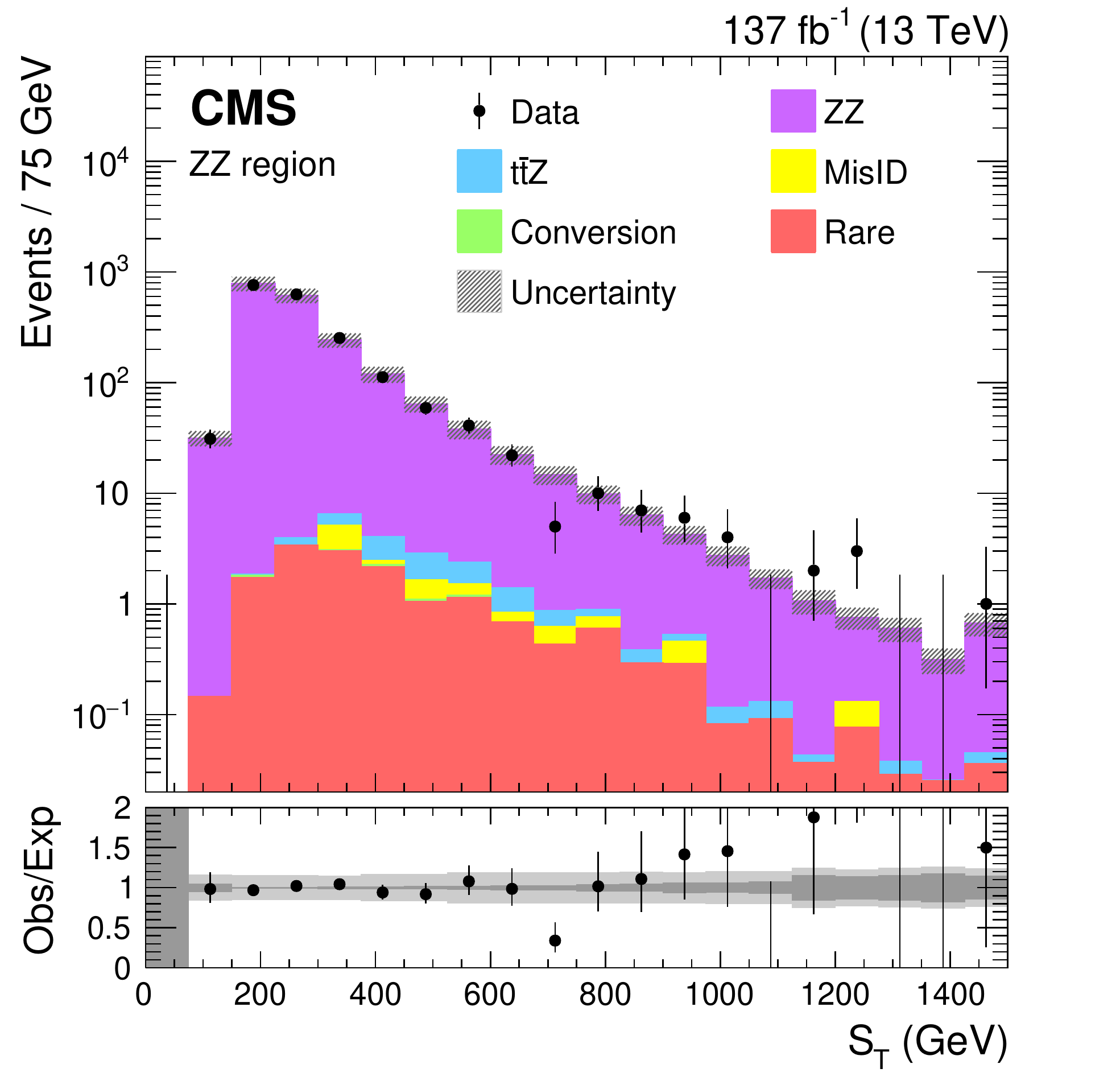} \hspace{.05\textwidth}
\includegraphics[width=.4\textwidth]{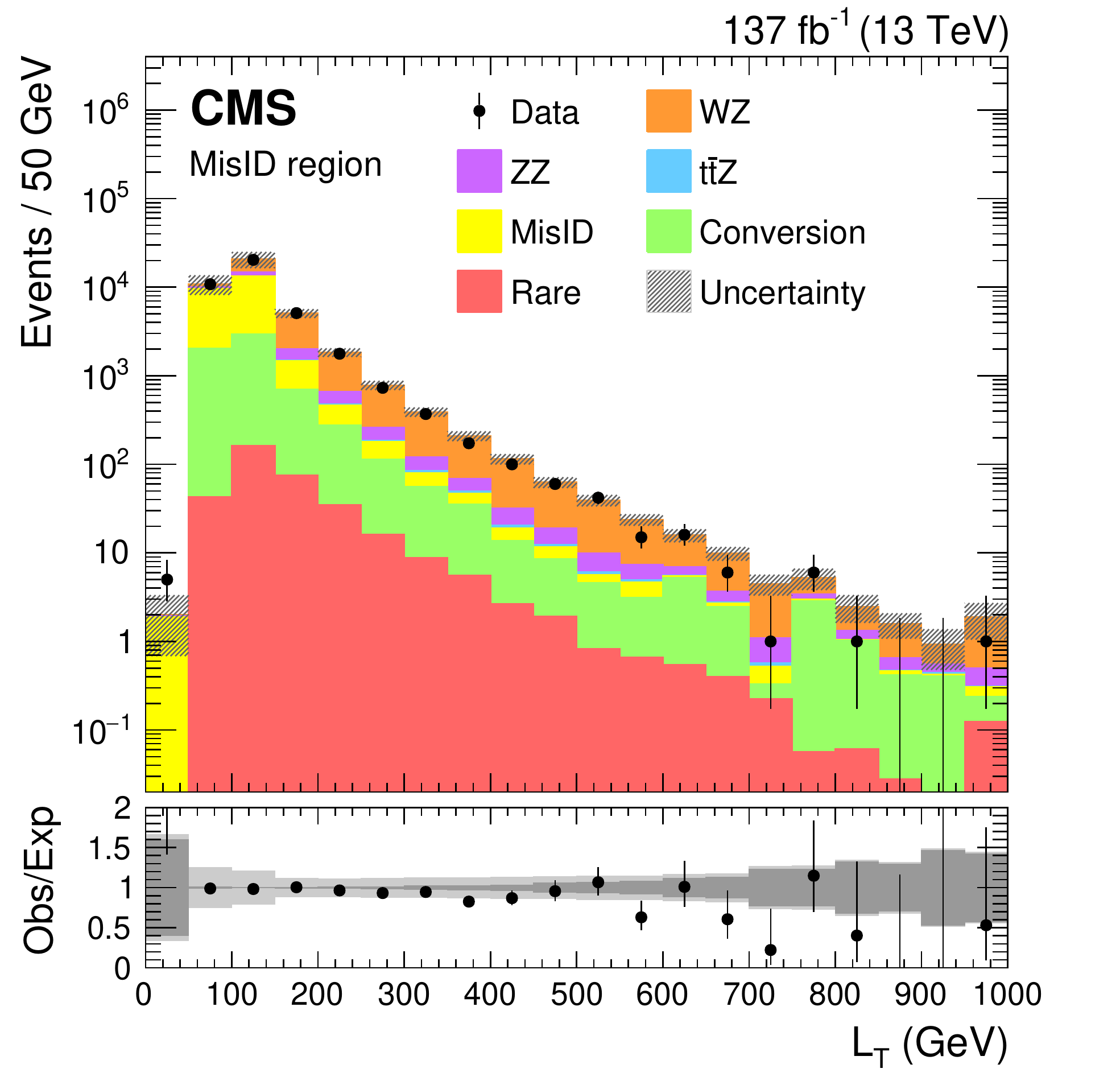}
\caption{The $\MTthree$ distribution in the $\PW\PZ$-enriched control region (upper left),
the $\LT$ distribution in the $\ttbar\PZ$-enriched control region (upper right),
the $\ST$ distribution in the $\PZ\PZ$-enriched control region (lower left),
and the $\LT$ distribution in the misidentified-lepton ($\PZ$+jets) enriched control region (lower right).
The lower panels show the ratio of observed to expected events.
The hatched gray bands in the upper panels and the light gray bands in the lower panels represent the total (systematic and statistical) uncertainty of the backgrounds in each bin, whereas the dark gray bands in the lower panels represent only the statistical uncertainty of the backgrounds.
The rightmost bins contain the overflow events in each distribution.
\label{fig:controlRegions}}
\end{figure}

For the $\PW\PZ$ and $\ttbar\PZ$ processes, we select events with exactly three leptons with an on-$\PZ$ OSSF pair, and the minimum lepton \pt is required to be above 20\GeV to increase the purity of these selections in the targeted process.
For the $\PW\PZ$-enriched selection, we require $50<\ptmiss<100\GeV$ and zero {\cPqb} tagged jets, whereas for the $\ttbar\PZ$-enriched selection we require $\ptmiss<100\GeV$, $\ST>350\GeV$, and at least one {\cPqb} tagged jet.
Similarly, for $\PZ\PZ$, we select events with exactly four leptons, $\ptmiss<100\GeV$, and two distinct on-$\PZ$ OSSF lepton pairs.
In the $\PW\PZ$- and $\PZ\PZ$-enriched selections, the simulated event yields are normalized to match those in the data in the 0--3 and 0--2 jet multiplicity bins including overflows, respectively, yielding normalization factor uncertainties in the range of 5--25\%, whereas an inclusive normalization is performed in the $\ttbar\PZ$-enriched selection, resulting in a $20\%$ uncertainty.
The various kinematic distributions in the $\PW\PZ$-, $\PZ\PZ$-, and $\ttbar\PZ$-enriched control regions, where the normalizations of these major irreducible backgrounds are performed, are illustrated in Fig.~\ref{fig:controlRegions} (top-left, top-right, bottom-left).

Similarly, a $\PZ\gamma$-enriched selection is created in three-lepton events with an OSSF lepton pair with mass below 76\GeV and trilepton mass within the $\PZ$ boson mass window, $91\pm15\GeV$.
This selection is dominated by $\PZ$+jets events with internal and external photon conversions originating from final-state radiation, and the normalization yields a relative uncertainty of $20\%$.
Conversion contributions from non-$\PZ\gamma$ processes play a subdominant role, and are estimated using simulated event samples.

Other irreducible backgrounds, such as $\ttbar\PW$, triboson, and Higgs boson processes, are estimated via simulation as well, using the cross sections obtained from the MC generation at NLO or higher accuracy, and are collectively referred to as `rare' backgrounds.
All rare and non-$\PZ\gamma$ conversion backgrounds, which are not normalized to data in dedicated control regions, are assigned a relative normalization uncertainty of $50\%$.

A small fraction of the irreducible backgrounds are due to misidentification of the charge of one or more prompt electrons. These backgrounds are also estimated using simulated event samples.
Following a study of same-sign dielectron events, in which the dielectron invariant mass is within a $\PZ$ boson mass window ($91\pm15\GeV$),
a relative uncertainty of $50\%$ is assigned to such contributions.
These constitute less than 35\% of the irreducible $\PW\PZ$, $\PZ\PZ$, and $\ttbar\PZ$ background contributions in the 3L~OSSF0, 4L~OSSF1, and 4L~OSSF0 signal regions, and are negligible in all other signal regions.

A category of systematic uncertainties in the simulated events is due to the corrections applied to background and signal simulation samples to account for differences with respect to data events.
These corrections are used in lepton reconstruction, identification, isolation, and trigger efficiencies, {\cPqb} tagging efficiencies, pileup modeling, as well as electron and muon resolution, and electron, muon, jet, and unclustered energy scale measurements.
The uncertainties due to such corrections typically correspond to a 1--10\% variation of the simulation-based irreducible background and signal yields across all signal regions. Therefore, they form a sub-dominant category of systematic uncertainties in the simulation-based background estimation.
Similarly, uncertainties due to choices of factorization and renormalization scales~\cite{Cacciari:2003fi} and PDFs~\cite{Ball:2017nwa} are also evaluated for signal and dominant irreducible background processes, yielding $<$10\% variation in signal regions.
The uncertainties in the integrated luminosity are in the range of 2.3--2.5\% in each year of data collection~\cite{CMS:2017sdi,CMS:2018elu,CMS:2019jhq}.

The reducible backgrounds are due to misidentified leptons (MisID) arising from events such as $\PZ$+jets and $\ttbar$+jets.
These are estimated using a three-dimensional implementation of a matrix method \cite{Khachatryan:2015bsa},
in which the rates at which prompt and misidentified leptons satisfying a loose lepton selection also pass a tight lepton selection are measured in dedicated signal-depleted selections of events in data.
The misidentification rates are measured in $\PZ$+jets and $\ttbar$+jets enriched trilepton (on-$\PZ$, $\ptmiss<50\GeV$) and same-sign dilepton (off-$\PZ$, $\ptmiss>50\GeV$, and with at least 3 jets) selections, respectively, whereas an on-$\PZ$ dilepton selection is used for the prompt rates.
The rates are parametrized as a function of lepton kinematic distributions and the multiplicity of tracks in the event.
A weighted average of these misidentification rates is used in the analysis, reflecting the approximate expected composition of the SM backgrounds in a given search region as obtained from simulated event samples.
The final uncertainty in the estimated background from misidentified leptons is obtained by varying the rates within the uncertainties as well as the differences in rates in $\PZ$+jets and $\ttbar$+jets events, and has a relative uncertainty of 30--40\%.
Figure~\ref{fig:controlRegions} (lower right) illustrates the misidentified-lepton background estimate as a function of $\LT$ in the trilepton selection used to measure the rates, where a misidentified lepton is produced in association with a $\PZ$ boson.

A summary of the uncertainty sources in this analysis, including the typical resultant variations on relevant background and signal processes, as well as the correlation model across the three different data taking periods, is given in Table~\ref{tab:systematics}. 
The quoted variations on affected processes, except those in the integrated luminosity, and the inclusive normalizations of the $\ttbar\PZ$, conversion and rare simulations, are calculated taking into account variations of the uncertainty sources as a function of object and event dependent parameters as appropriate, such as lepton momenta, or jet multiplicity. Thus, these uncertainties also include bin-to-bin correlations across the search regions.
The overall uncertainties in the total expected backgrounds are largely dominated by those in the irreducible $\PW\PZ$, $\PZ\PZ$, and $\ttbar\PZ$ processes, as well as the misidentified-lepton contributions, whereas the relatively large uncertainties in rare and conversion contributions and those due to electron charge misidentification are subdominant and have a negligible effect on the results across different signal regions.

\begin{table}
\centering
\topcaption{Sources of systematic uncertainties, affected background and signal processes, relative variations of the affected processes, and presence or otherwise of correlation between years in signal regions.}
\label{tab:systematics}
\resizebox{\textwidth}{!}{
\begin{tabular}{l l c c}
\hline
Uncertainty source                    & Signal/Background process & Variation (\%)    & Correlation \\ \hline
Integrated Luminosity                            & Signal/Rare/Non-$\PZ\gamma$ conversion  & 2.3--2.5 & No  \\
Lepton reconstruction, identification,& \multirow{2}{*}{Signal/Background$^{\star}$} & \multirow{2}{*}{4--5}  & \multirow{2}{*}{No}  \\
           and isolation efficiency   &                                             &        & \\
Lepton displacement efficiency (only in 3L)   & Signal/Background$^{\star}$ & 3--5     & Yes \\
Trigger efficiency                    & Signal/Background$^{\star}$ & $<$3    & No  \\
{\cPqb} tagging efficiency                      & Signal/Background$^{\star}$ & $<$5    & No  \\
Pileup modeling                       & Signal/Background$^{\star}$ & $<$3    & Yes \\
Factorization/renormalization scales $\&$ PDF & Signal/Background$^{\star}$ & $<$10   & Yes \\
Jet energy scale                      & Signal/Background$^{\star}$ & $<$5    & Yes \\
Unclustered energy scale              & Signal/Background$^{\star}$ & $<$5    & Yes \\
Muon energy scale and resolution      & Signal/Background$^{\star}$ & $<$5    & Yes \\
Electron energy scale and resolution  & Signal/Background$^{\star}$ & $<$2    & Yes \\
[\cmsTabSkip]
$\PW\PZ$ normalization (0/1/2/$\geq$3 jets) & $\PW\PZ$                      & 5--10    & Yes \\
$\PZ\PZ$ normalization (0/1/$\geq$2 jets)   & $\PZ\PZ$                      & 5--10    & Yes \\
$\ttbar\PZ$ normalization                 & $\ttbar\PZ$                 & 15--20   & Yes \\
Conversion normalization              & Conversion              & 20--50   & Yes \\
Rare normalization                    & Rare                    & 50      & Yes \\
Lepton misidentification rates        & Misidentified lepton    & 30--40   & Yes \\
Electron charge misidentification     & $\PW\PZ$/$\PZ\PZ$$^{\dagger}$         & $<$20   & No  \\[\cmsTabSkip]
&\multicolumn{3}{l}{$^{\star} \PW\PZ,~\PZ\PZ,~\ttbar\PZ$, rare, and conversion background processes.} \\
&\multicolumn{3}{l}{$^{\dagger}$Only in 3L OSSF0, 4L OSSF0, and 4L OSSF1 signal regions.} \\  \hline
\end{tabular}}
\end{table}

\section{Results}

\begin{figure}[!htp]
\centering
\includegraphics[width=.4\textwidth]{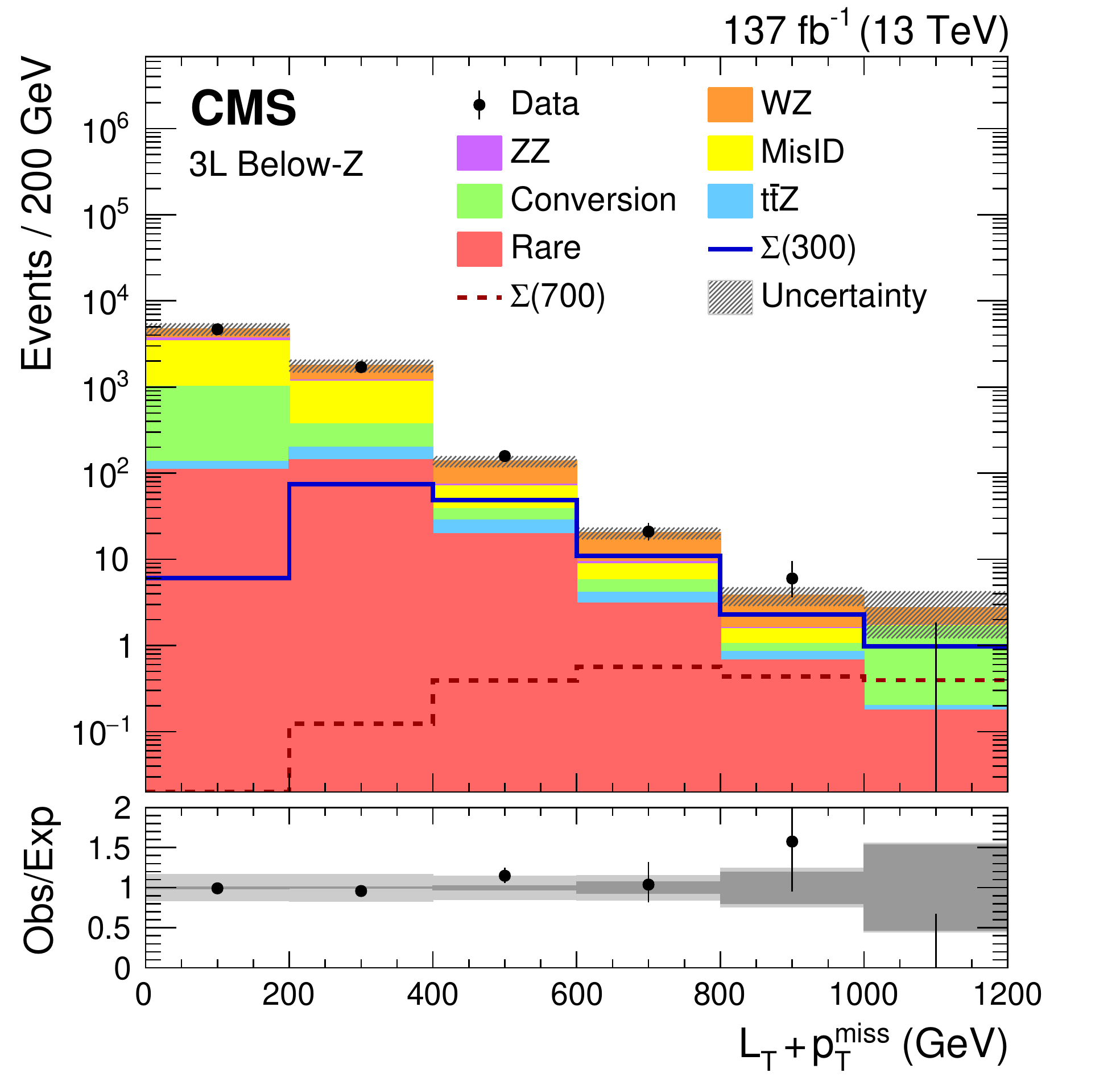} \hspace{.05\textwidth}
\includegraphics[width=.4\textwidth]{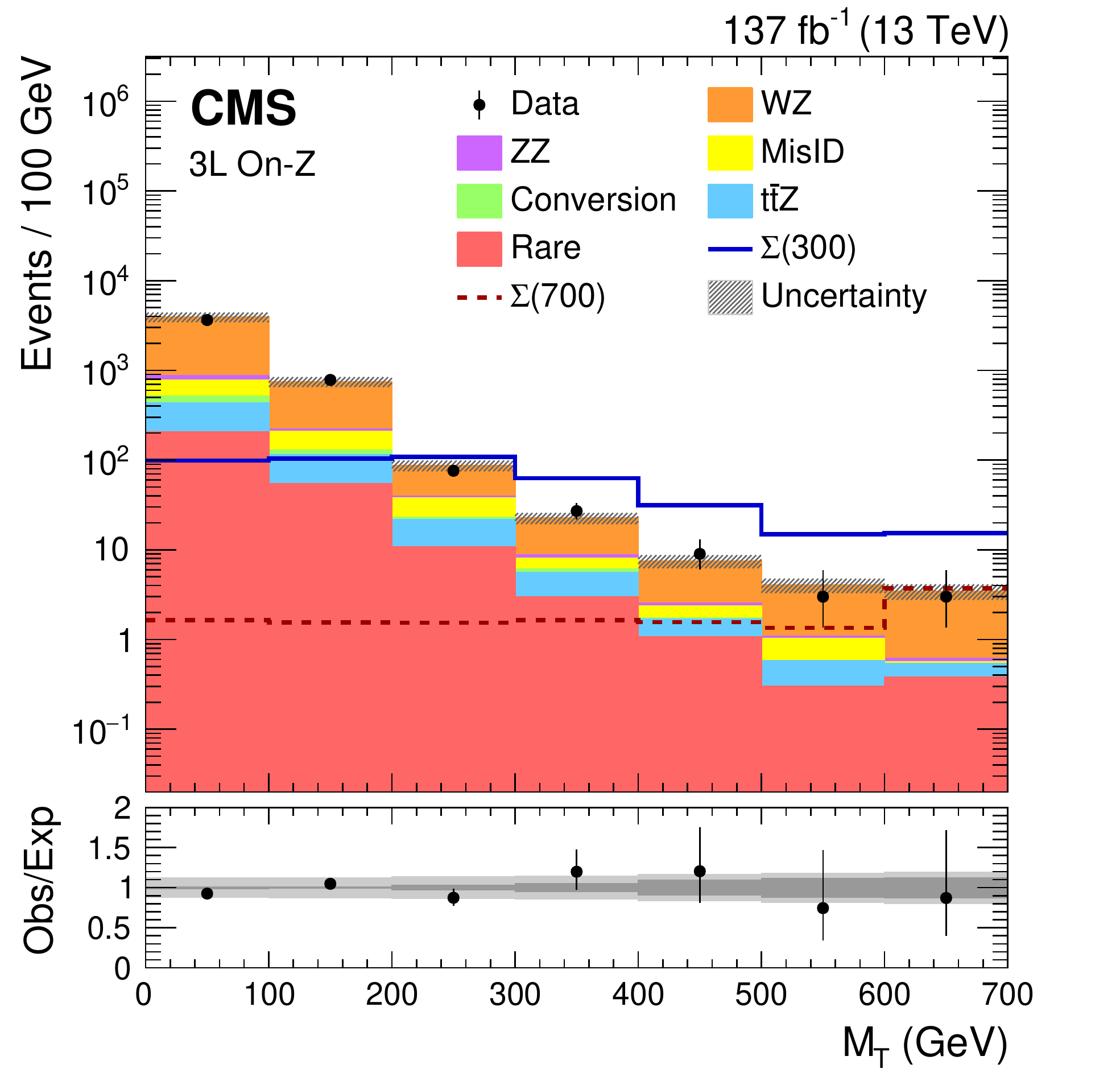}
\includegraphics[width=.4\textwidth]{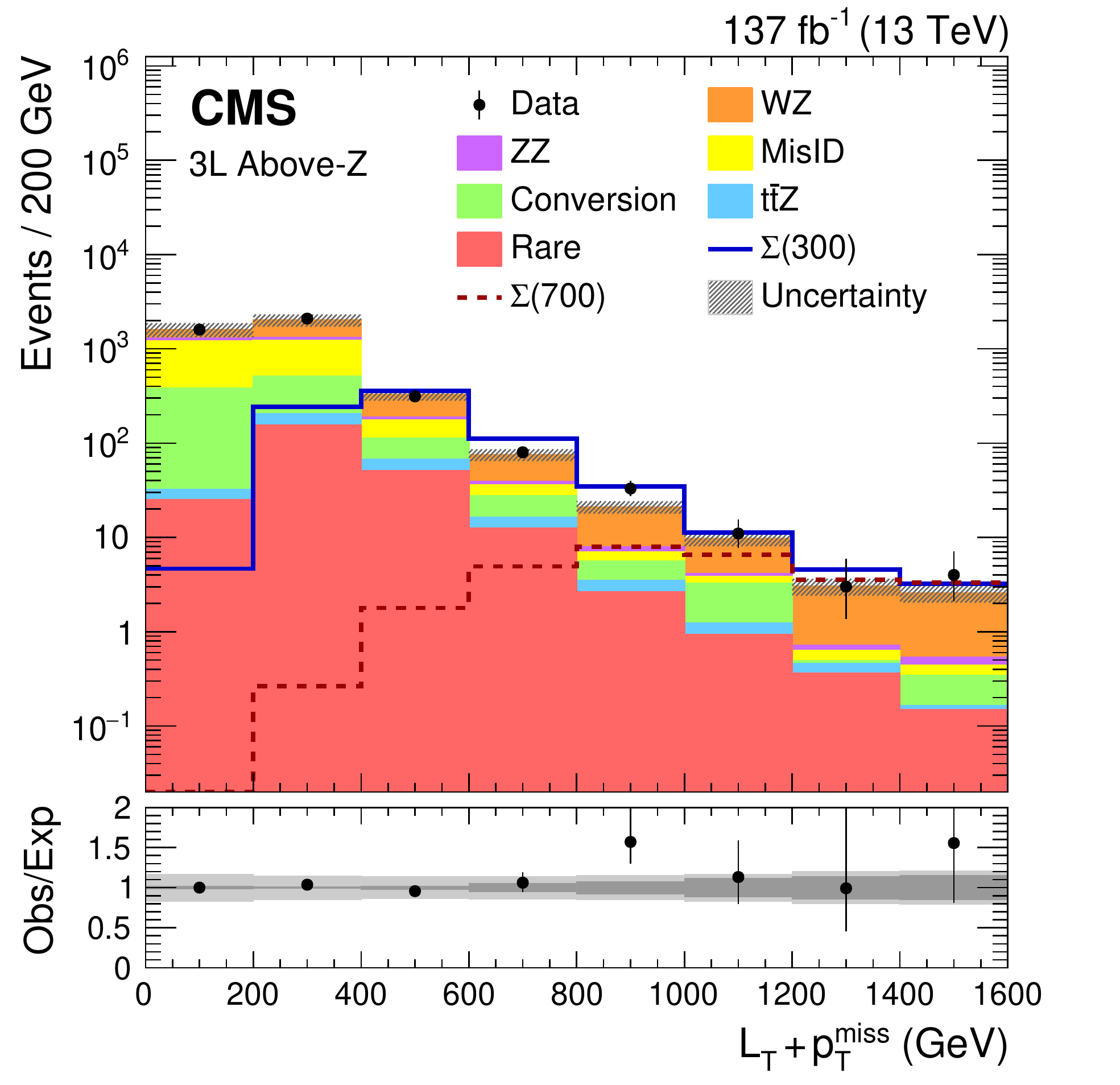} \hspace{.05\textwidth}
\includegraphics[width=.4\textwidth]{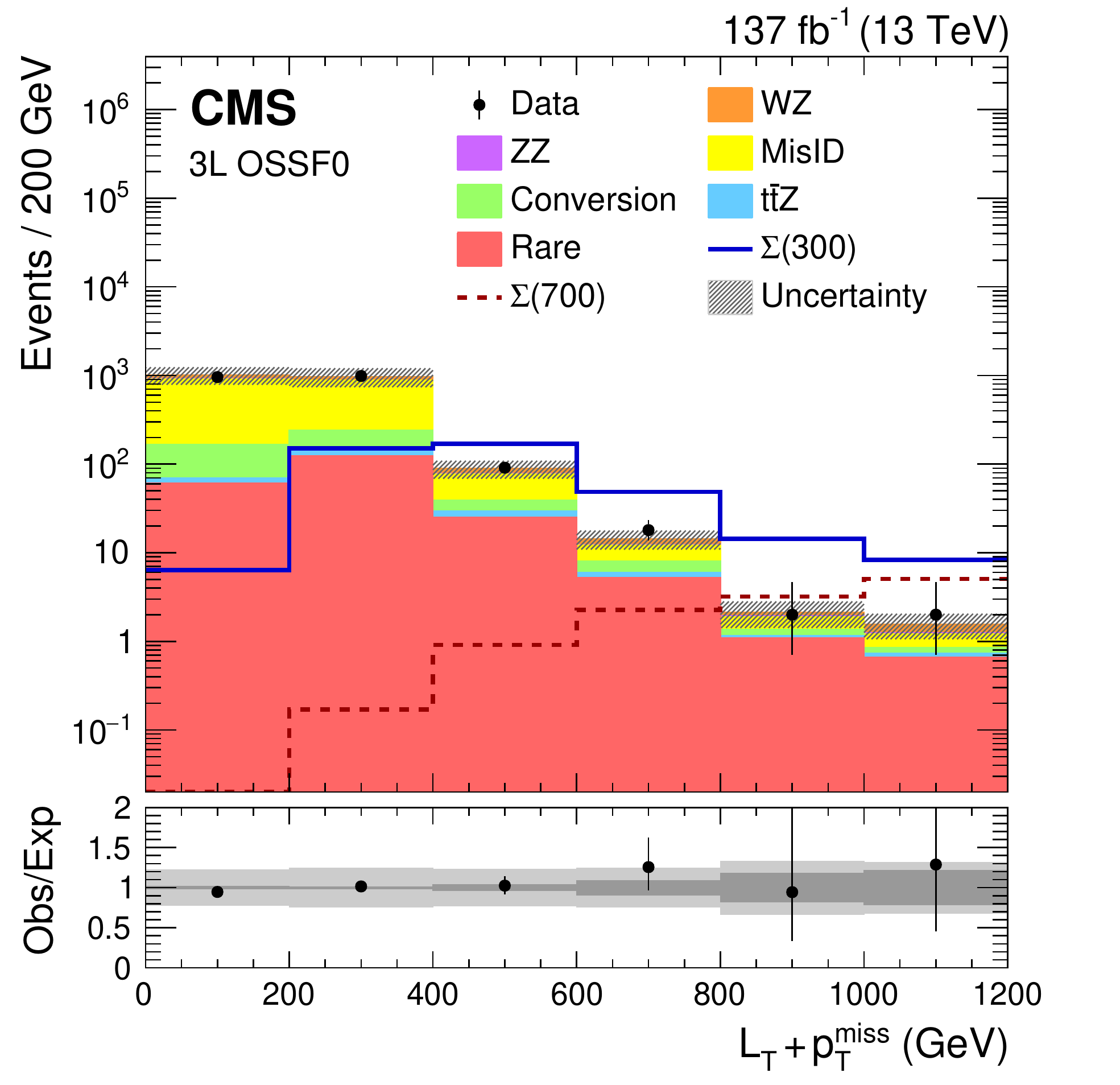}
\caption{
Type-III seesaw signal regions in 3L below-$\PZ$ (upper left), on-$\PZ$ (upper right), above-$\PZ$ (lower left), and OSSF0 (lower right) events.
The total SM background is shown as a stacked histogram of all contributing processes.
The predictions for type-III seesaw models with $\Sigma$ masses of 300 and 700\GeV in the flavor-democratic scenario are also shown.
The lower panels show the ratio of observed to expected events.
The hatched gray bands in the upper panels and the light gray bands in the lower panels represent the total (systematic and statistical) uncertainty of the backgrounds in each bin, whereas the dark gray bands in the lower panels represent only the statistical uncertainty of the backgrounds.
The rightmost bins contain the overflow events in each distribution.
\label{fig:Seesaw3LSR}}
\end{figure}

\begin{figure}[!htp]
\centering
\includegraphics[width=.4\textwidth]{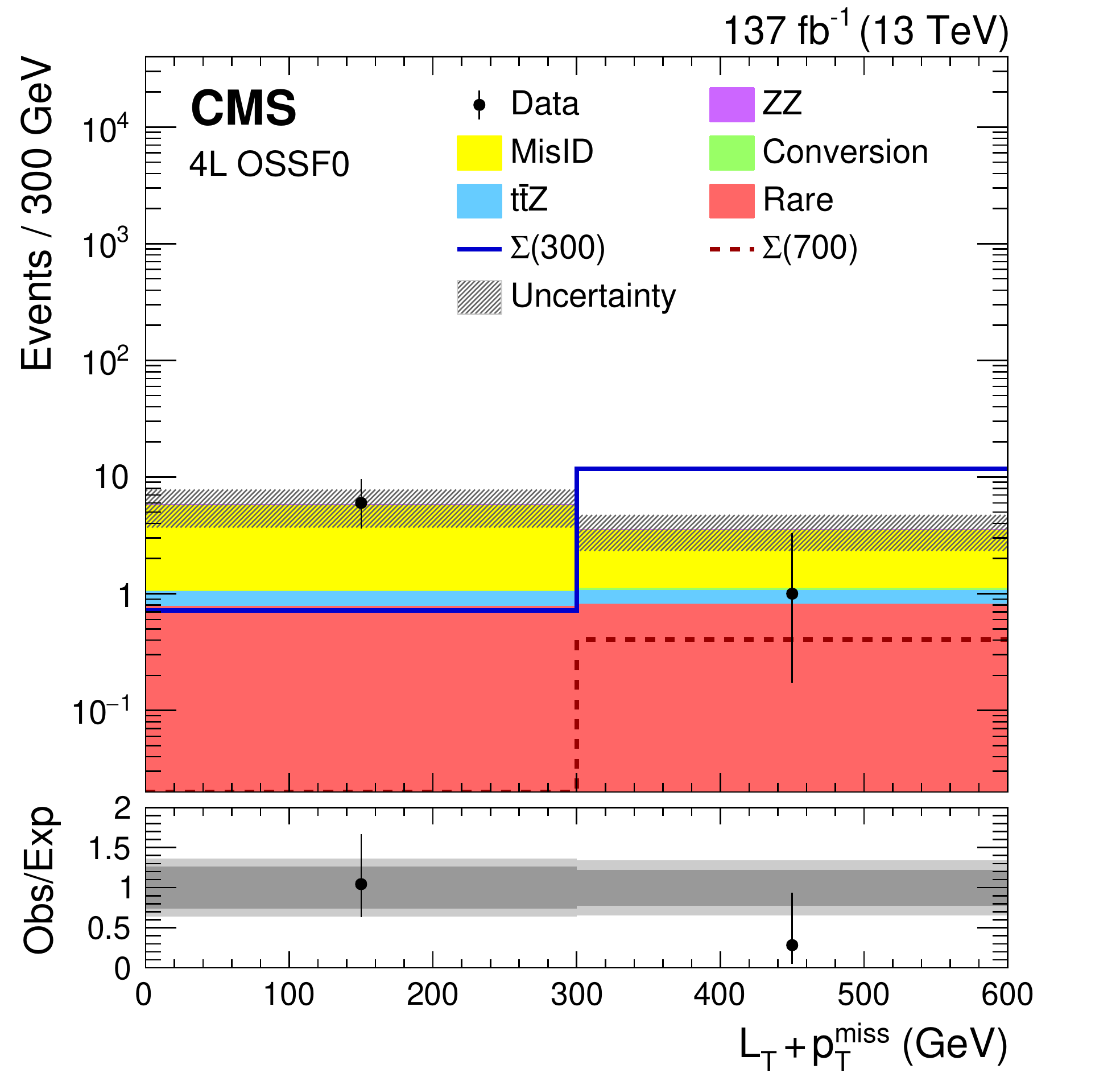} \hspace{.05\textwidth}
\includegraphics[width=.4\textwidth]{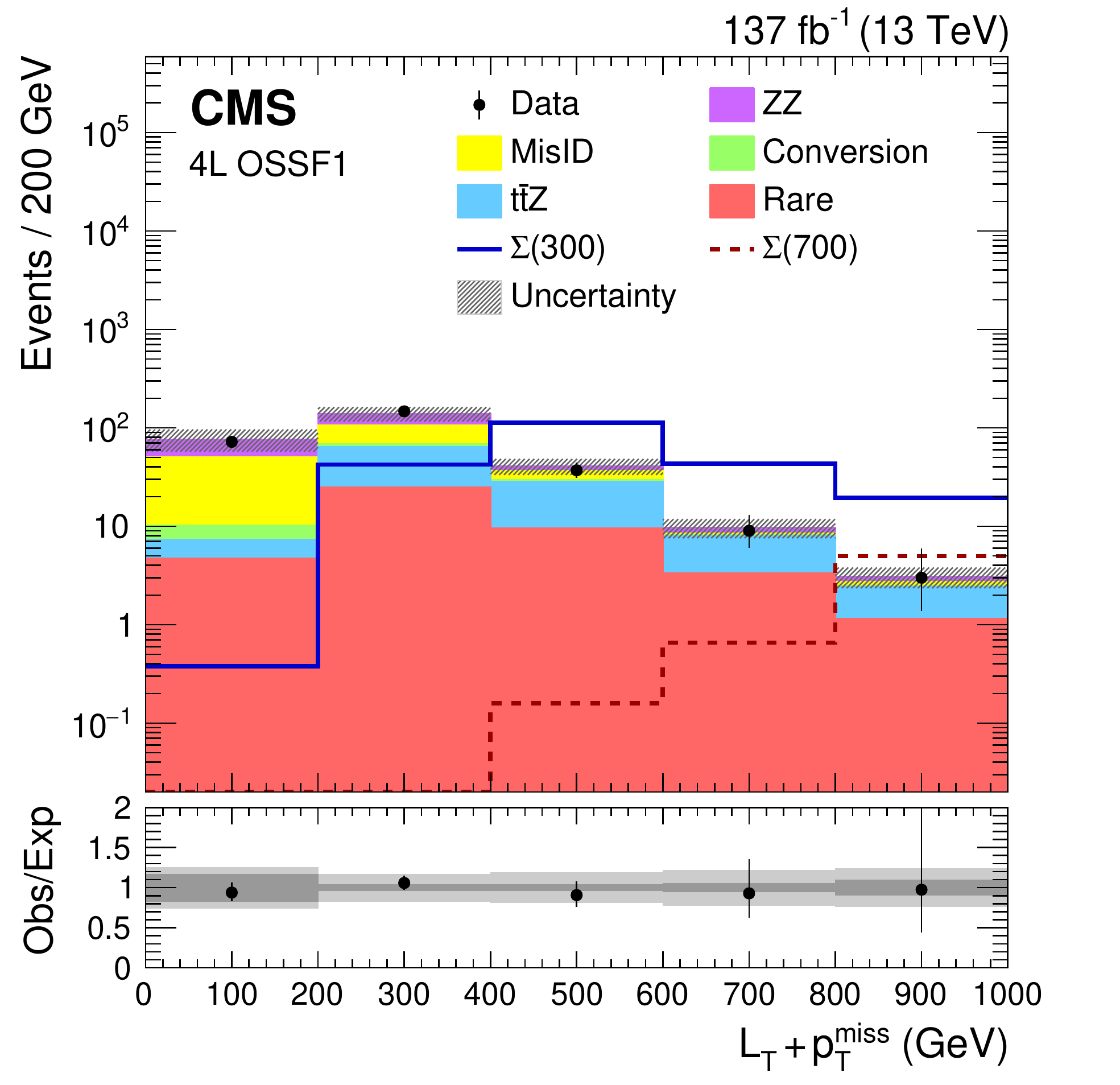}
\includegraphics[width=.4\textwidth]{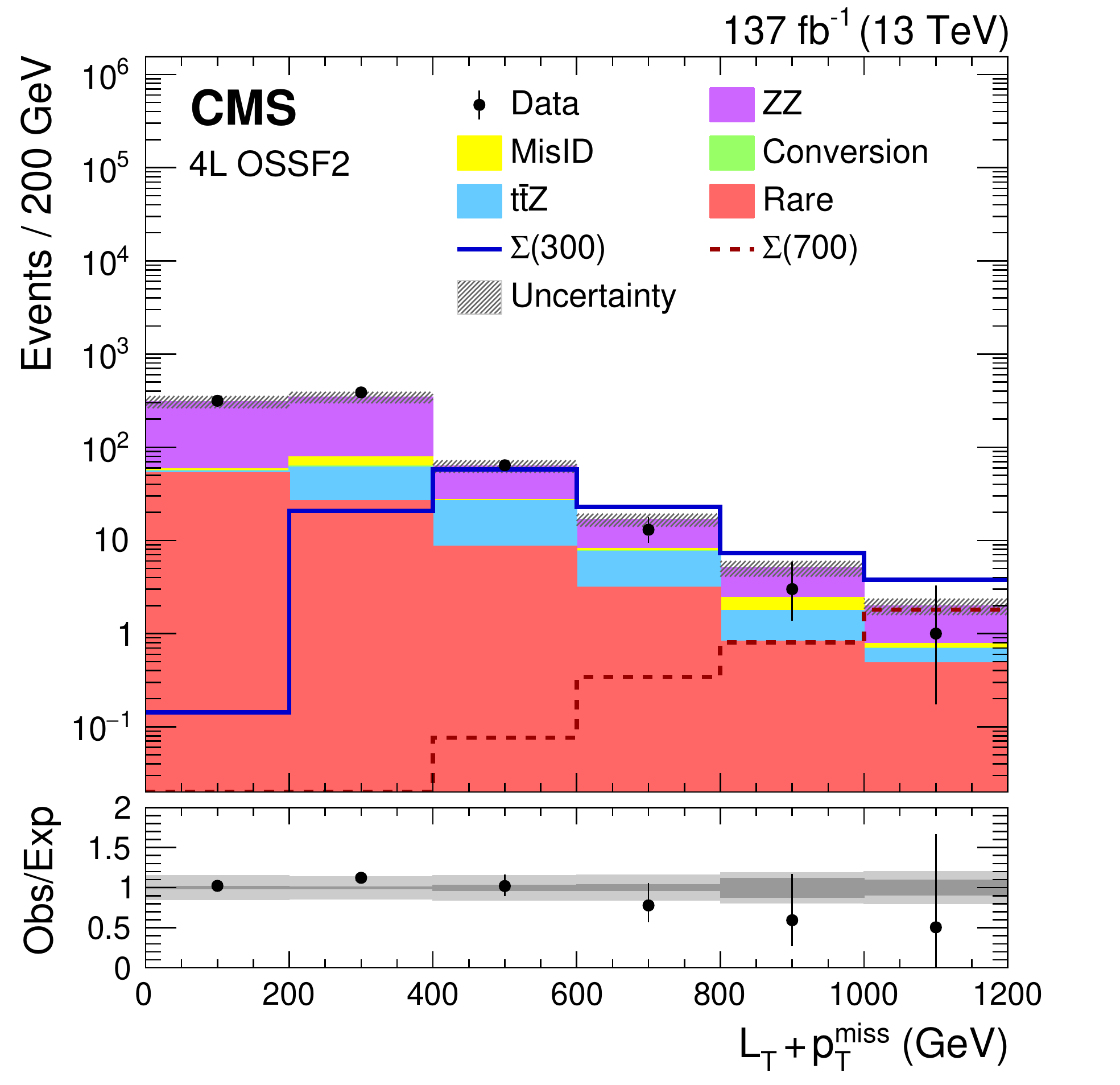}
\caption{
Type-III seesaw signal regions in 4L OSSF0 (upper left), OSSF1 (upper right), and OSSF2 (lower) events.
The total SM background is shown as a stacked histogram of all contributing processes.
The predictions for type-III seesaw models with $\Sigma$ masses of 300 and 700\GeV in the flavor-democratic scenario are also shown.
The lower panels show the ratio of observed to expected events.
The hatched gray bands in the upper panels and the light gray bands in the lower panels represent the total (systematic and statistical) uncertainty of the backgrounds in each bin, whereas the dark gray bands in the lower panels represent only the statistical uncertainty of the backgrounds.
The rightmost bins contain the overflow events in each distribution.
\label{fig:Seesaw4LSR}}
\end{figure}

\begin{figure}[!htp]
\centering
\includegraphics[width=.4\textwidth]{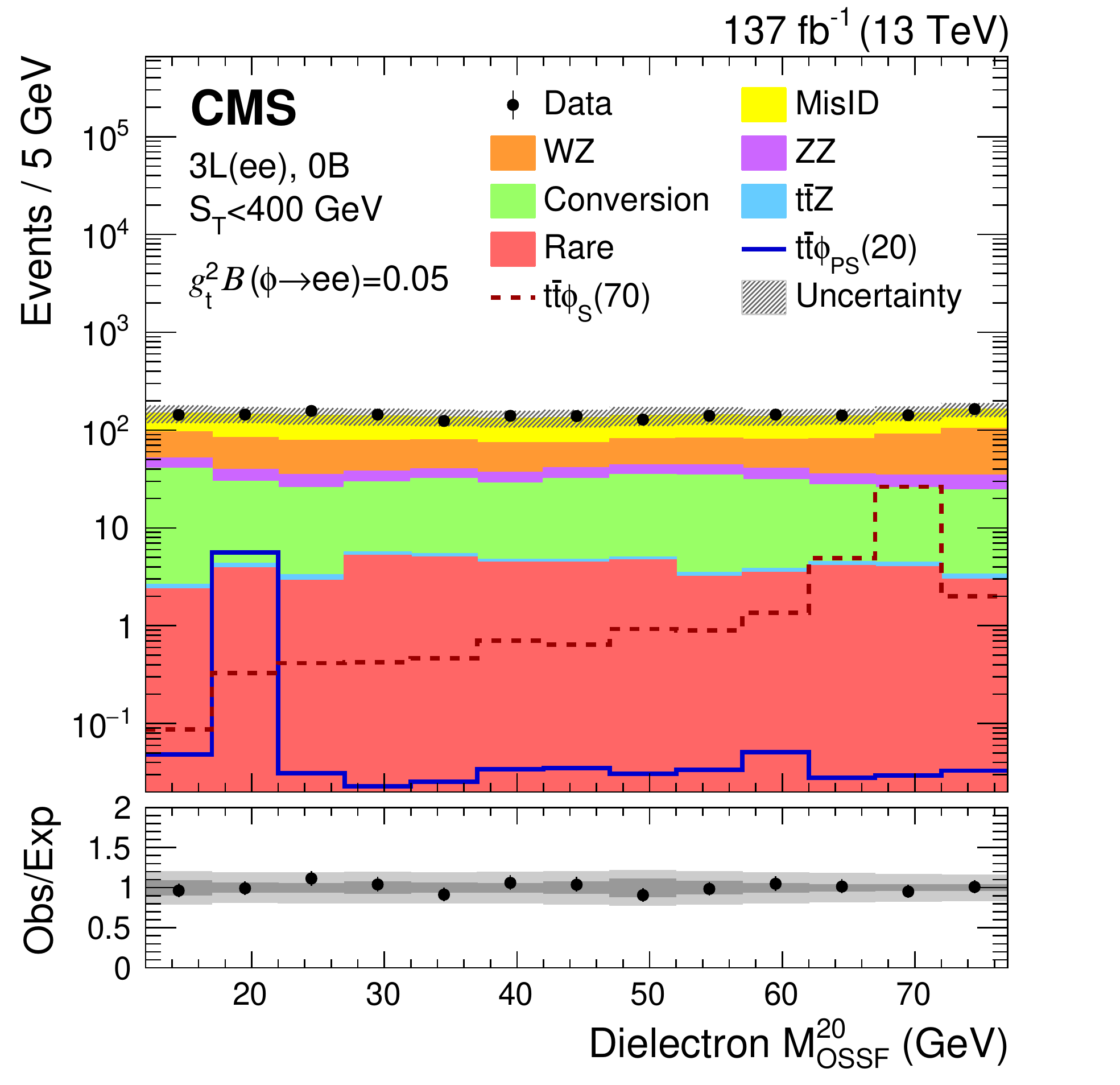} \hspace{.05\textwidth}
\includegraphics[width=.4\textwidth]{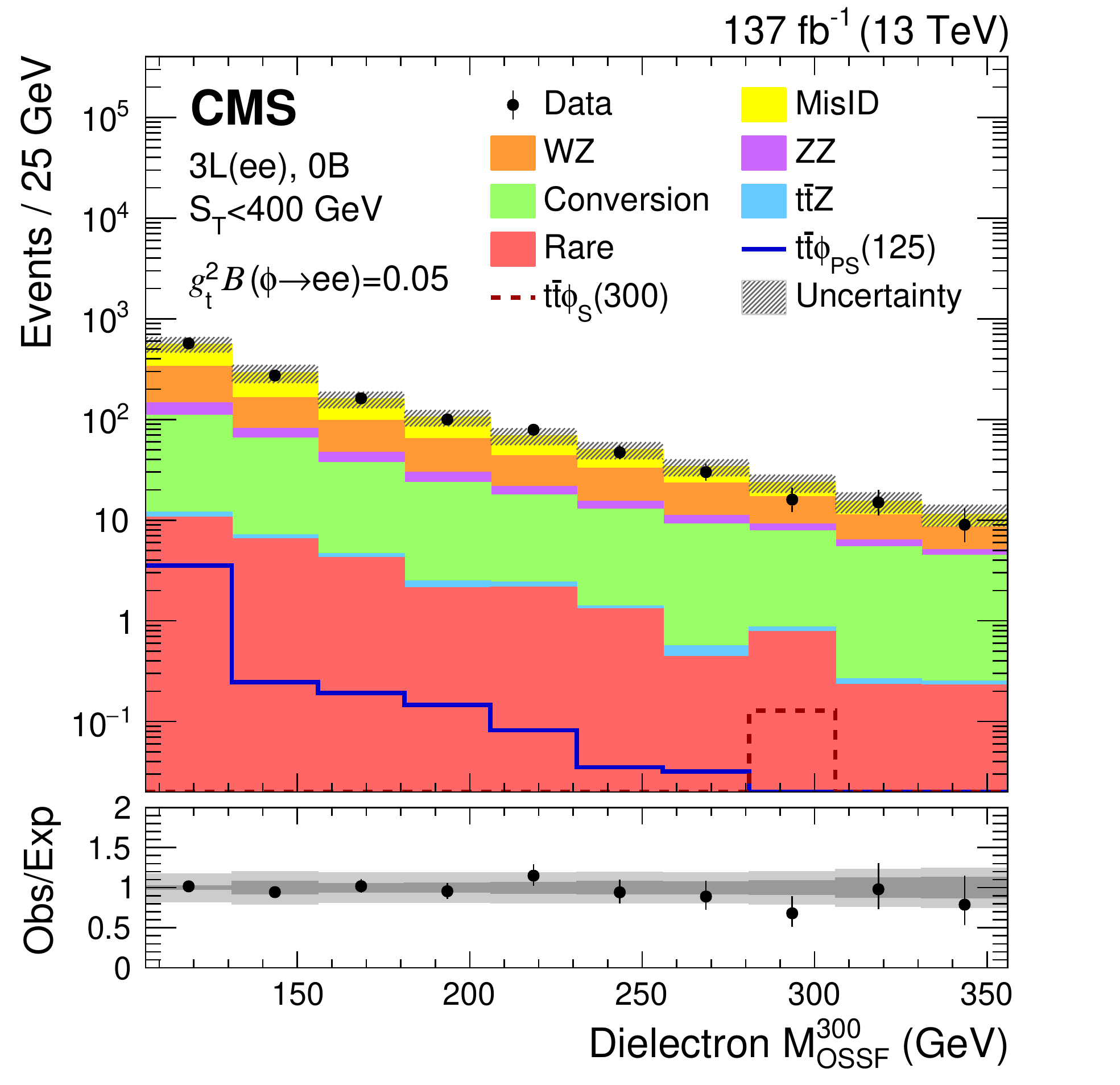}
\includegraphics[width=.4\textwidth]{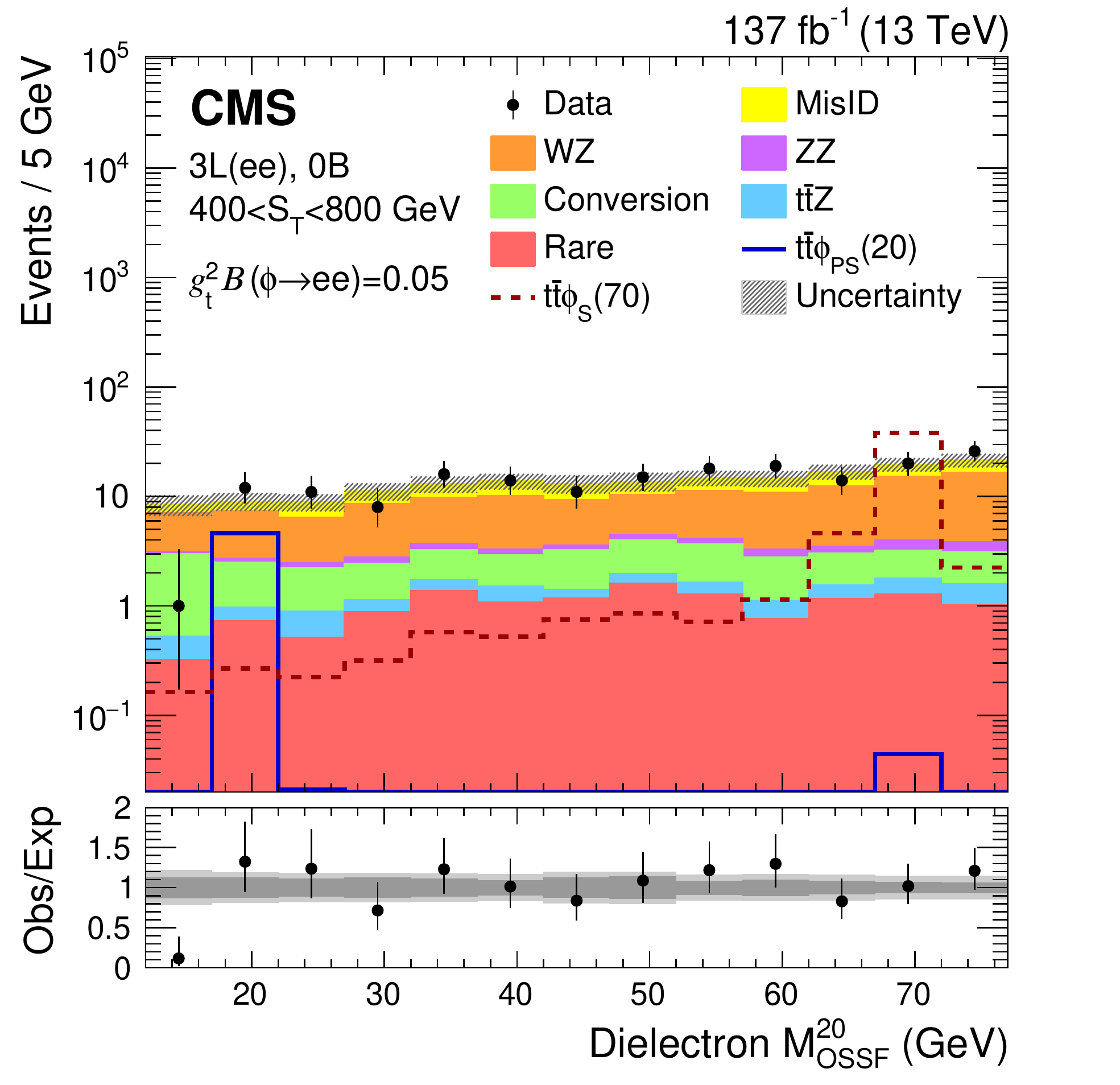} \hspace{.05\textwidth}
\includegraphics[width=.4\textwidth]{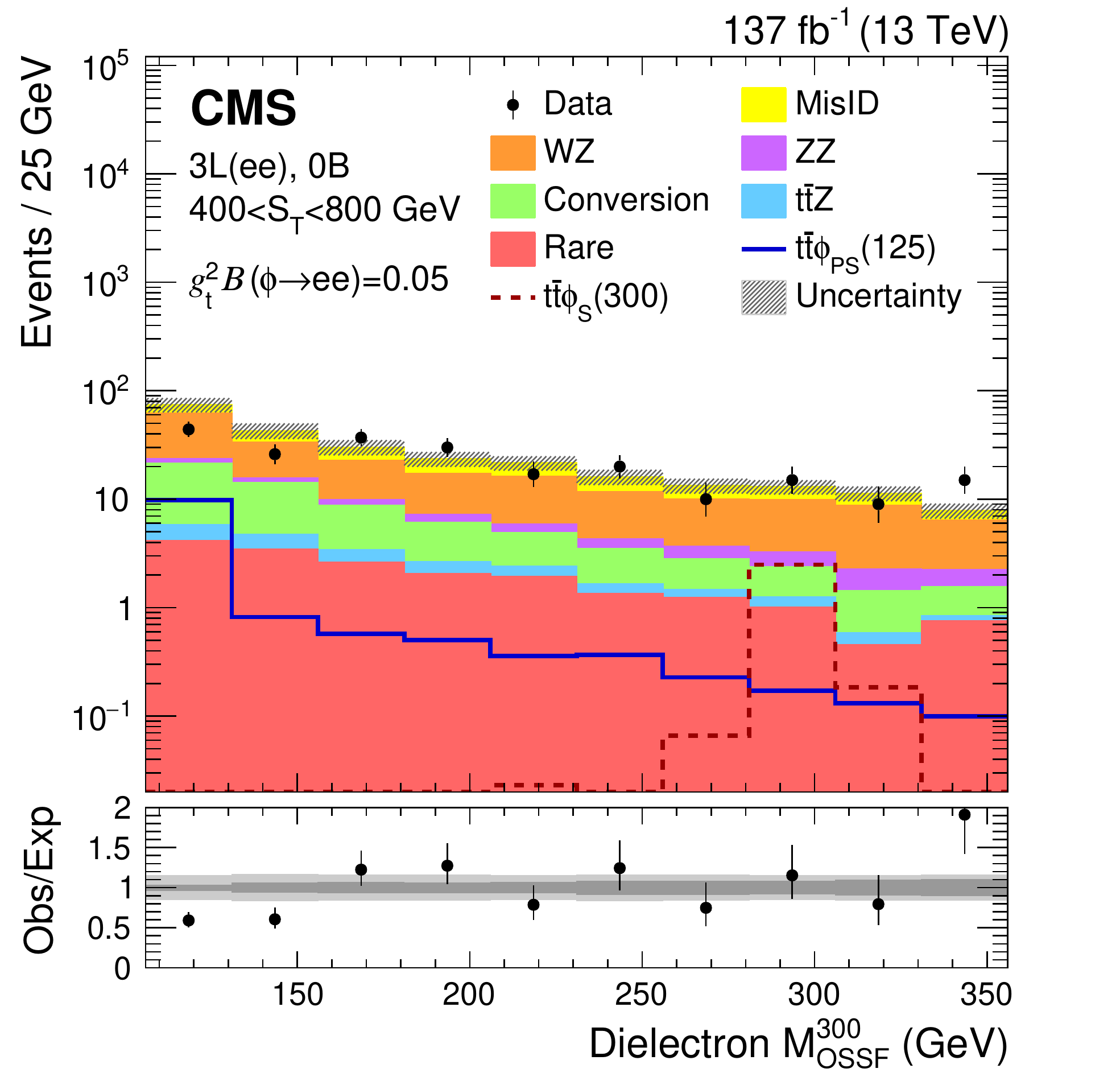}
\includegraphics[width=.4\textwidth]{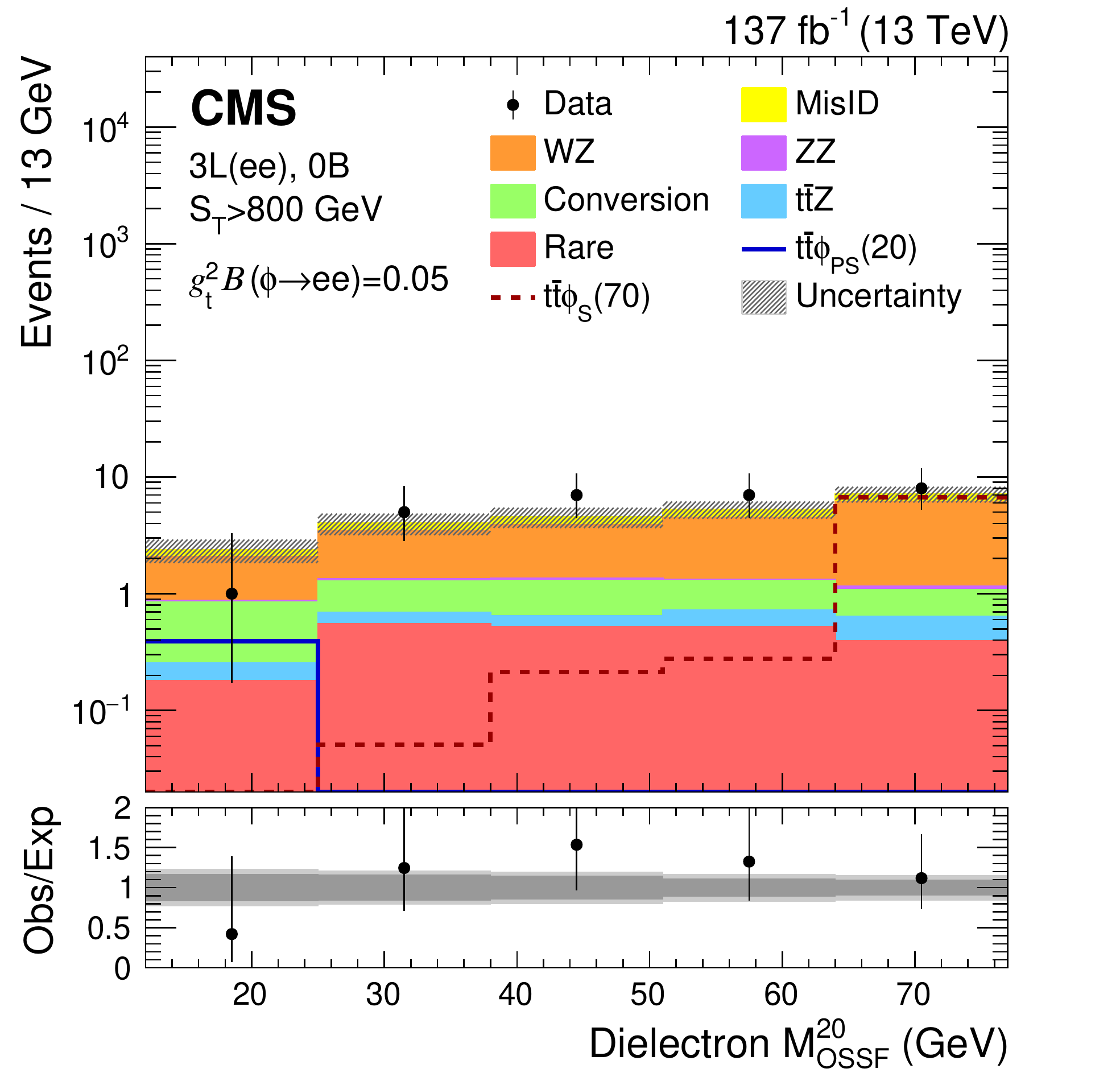} \hspace{.05\textwidth}
\includegraphics[width=.4\textwidth]{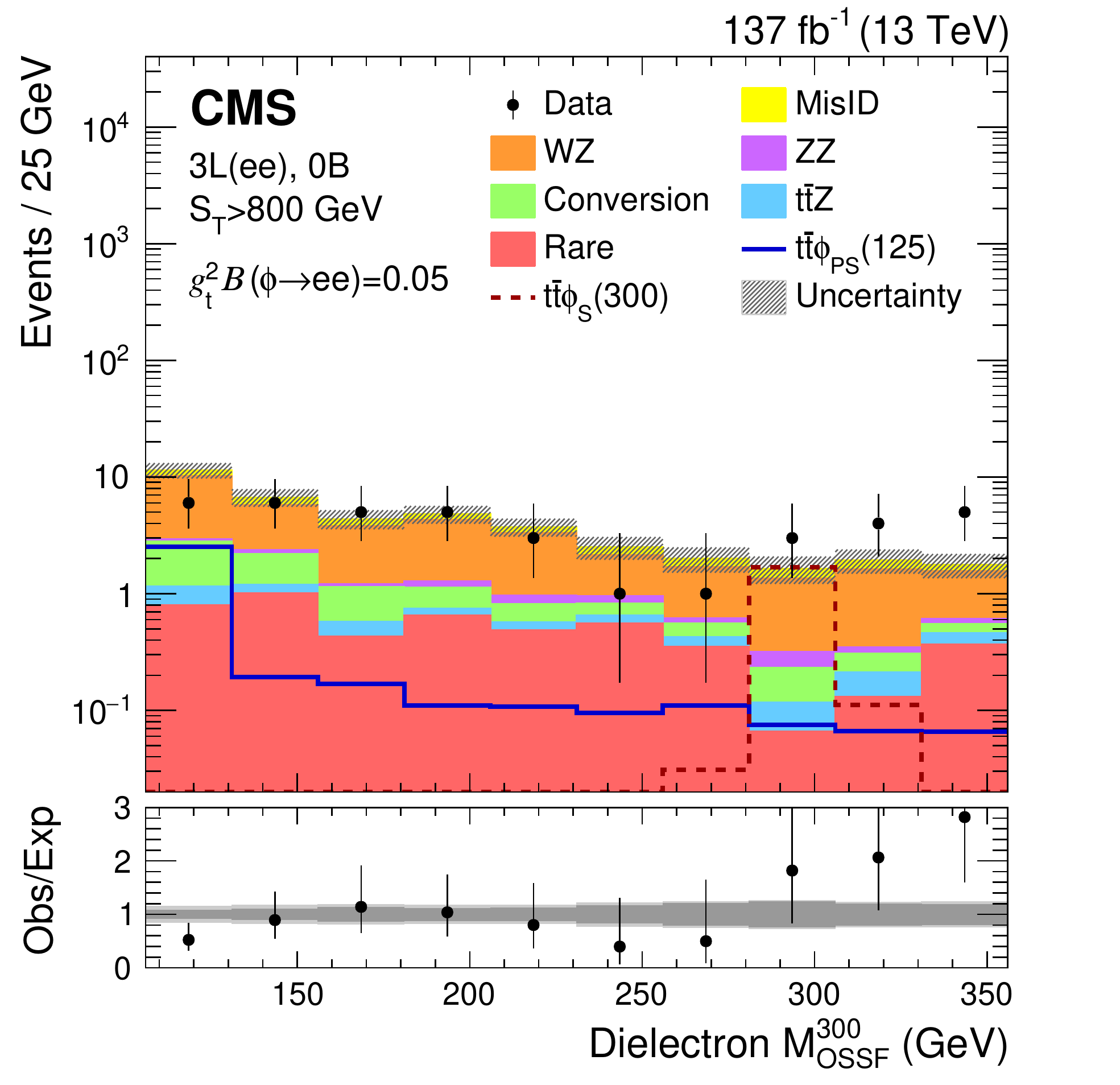}
\caption{
Dielectron $\mossf^{20}$ (left column) and $\mossf^{300}$ (right column) distributions in the 3L($\Pe\Pe$) 0B $\ttphi$ signal regions.
Upper, center, and lower plots are for $\ST<400\GeV$, $400<\ST<800\GeV$, and $\ST>800\GeV$, respectively.
The total SM background is shown as a stacked histogram of all contributing processes.
The predictions for $\ttphi(\to{\Pe\Pe}$) models with a pseudoscalar (scalar) $\phi$ of 20 and 125 (70 and 300)\GeV mass assuming $g_{\PQt}^2\mathcal{B}(\phi\to\Pe\Pe)=0.05$ are also shown.
The lower panels show the ratio of observed to expected events.
The hatched gray bands in the upper panels and the light gray bands in the lower panels represent the total (systematic and statistical) uncertainty of the backgrounds in each bin, whereas the dark gray bands in the lower panels represent only the statistical uncertainty of the backgrounds.
The rightmost bins do not contain the overflow events as these are outside the probed mass ranges.
\label{fig:ttPhiEle0B3LSR}}
\end{figure}

\begin{figure}[!htp]
\centering
\includegraphics[width=.4\textwidth]{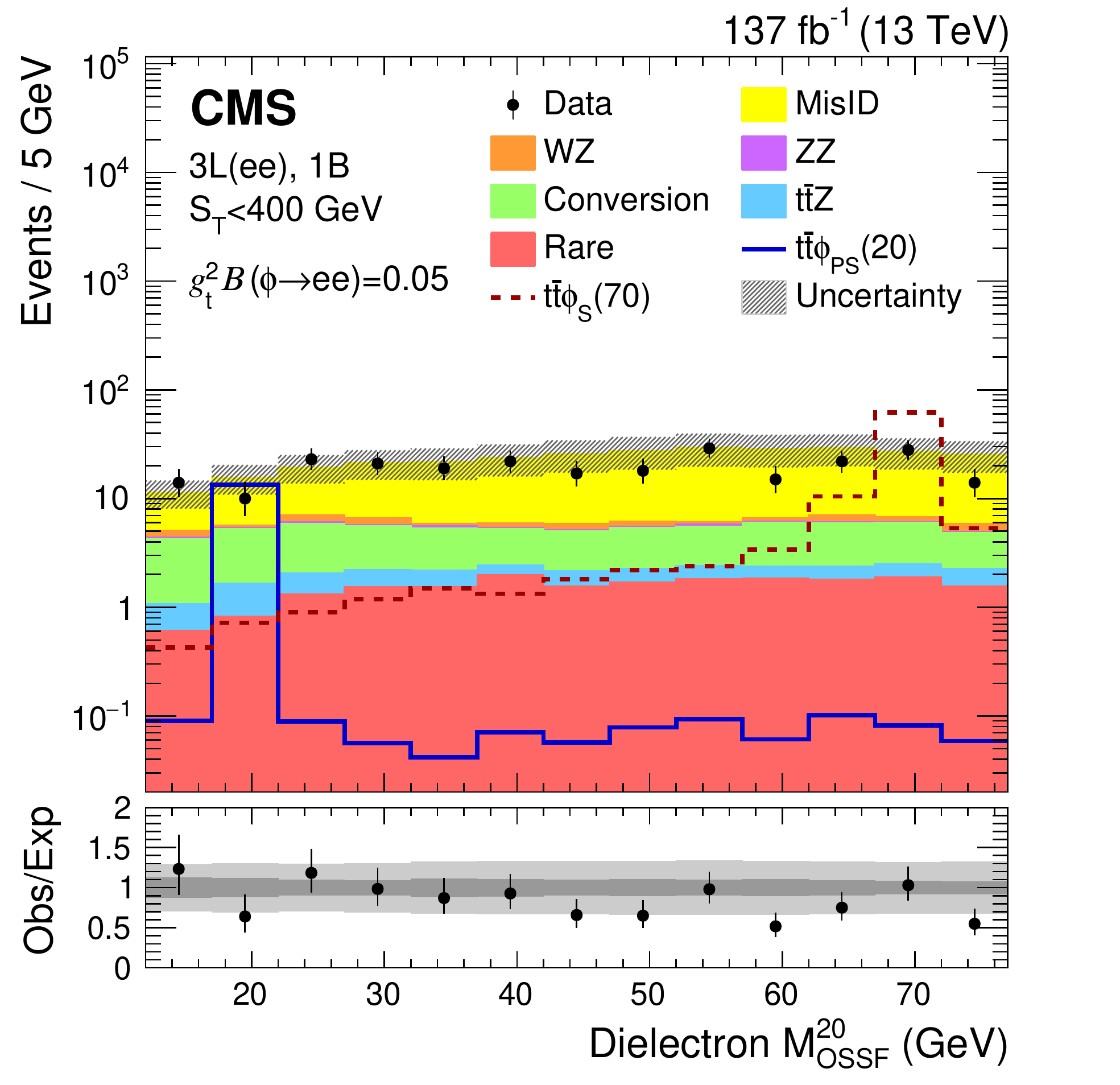} \hspace{.05\textwidth}
\includegraphics[width=.4\textwidth]{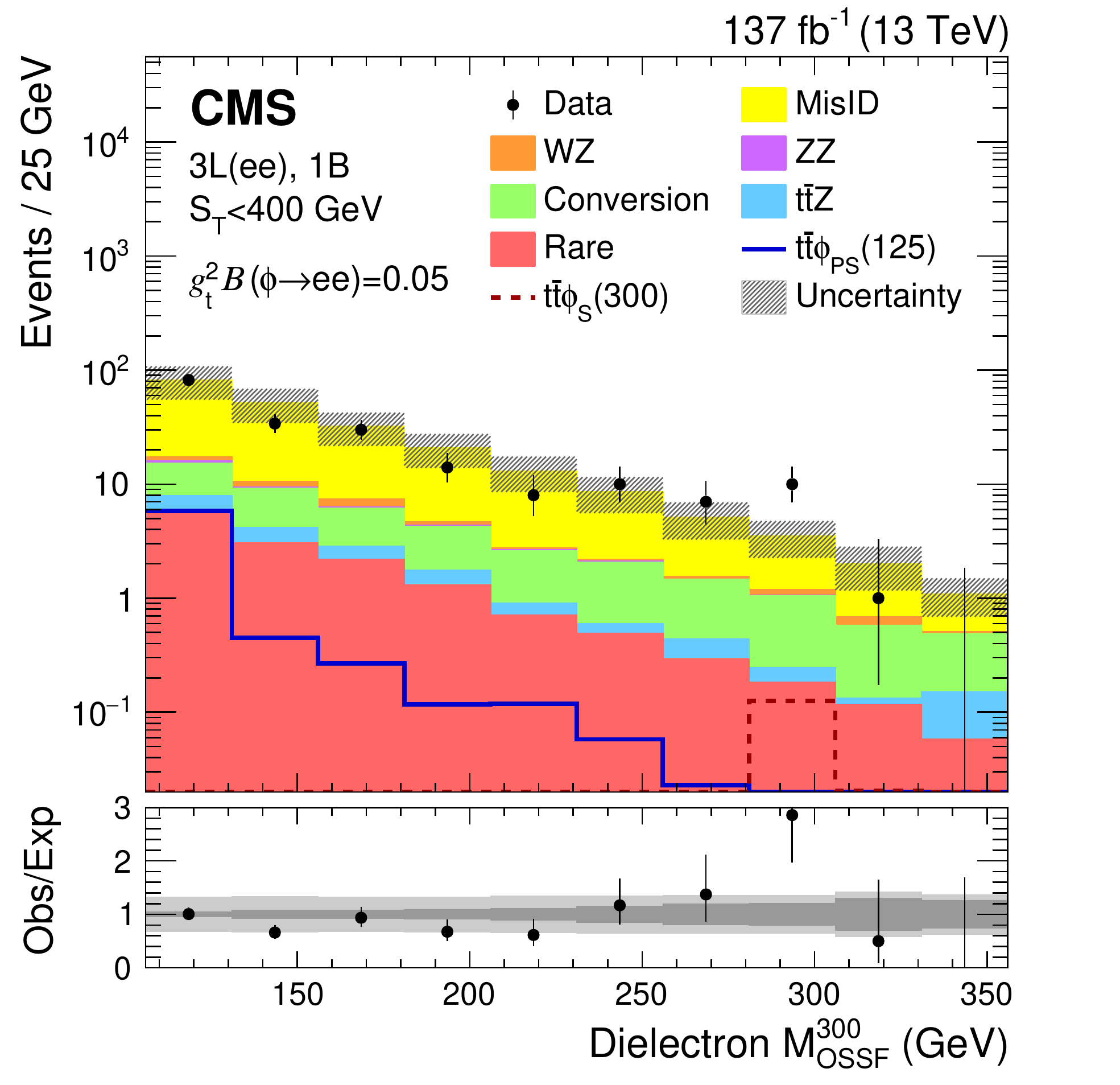}
\includegraphics[width=.4\textwidth]{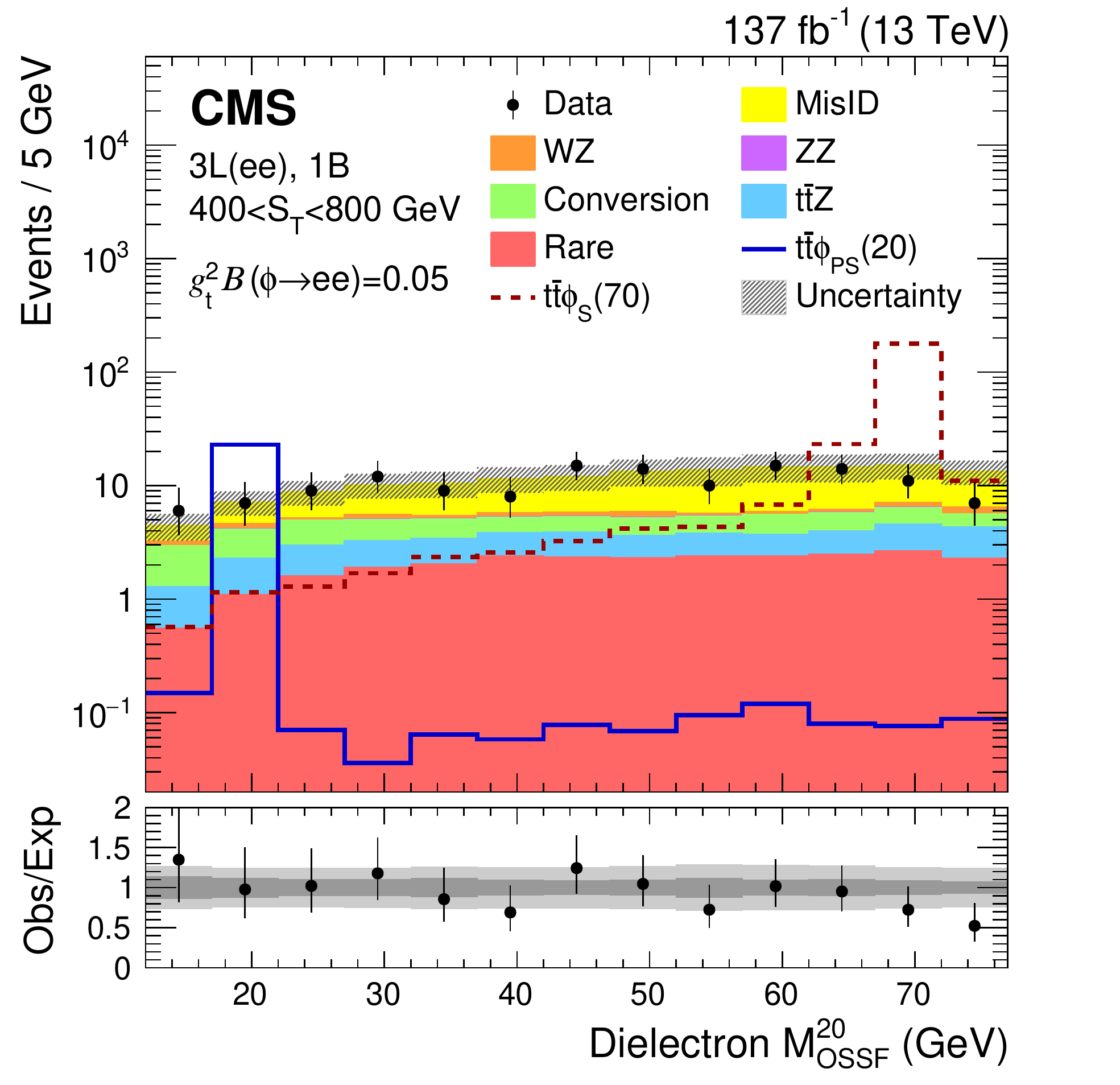} \hspace{.05\textwidth}
\includegraphics[width=.4\textwidth]{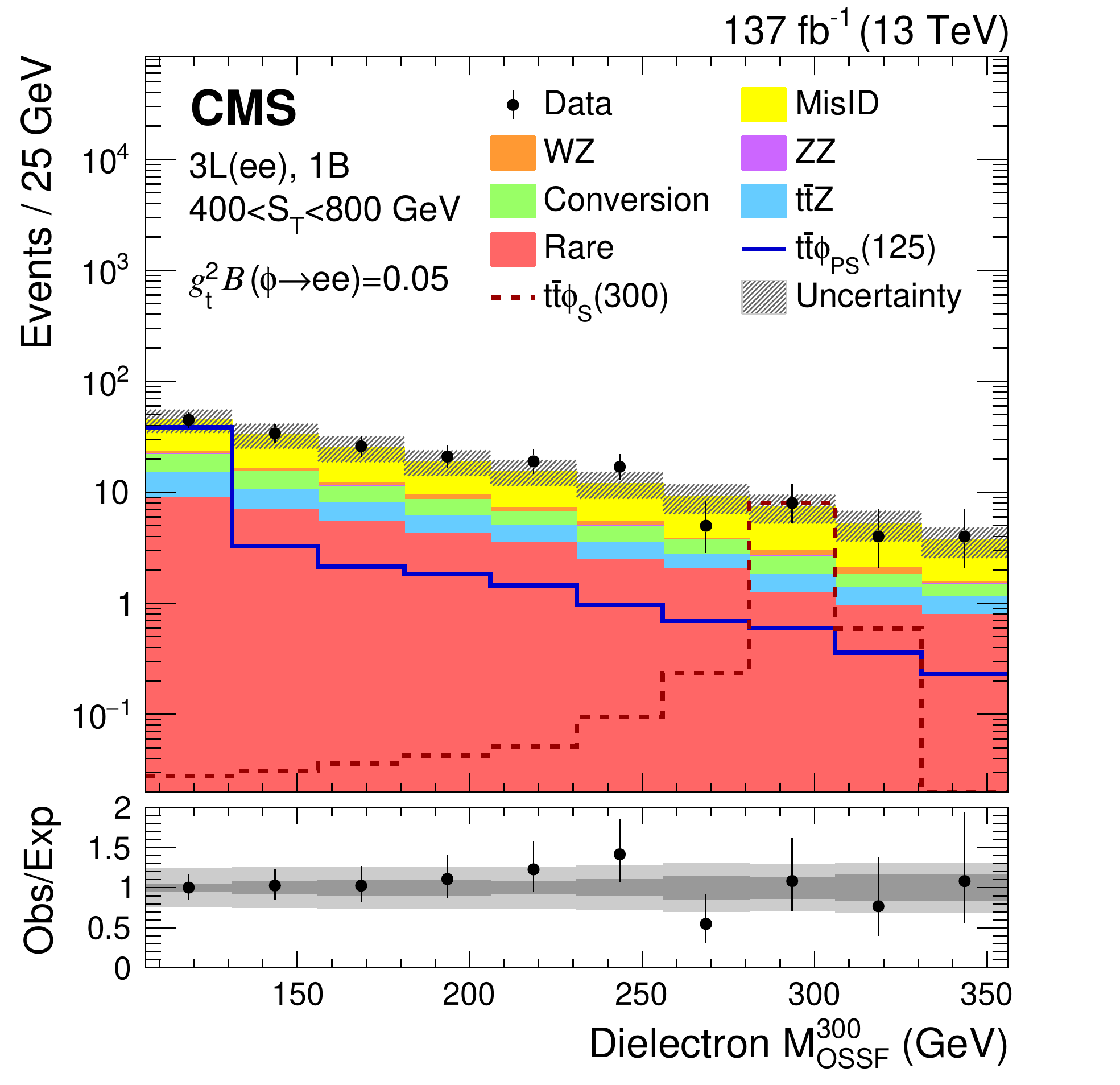}
\includegraphics[width=.4\textwidth]{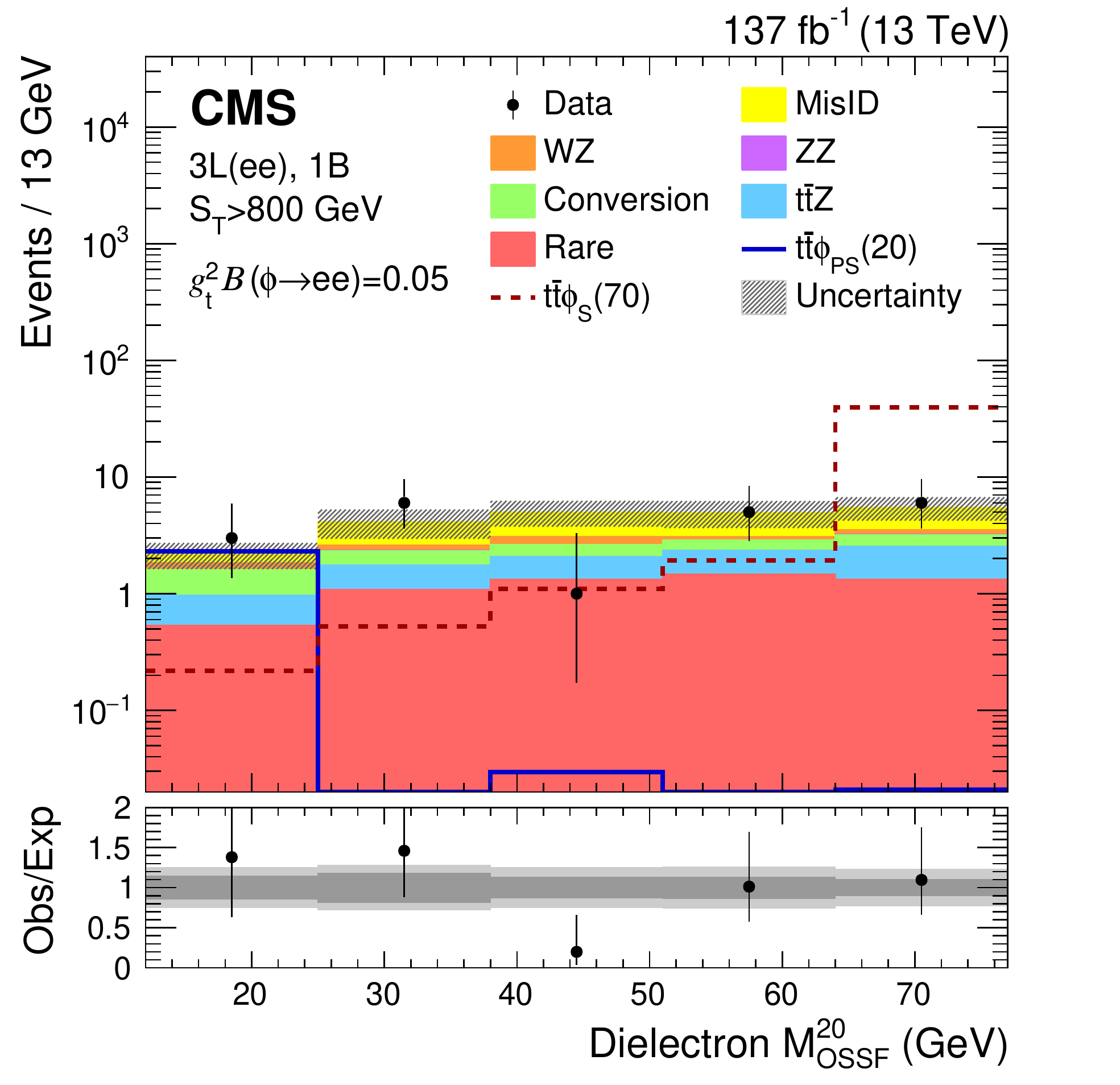} \hspace{.05\textwidth}
\includegraphics[width=.4\textwidth]{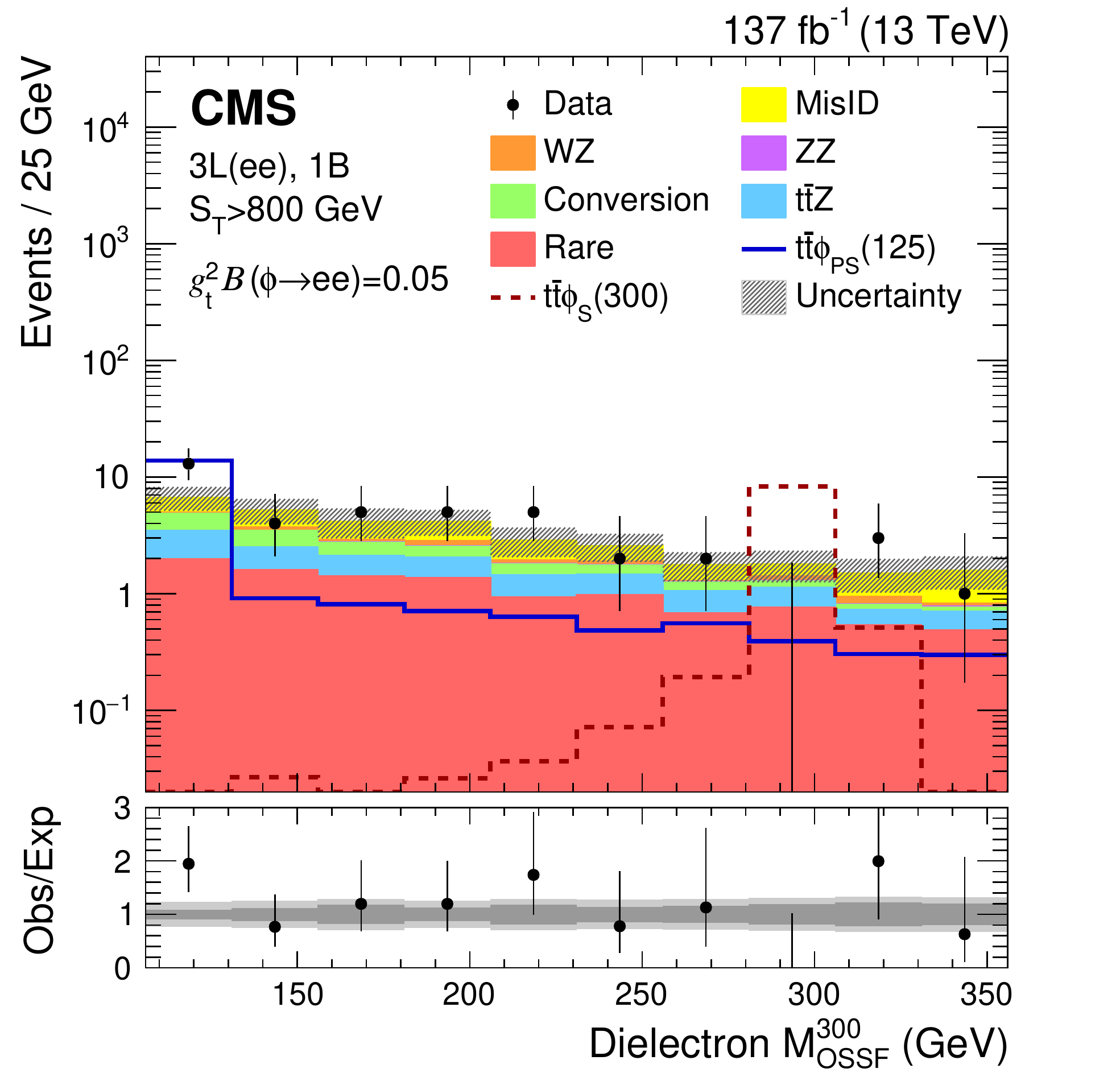}
\caption{
Dielectron $\mossf^{20}$ (left column) and $\mossf^{300}$ (right column) distributions in the 3L($\Pe\Pe$) 1B $\ttphi$ signal regions.
Upper, center, and lower plots are for $\ST<400\GeV$, $400<\ST<800\GeV$, and $\ST>800\GeV$, respectively.
The total SM background is shown as a stacked histogram of all contributing processes.
The predictions for $\ttphi(\to{\Pe\Pe}$) models with a pseudoscalar (scalar) $\phi$ of 20 and 125 (70 and 300)\GeV mass assuming $g_{\PQt}^2\mathcal{B}(\phi\to\Pe\Pe)=0.05$ are also shown.
The lower panels show the ratio of observed to expected events.
The hatched gray bands in the upper panels and the light gray bands in the lower panels represent the total (systematic and statistical) uncertainty of the backgrounds in each bin, whereas the dark gray bands in the lower panels represent only the statistical uncertainty of the backgrounds.
The rightmost bins do not contain the overflow events as these are outside the probed mass ranges.
\label{fig:ttPhiEle1B3LSR}}
\end{figure}

\begin{figure}[!htp]
\centering
\includegraphics[width=.4\textwidth]{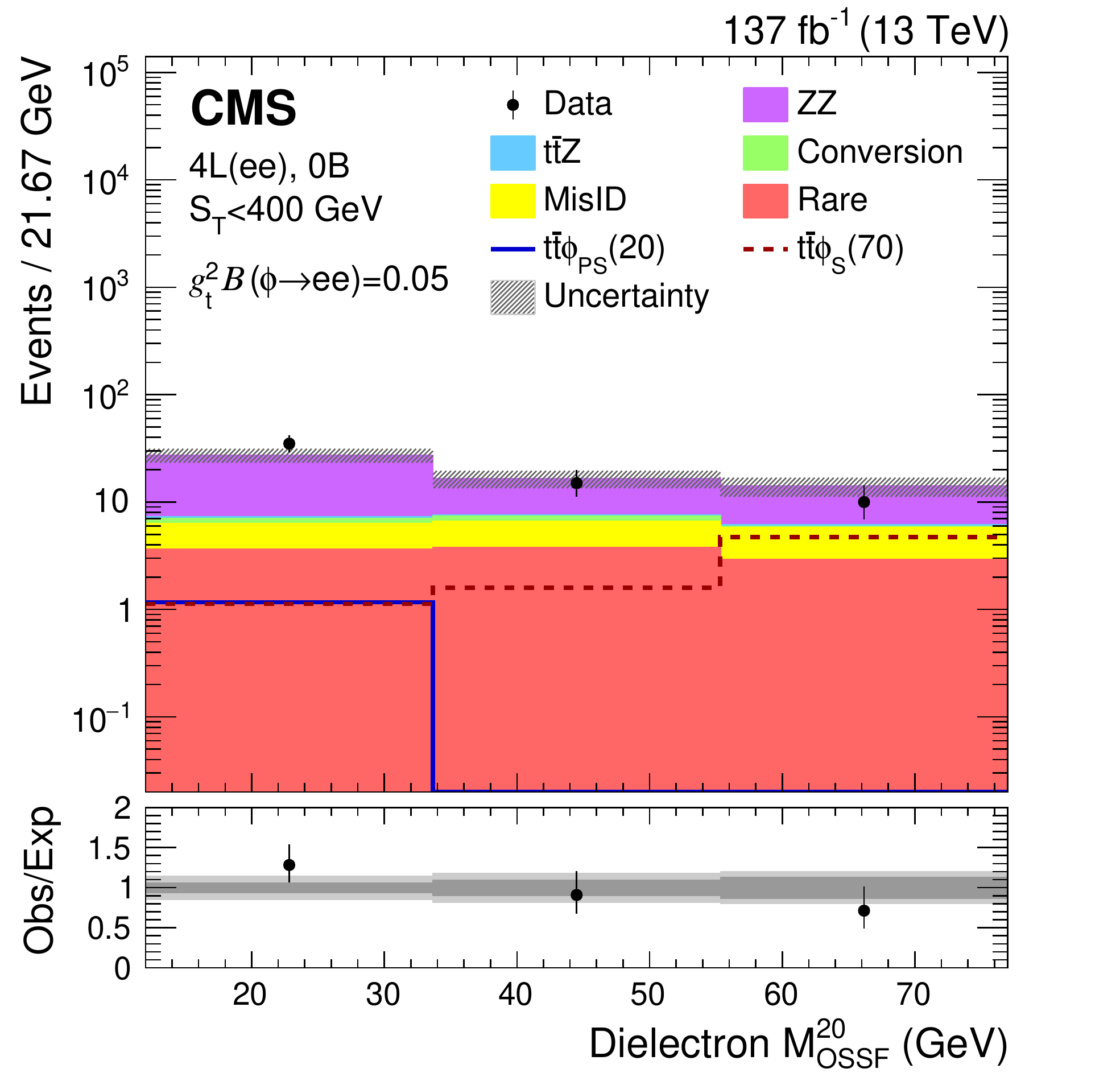} \hspace{.05\textwidth}
\includegraphics[width=.4\textwidth]{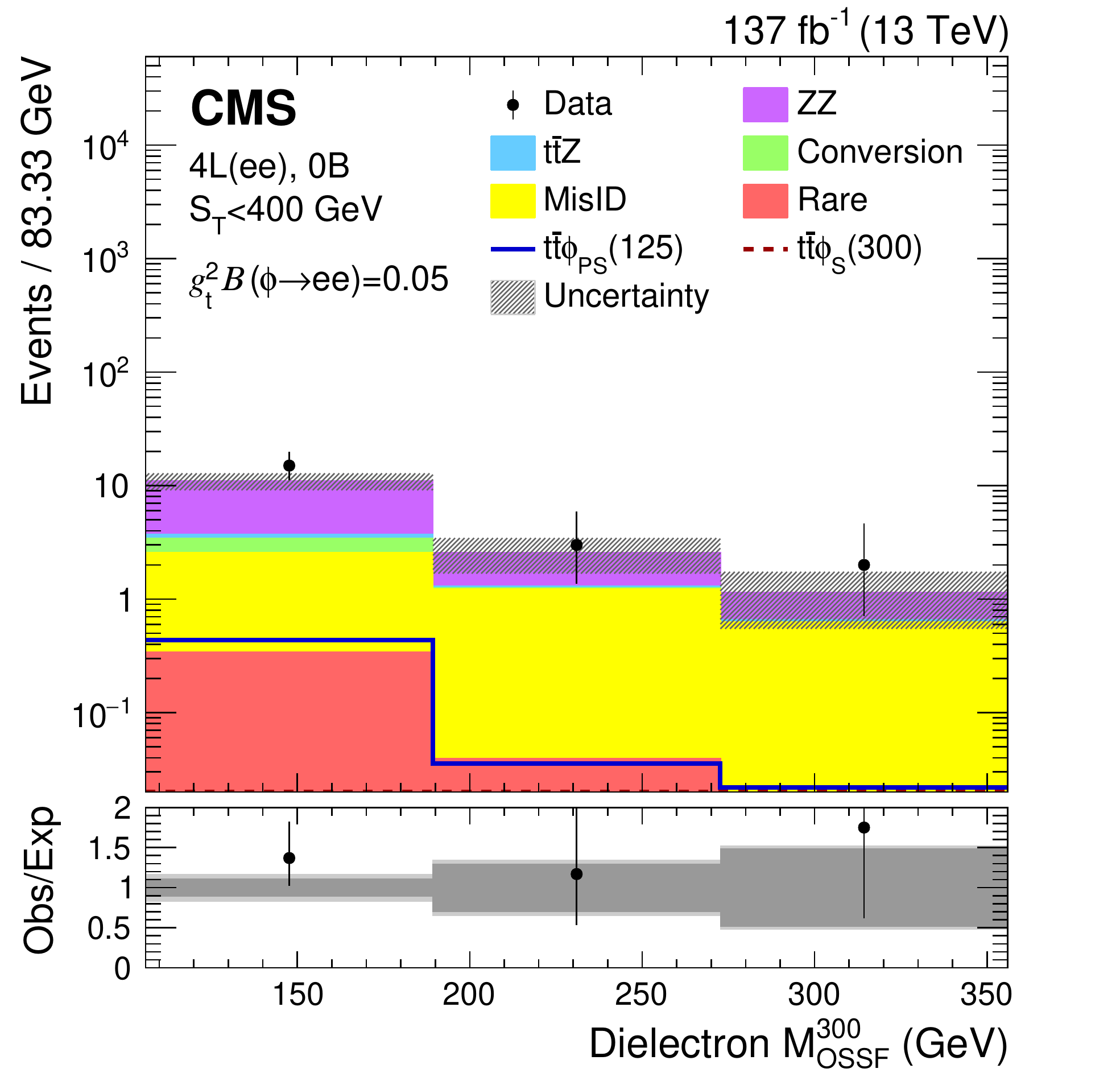}
\includegraphics[width=.4\textwidth]{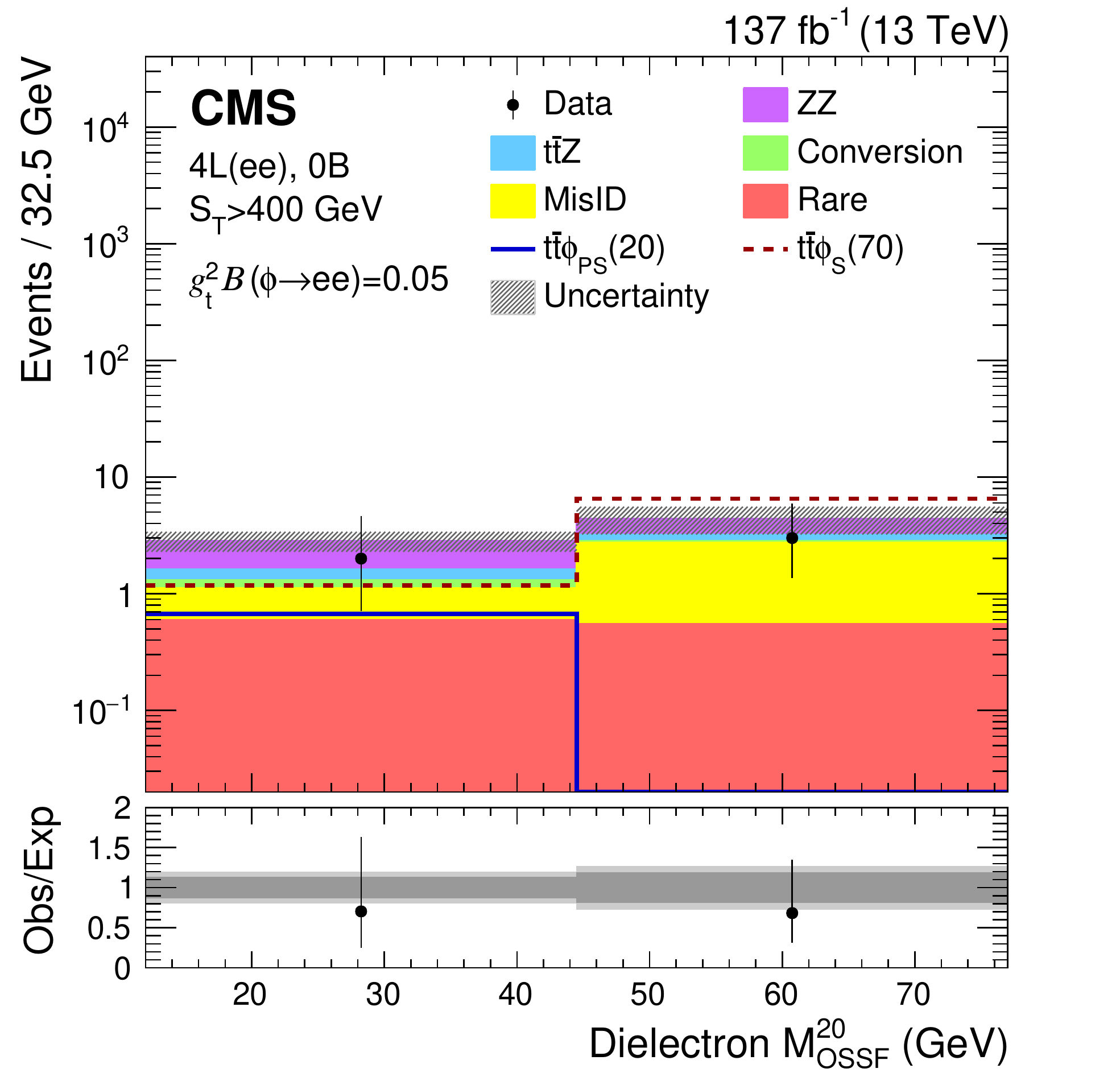} \hspace{.05\textwidth}
\includegraphics[width=.4\textwidth]{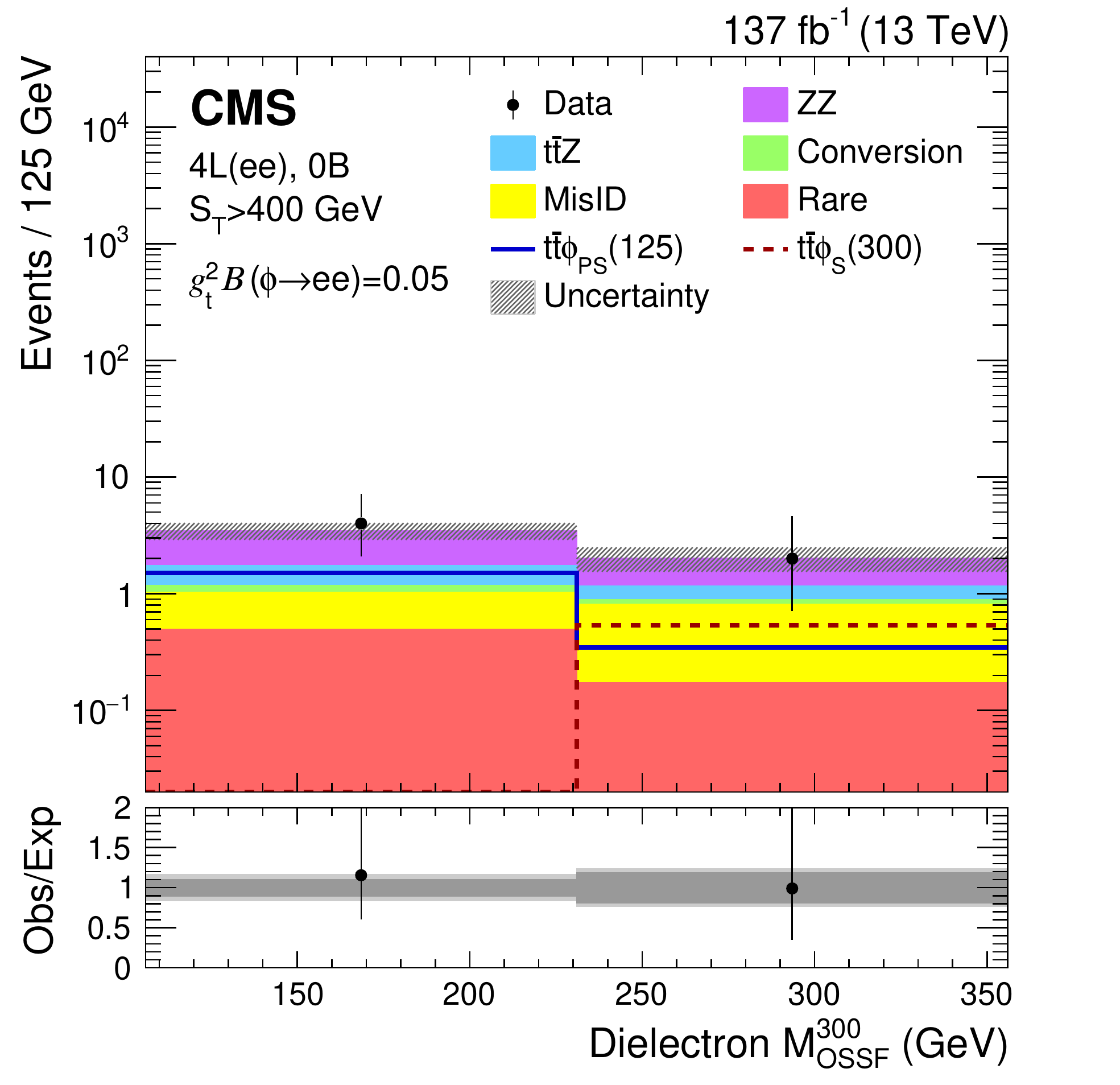}
\includegraphics[width=.4\textwidth]{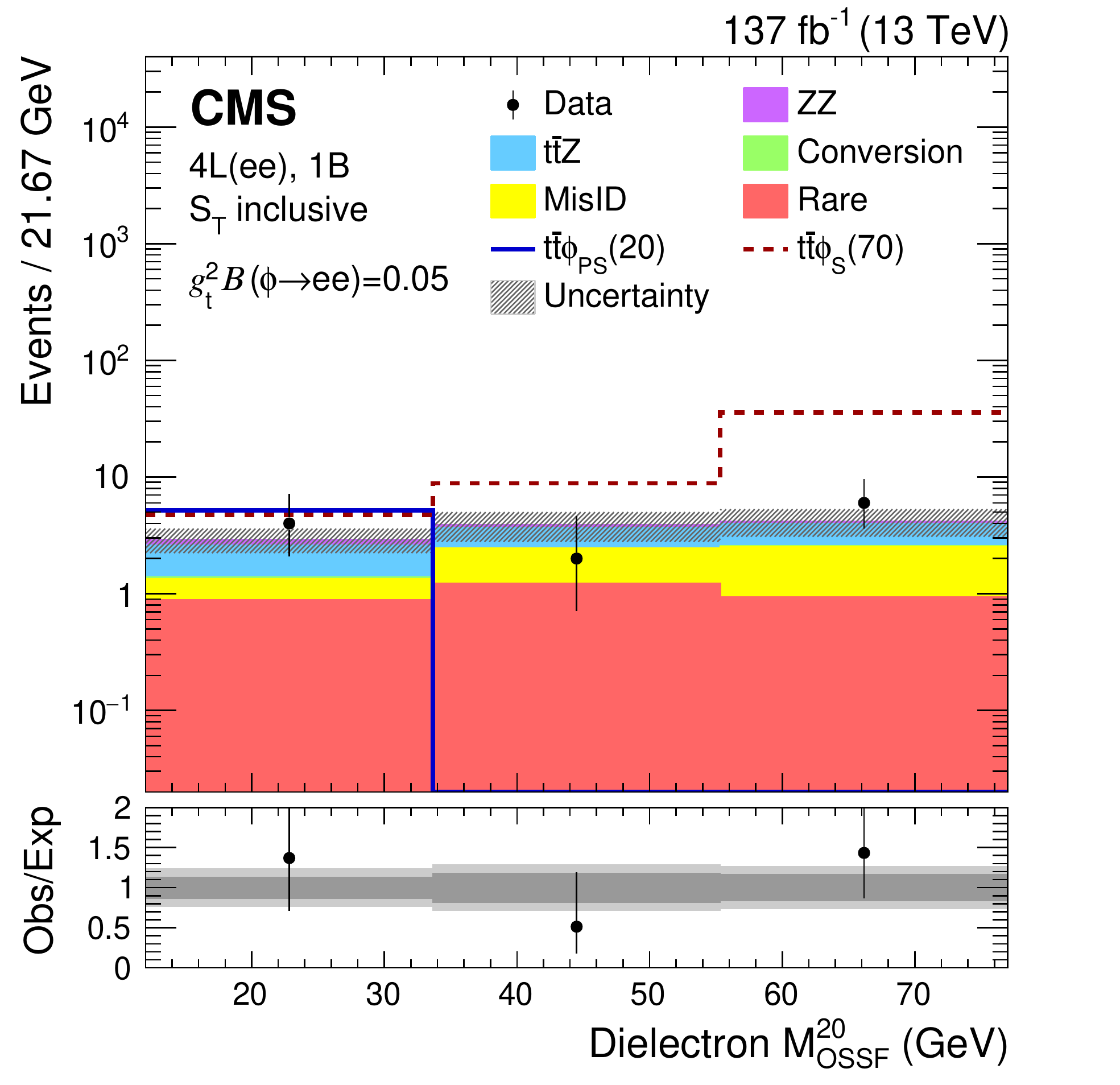} \hspace{.05\textwidth}
\includegraphics[width=.4\textwidth]{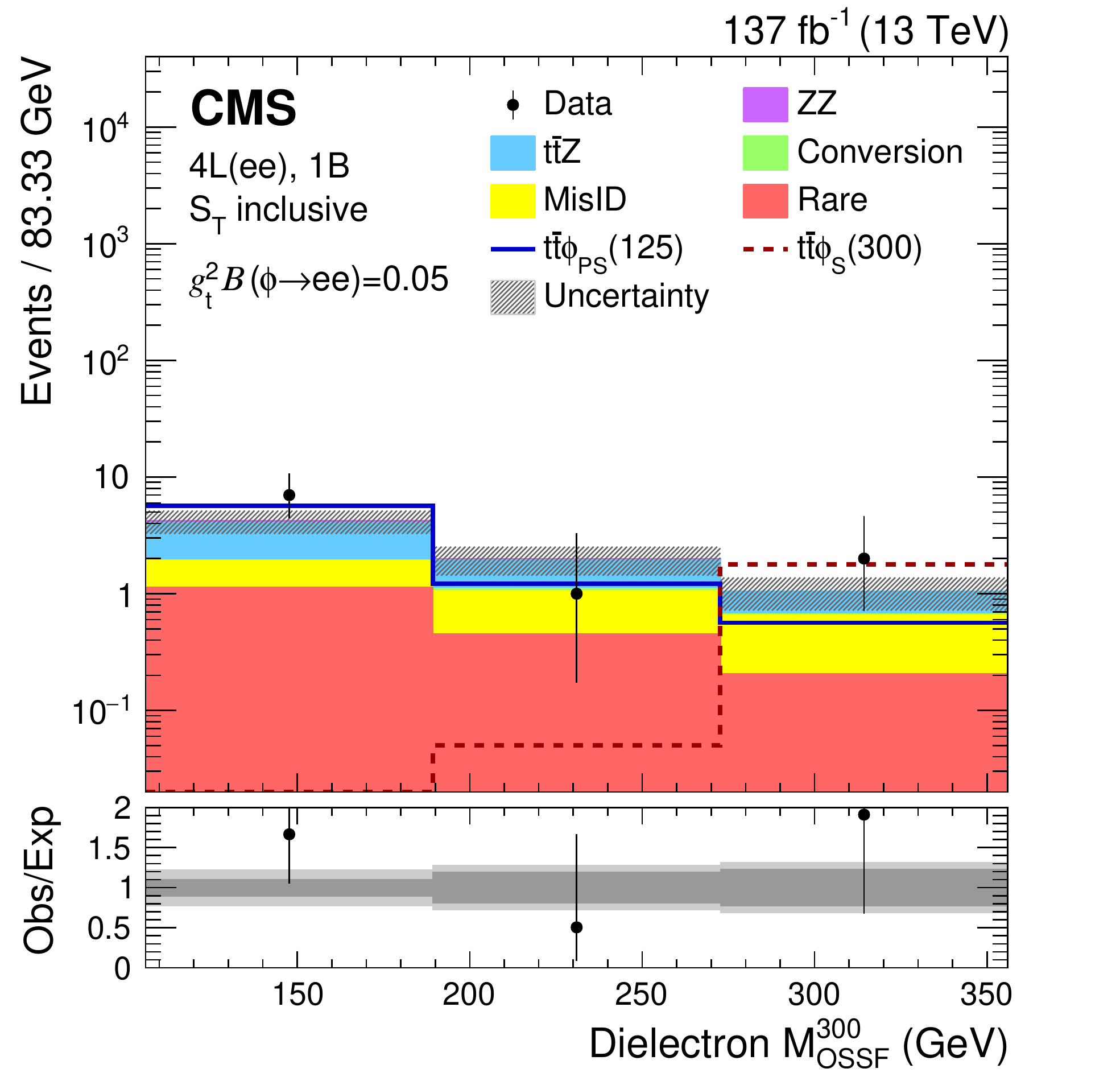}
\caption{
Dielectron $\mossf^{20}$ (left column) and $\mossf^{300}$ (right column) distributions in the 4L($\Pe\Pe$) $\ttphi$ signal regions.
Upper, center, and lower plots are for 0B $\ST<400\GeV$, 0B $\ST>400\GeV$, and 1B $\ST$-inclusive, respectively.
The total SM background is shown as a stacked histogram of all contributing processes.
The predictions for $\ttphi(\to{\Pe\Pe}$) models with a pseudoscalar (scalar) $\phi$ of 20 and 125 (70 and 300)\GeV mass assuming $g_{\PQt}^2\mathcal{B}(\phi\to\Pe\Pe)=0.05$ are also shown.
The lower panels show the ratio of observed to expected events.
The hatched gray bands in the upper panels and the light gray bands in the lower panels represent the total (systematic and statistical) uncertainty of the backgrounds in each bin, whereas the dark gray bands in the lower panels represent only the statistical uncertainty of the backgrounds.
The rightmost bins do not contain the overflow events as these are outside the probed mass range.
\label{fig:ttPhiEle4LSR}}
\end{figure}

\begin{figure}[!htp]
\centering
\includegraphics[width=.4\textwidth]{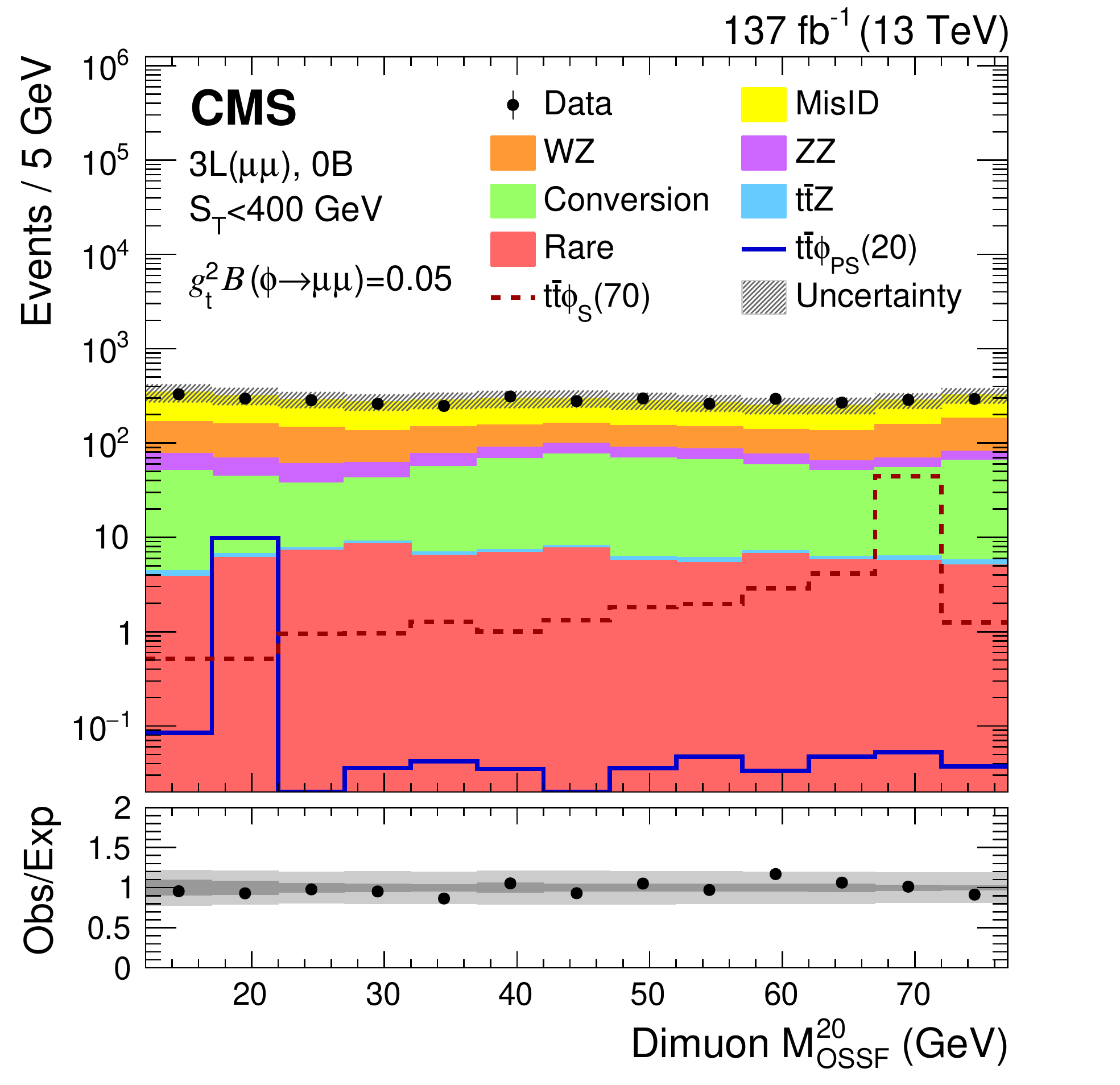} \hspace{.05\textwidth}
\includegraphics[width=.4\textwidth]{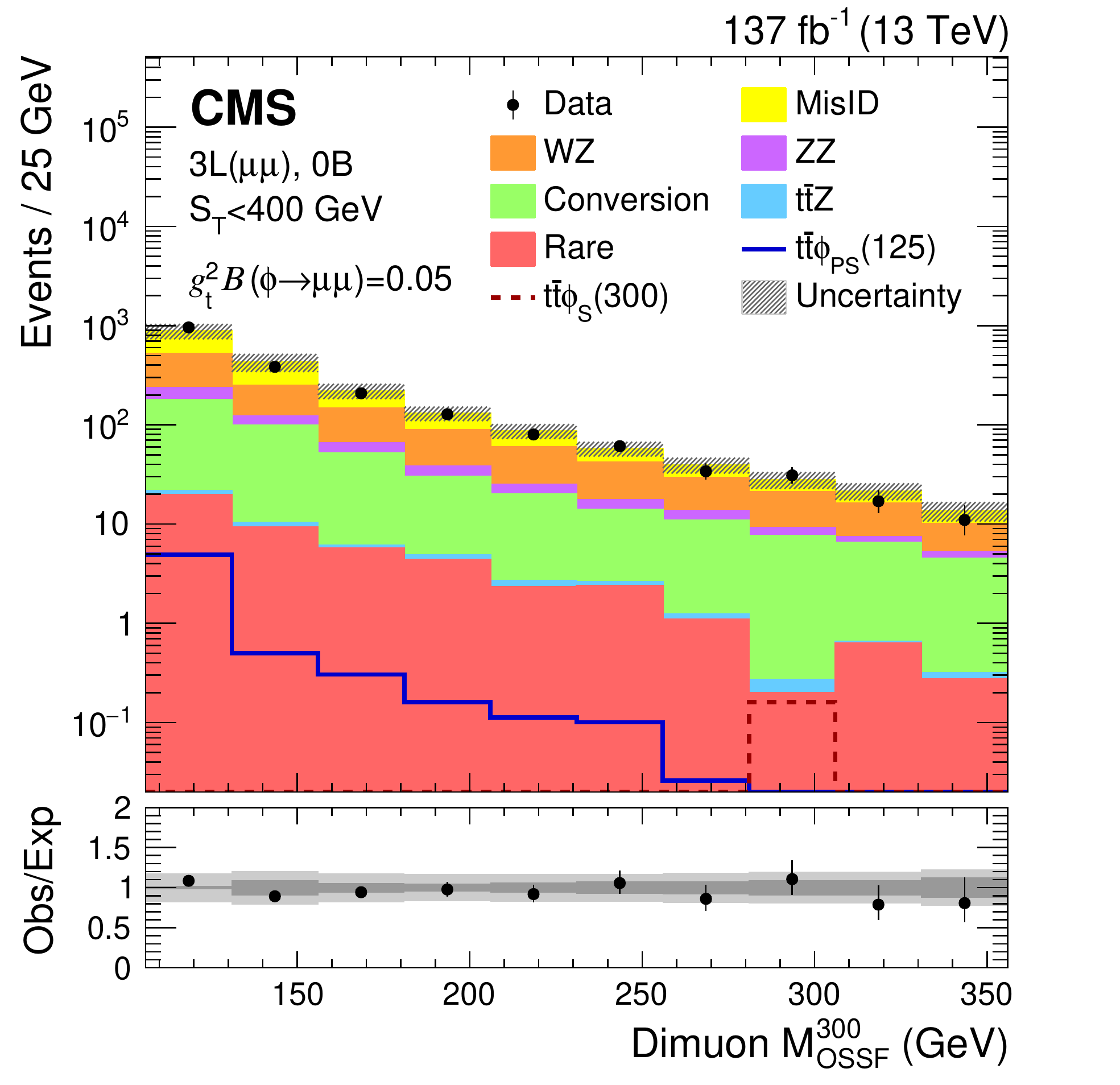}
\includegraphics[width=.4\textwidth]{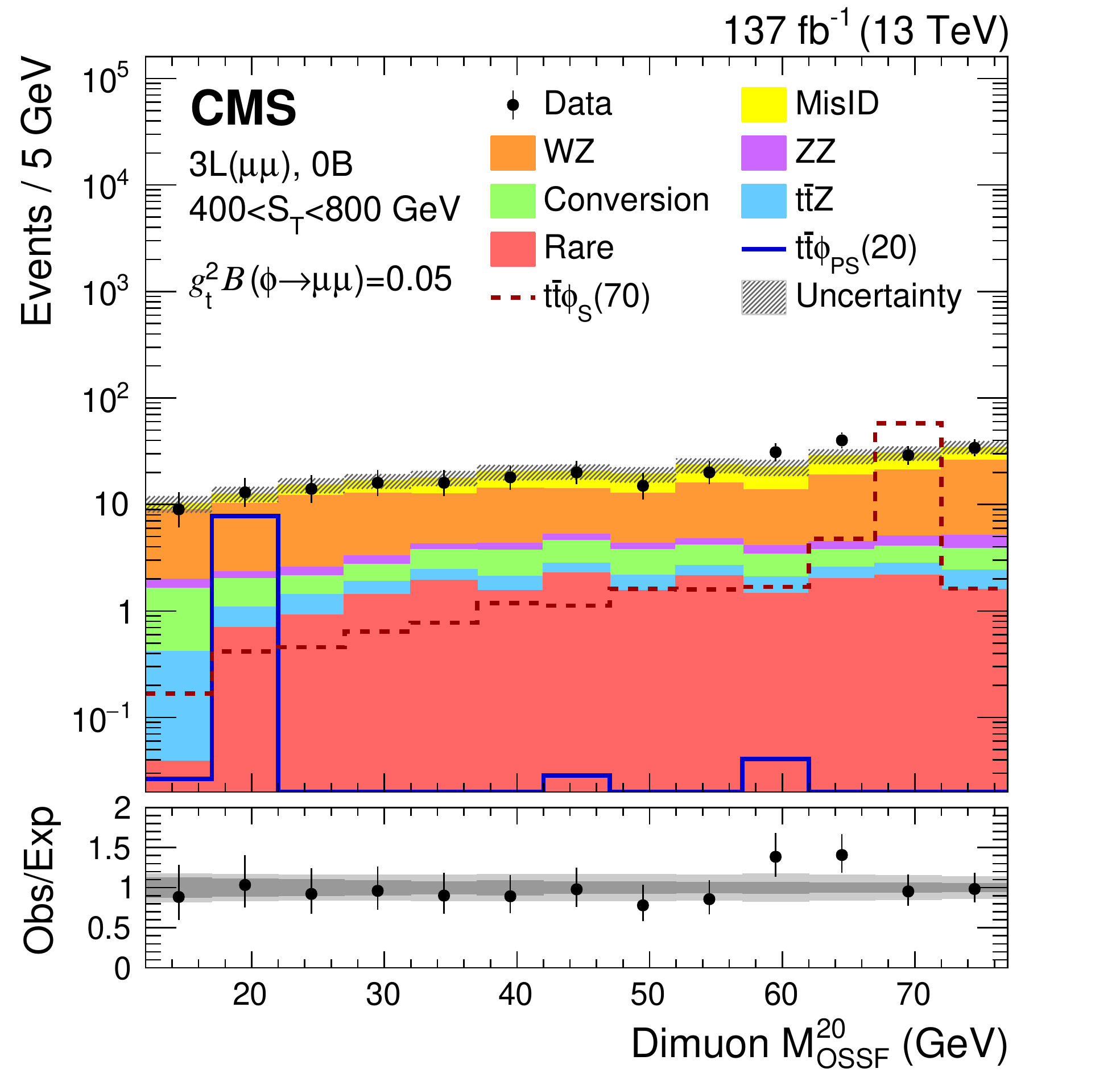} \hspace{.05\textwidth}
\includegraphics[width=.4\textwidth]{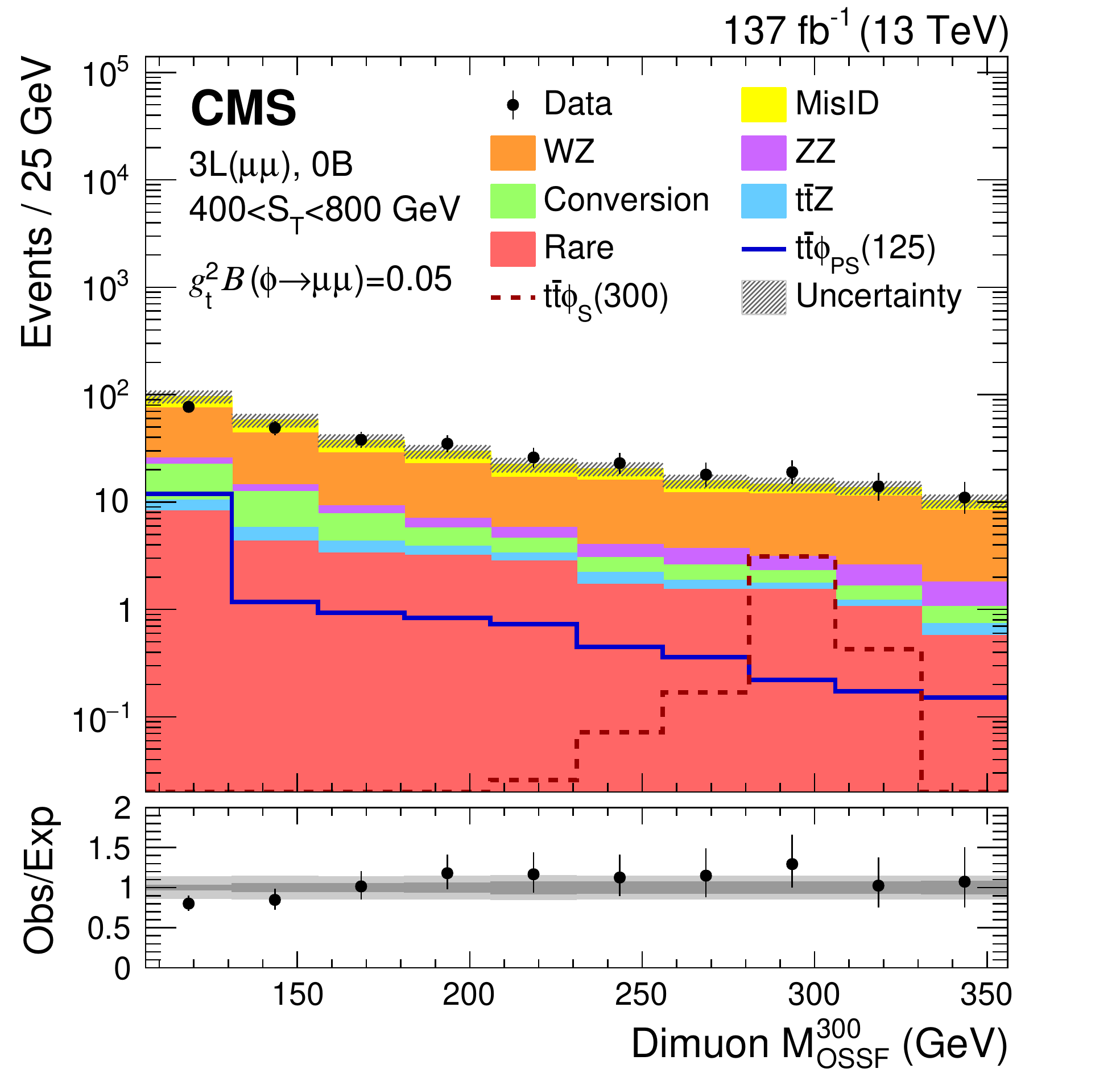}
\includegraphics[width=.4\textwidth]{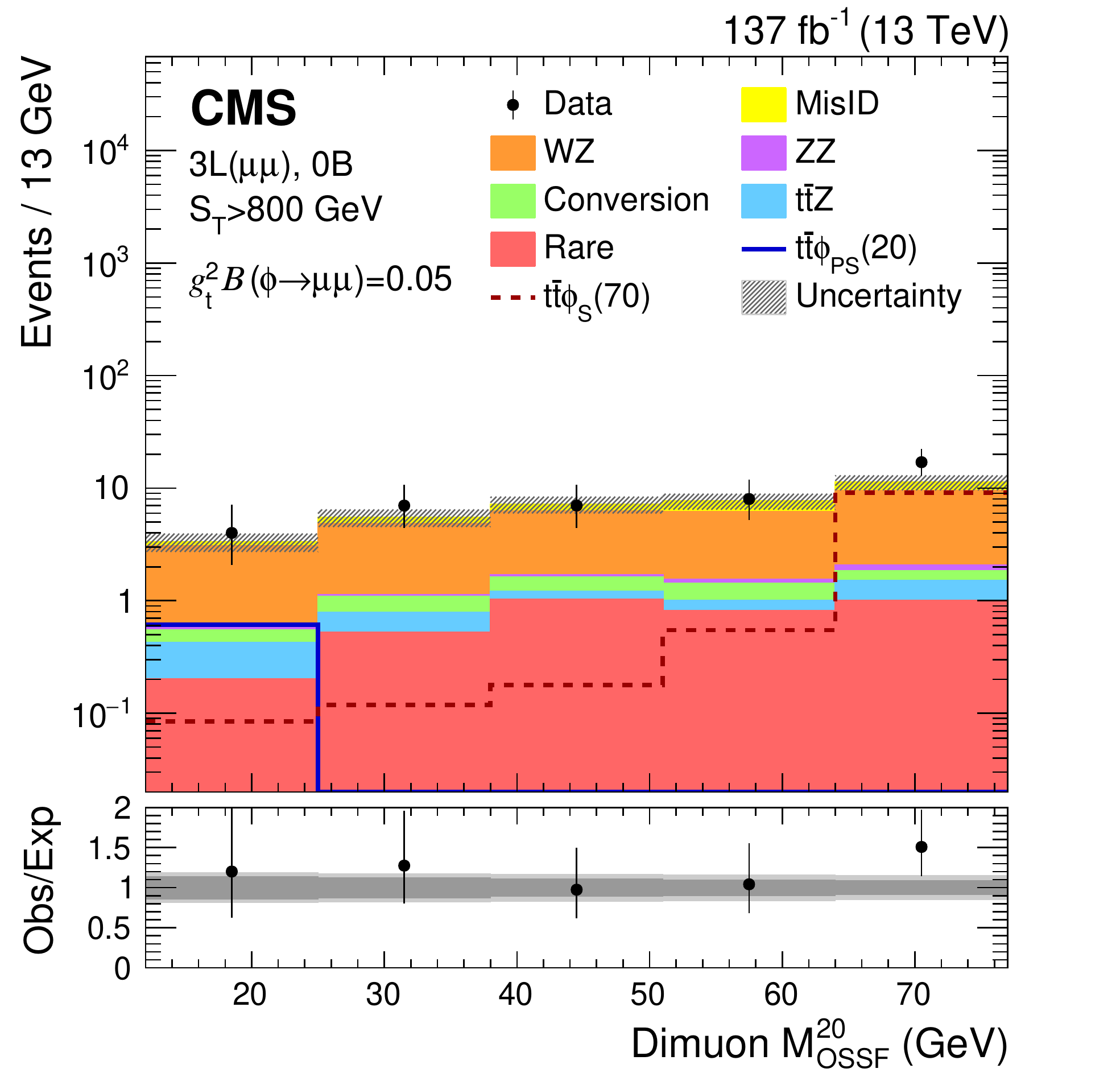} \hspace{.05\textwidth}
\includegraphics[width=.4\textwidth]{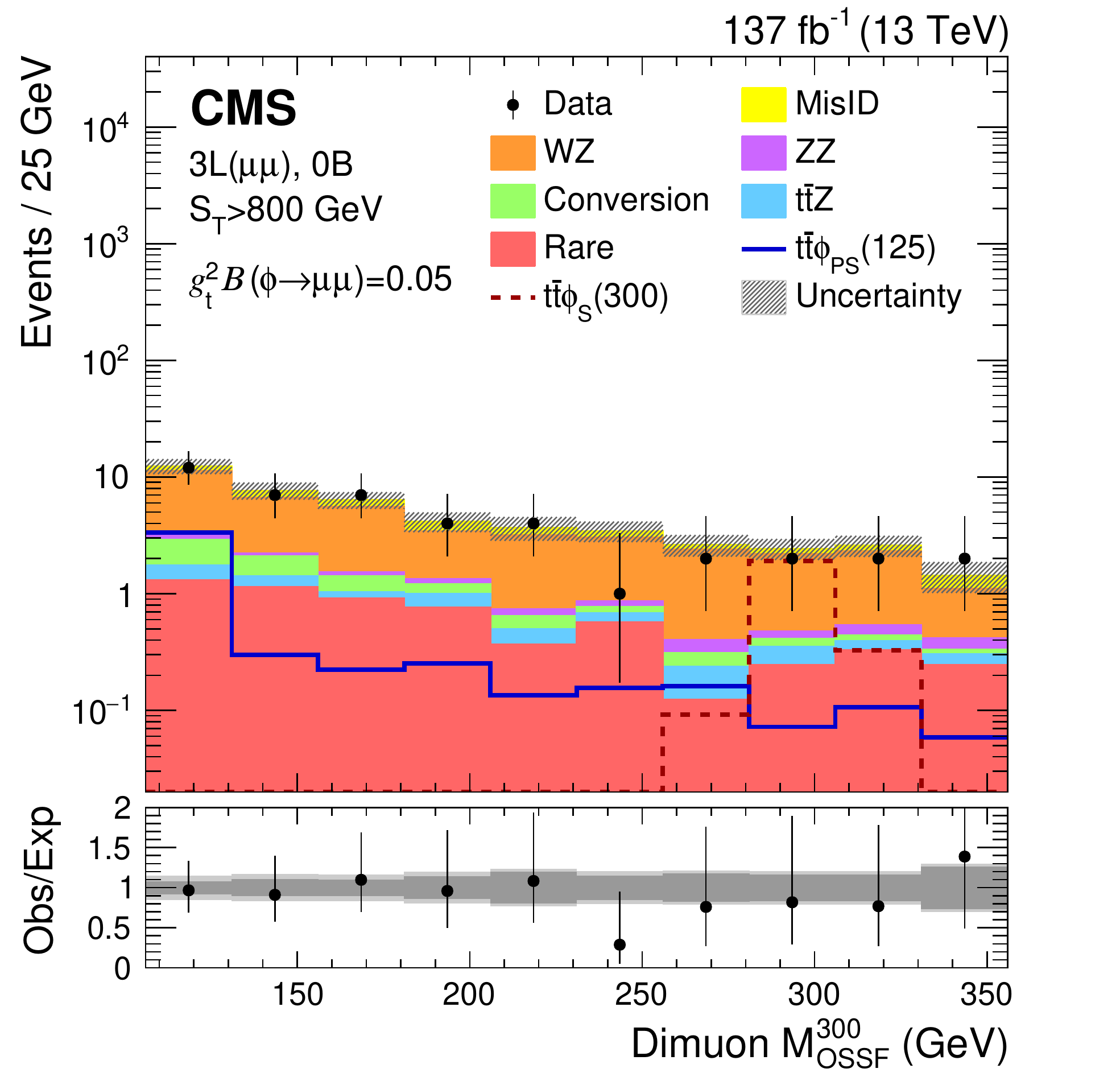}
\caption{
Dimuon $\mossf^{20}$ (left column) and $\mossf^{300}$ (right column) distributions in the 3L($\mu\mu$) 0B $\ttphi$ signal regions.
Upper, center, and lower plots are for $\ST<400\GeV$, $400<\ST<800\GeV$, and $\ST>800\GeV$, respectively.
The total SM background is shown as a stacked histogram of all contributing processes.
The predictions for $\ttphi(\to\mu\mu$) models with a pseudoscalar (scalar) $\phi$ of 20 and 125 (70 and 300)\GeV mass assuming $g_{\PQt}^2\mathcal{B}(\phi\to\mu\mu)=0.05$ are also shown.
The lower panels show the ratio of observed to expected events.
The hatched gray bands in the upper panels and the light gray bands in the lower panels represent the total (systematic and statistical) uncertainty of the backgrounds in each bin, whereas the dark gray bands in the lower panels represent only the statistical uncertainty of the backgrounds.
The rightmost bins do not contain the overflow events as these are outside the probed mass range.
\label{fig:ttPhiMu0B3LSR}}
\end{figure}

\begin{figure}[!htp]
\centering
\includegraphics[width=.4\textwidth]{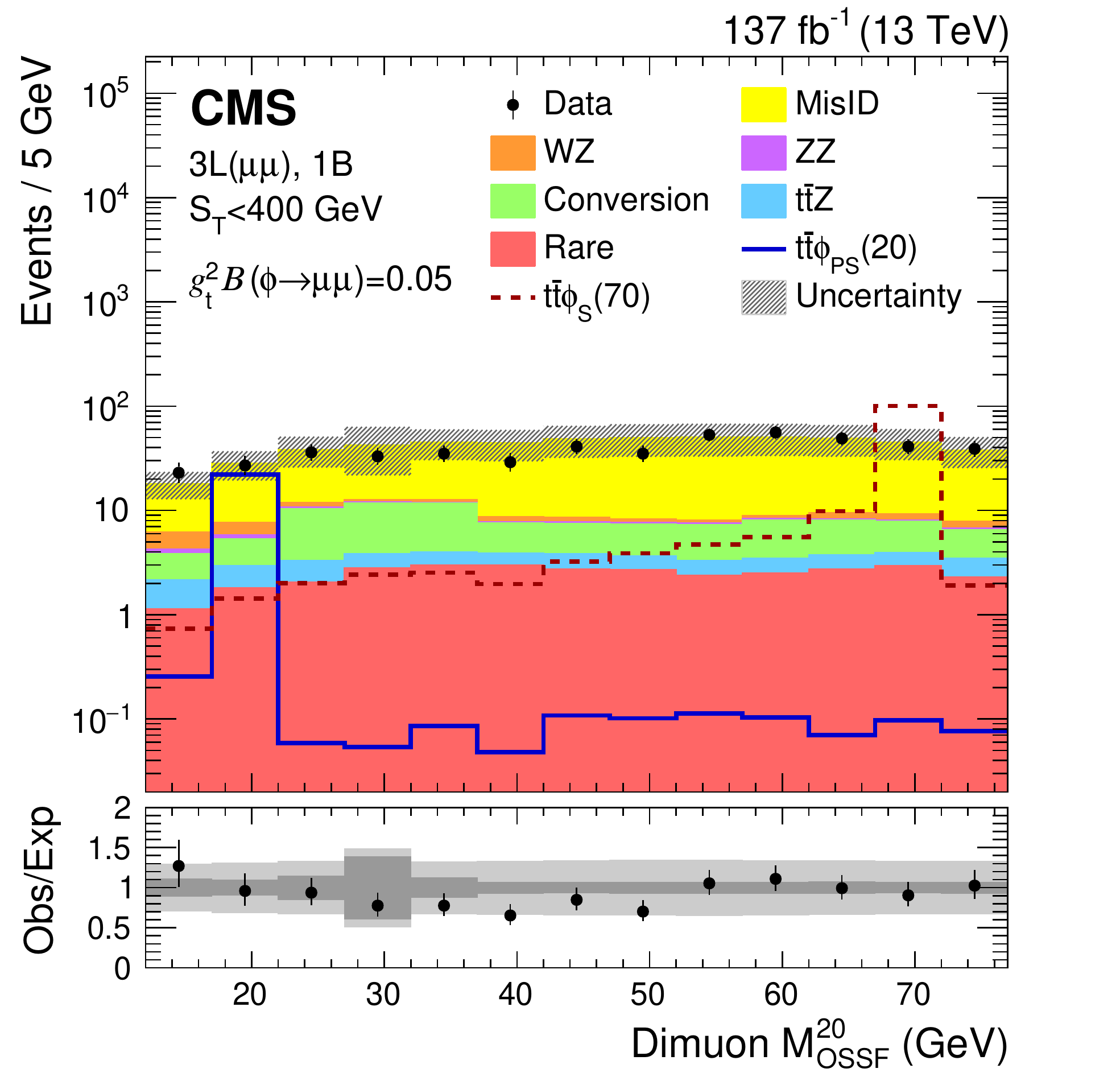} \hspace{.05\textwidth}
\includegraphics[width=.4\textwidth]{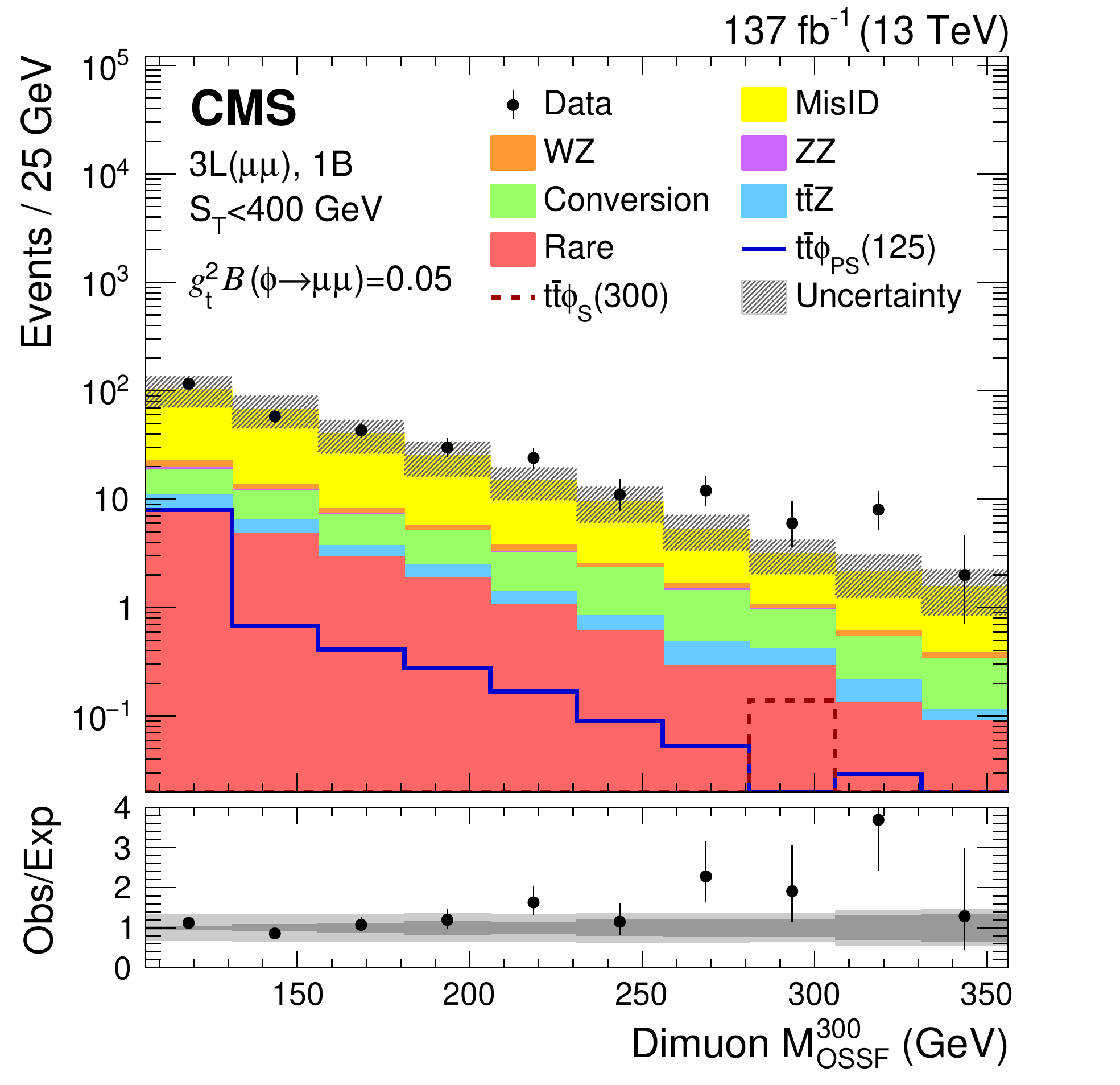}
\includegraphics[width=.4\textwidth]{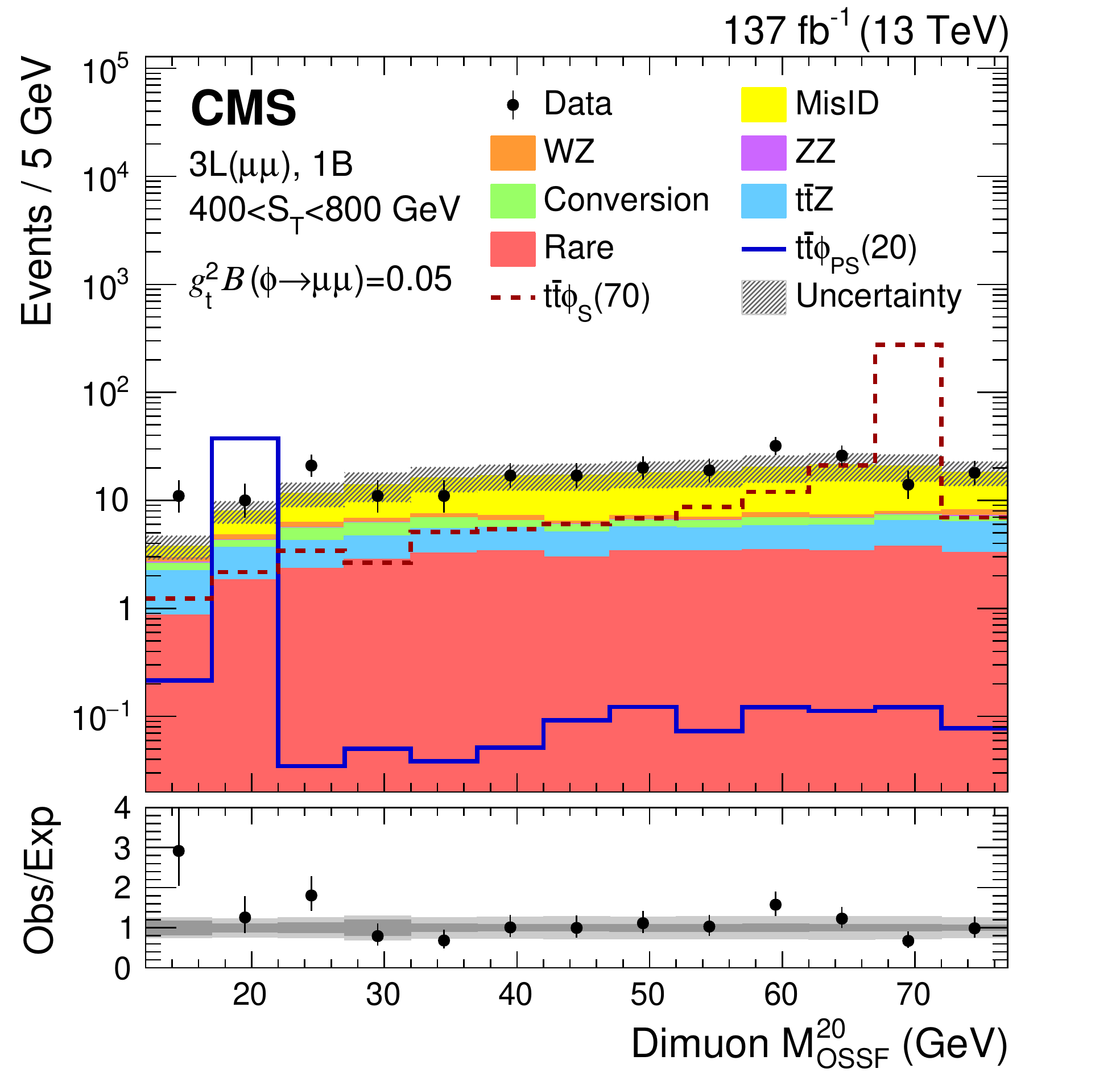} \hspace{.05\textwidth}
\includegraphics[width=.4\textwidth]{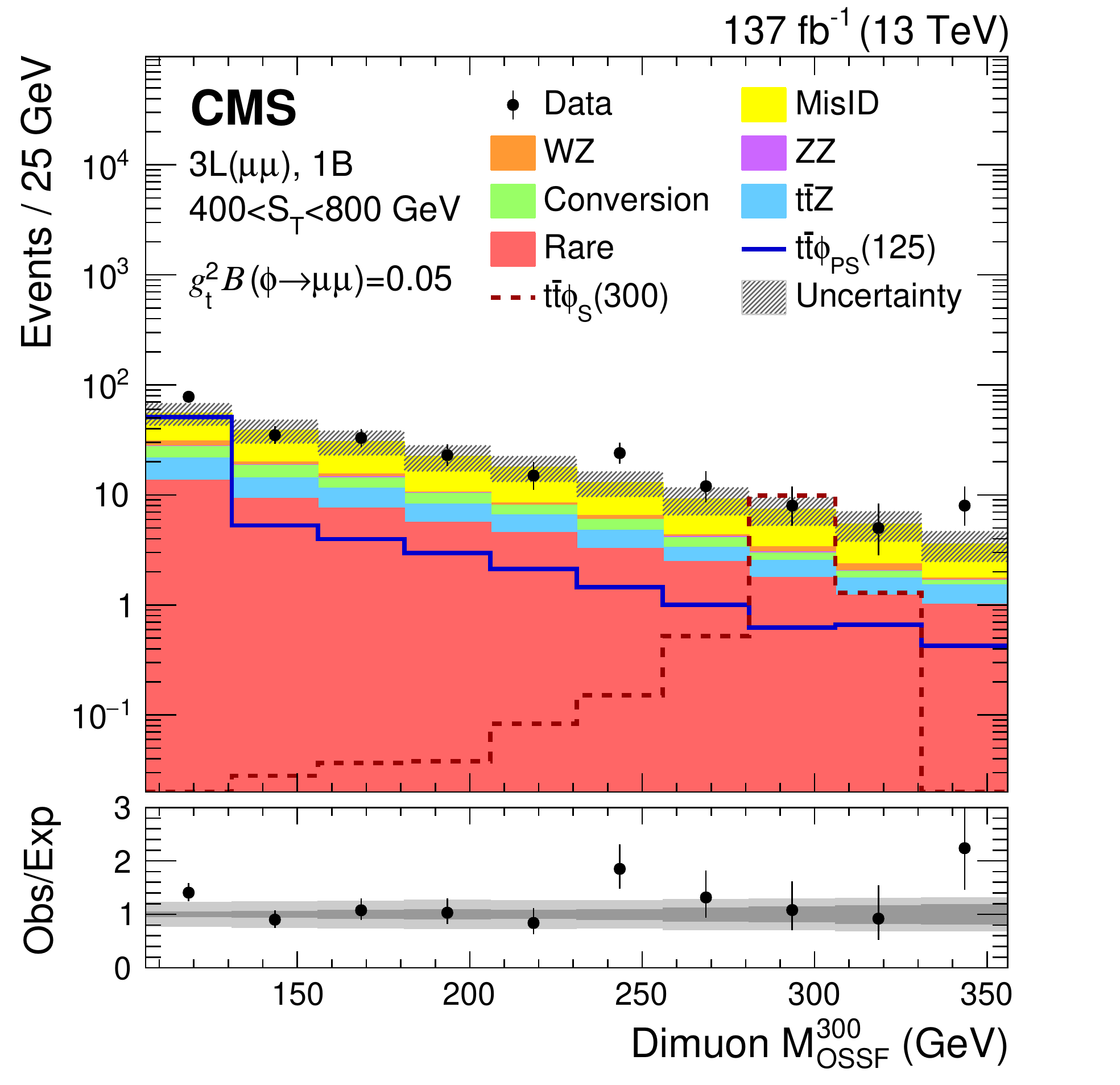}
\includegraphics[width=.4\textwidth]{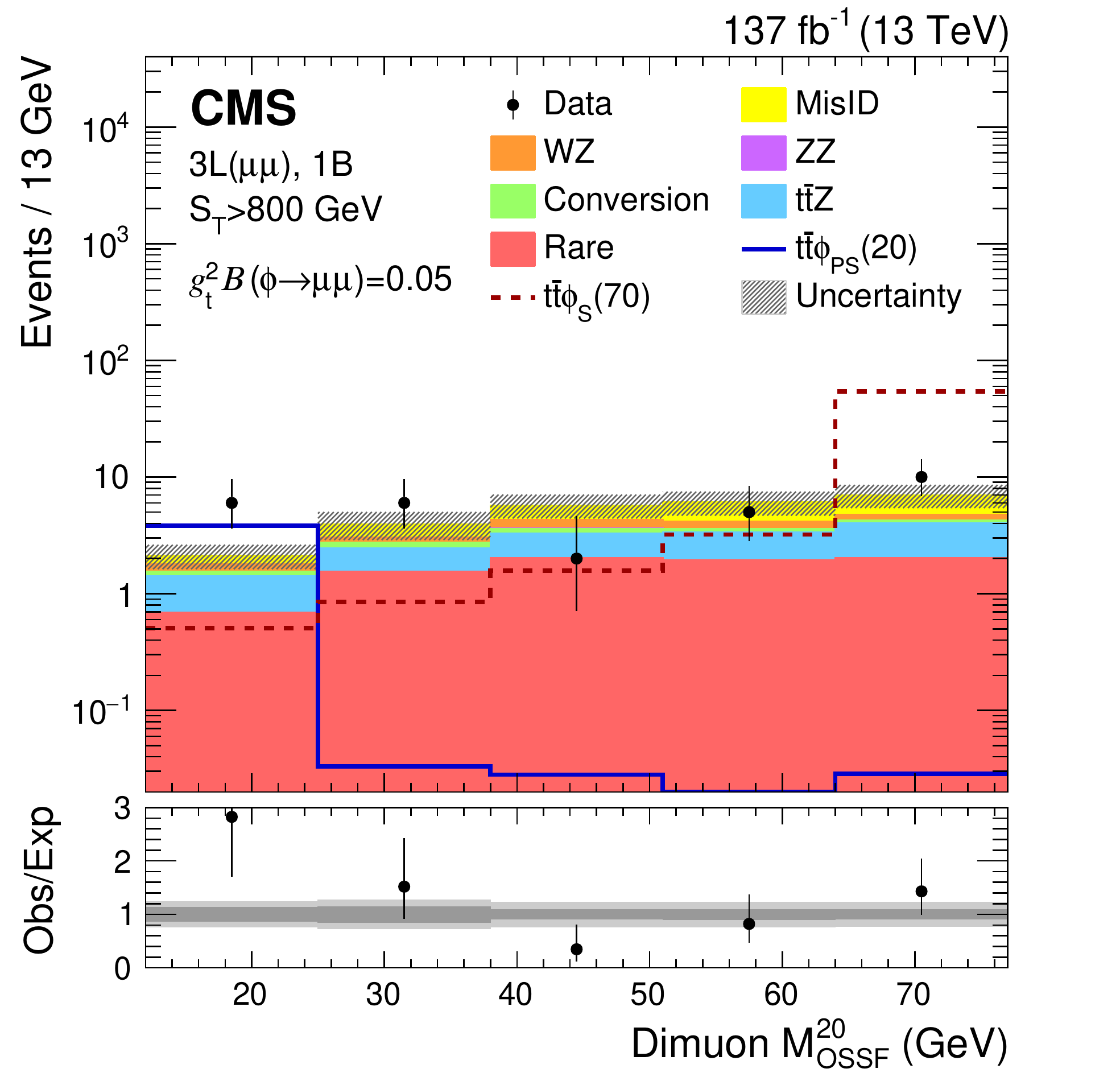} \hspace{.05\textwidth}
\includegraphics[width=.4\textwidth]{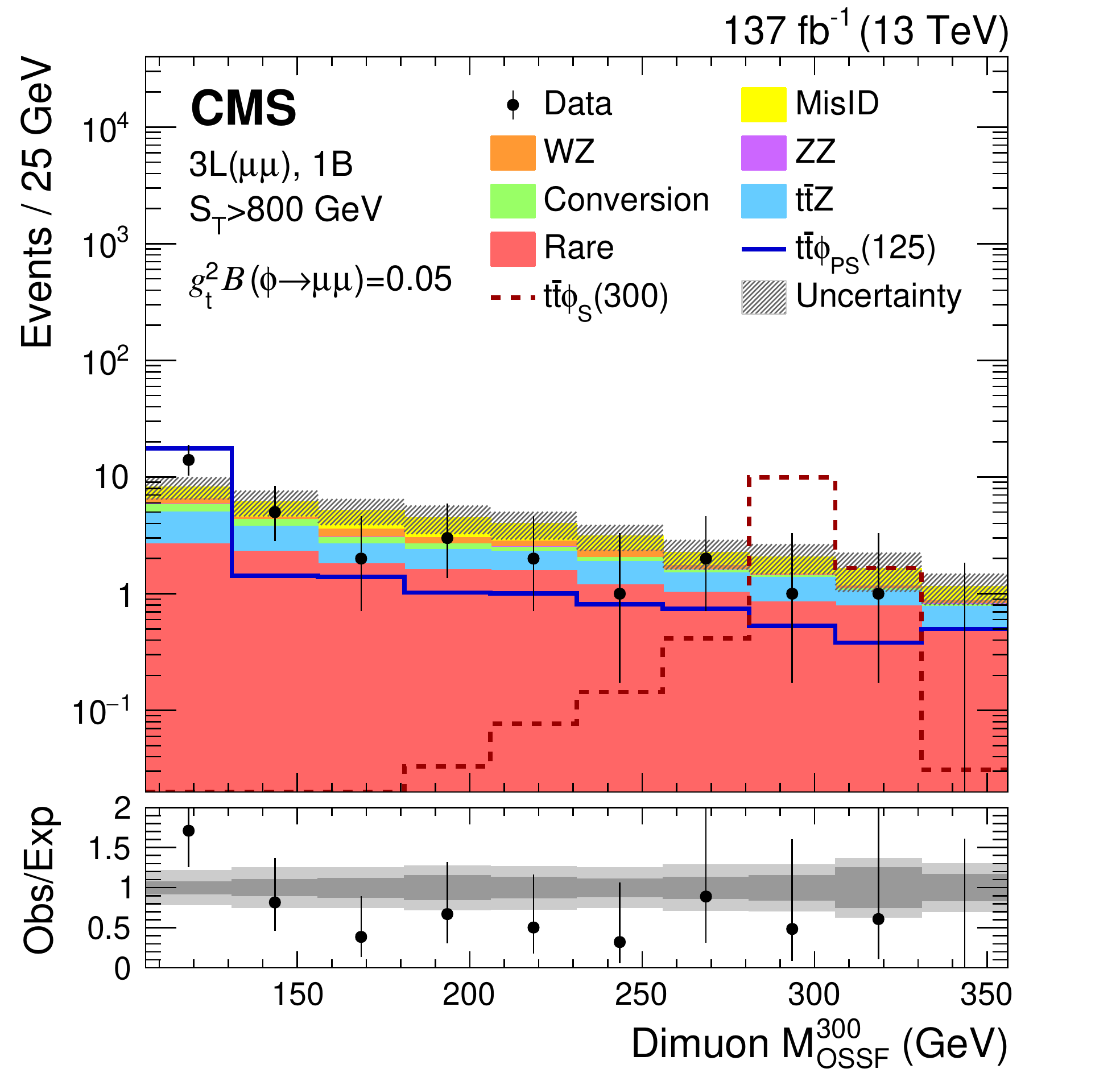}
\caption{
Dimuon $\mossf^{20}$ (left column) and $\mossf^{300}$ (right column) distributions in the 3L($\mu\mu$) 1B $\ttphi$ signal regions.
Upper, center, and lower plots are for $\ST<400\GeV$, $400<\ST<800\GeV$, and $\ST>800\GeV$, respectively.
The total SM background is shown as a stacked histogram of all contributing processes.
The predictions for $\ttphi(\to\mu\mu$) models with a pseudoscalar (scalar) $\phi$ of 20 and 125 (70 and 300)\GeV mass assuming $g_{\PQt}^2\mathcal{B}(\phi\to\mu\mu)=0.05$ are also shown.
The lower panels show the ratio of observed to expected events.
The hatched gray bands in the upper panels and the light gray bands in the lower panels represent the total (systematic and statistical) uncertainty of the backgrounds in each bin, whereas the dark gray bands in the lower panels represent only the statistical uncertainty of the backgrounds.
The rightmost bins do not contain the overflow events as these are outside the probed mass range.
\label{fig:ttPhiMu1B3LSR}}
\end{figure}

\begin{figure}[!htp]
\centering
\includegraphics[width=.4\textwidth]{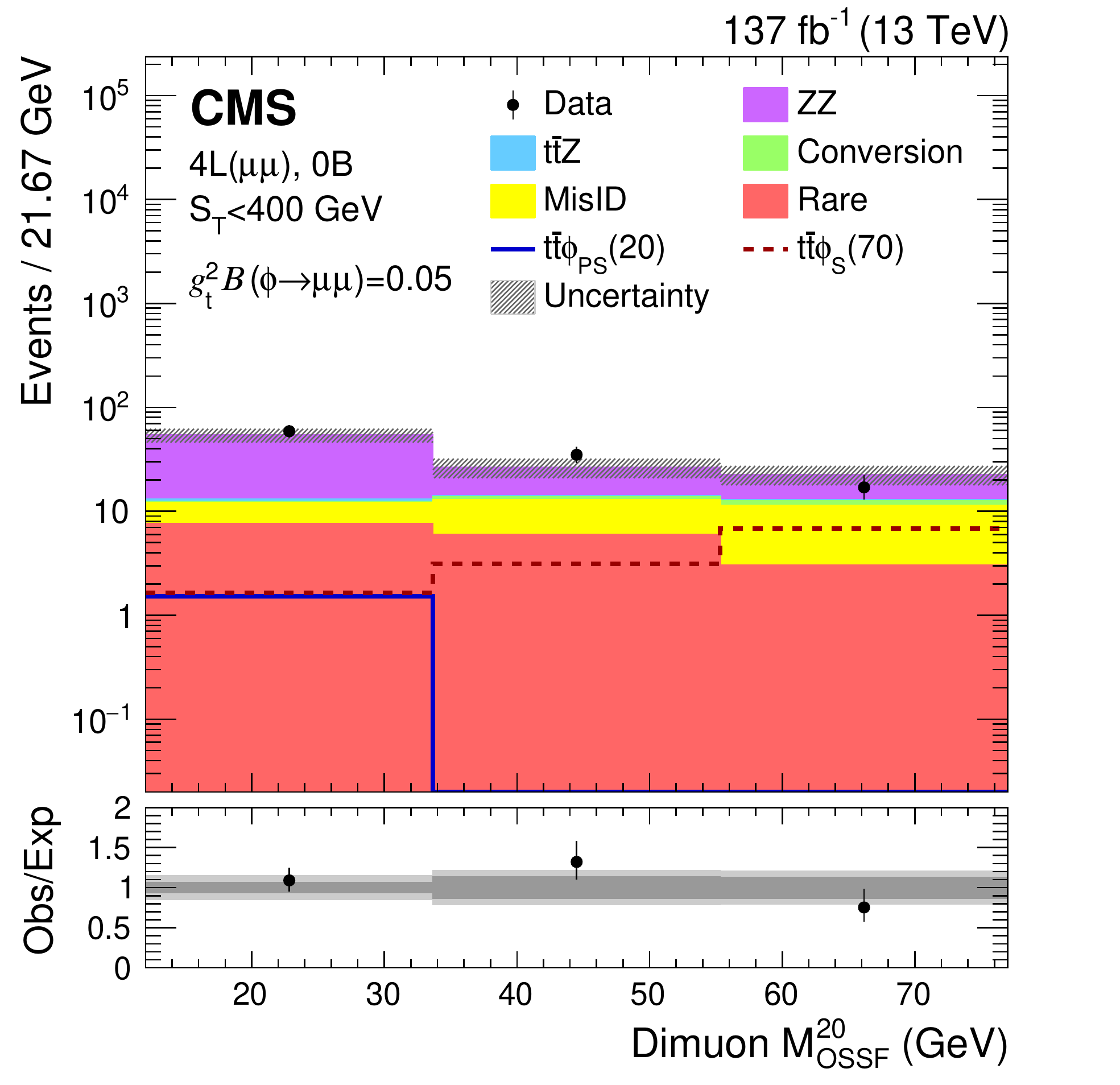} \hspace{.05\textwidth}
\includegraphics[width=.4\textwidth]{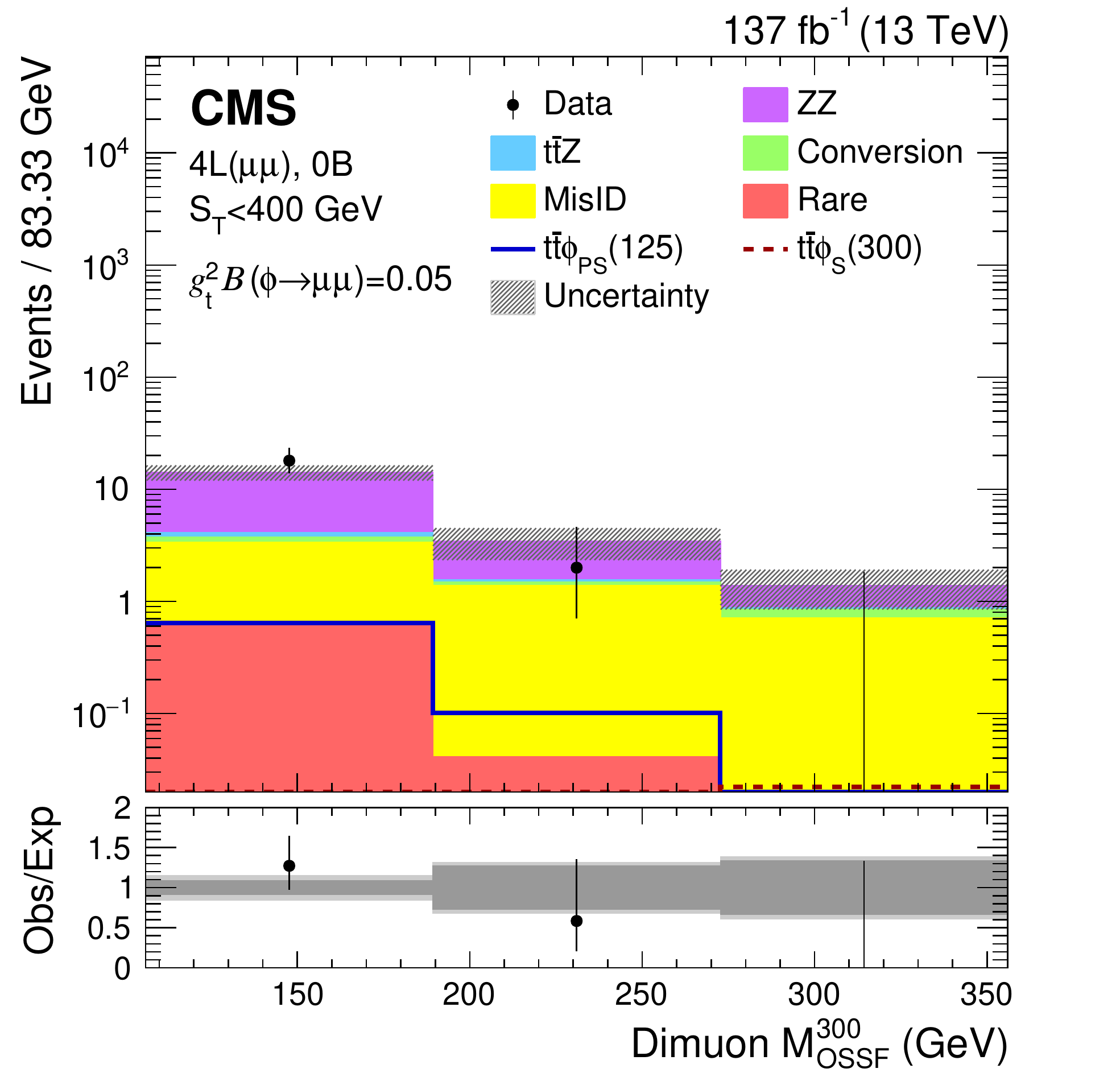}
\includegraphics[width=.4\textwidth]{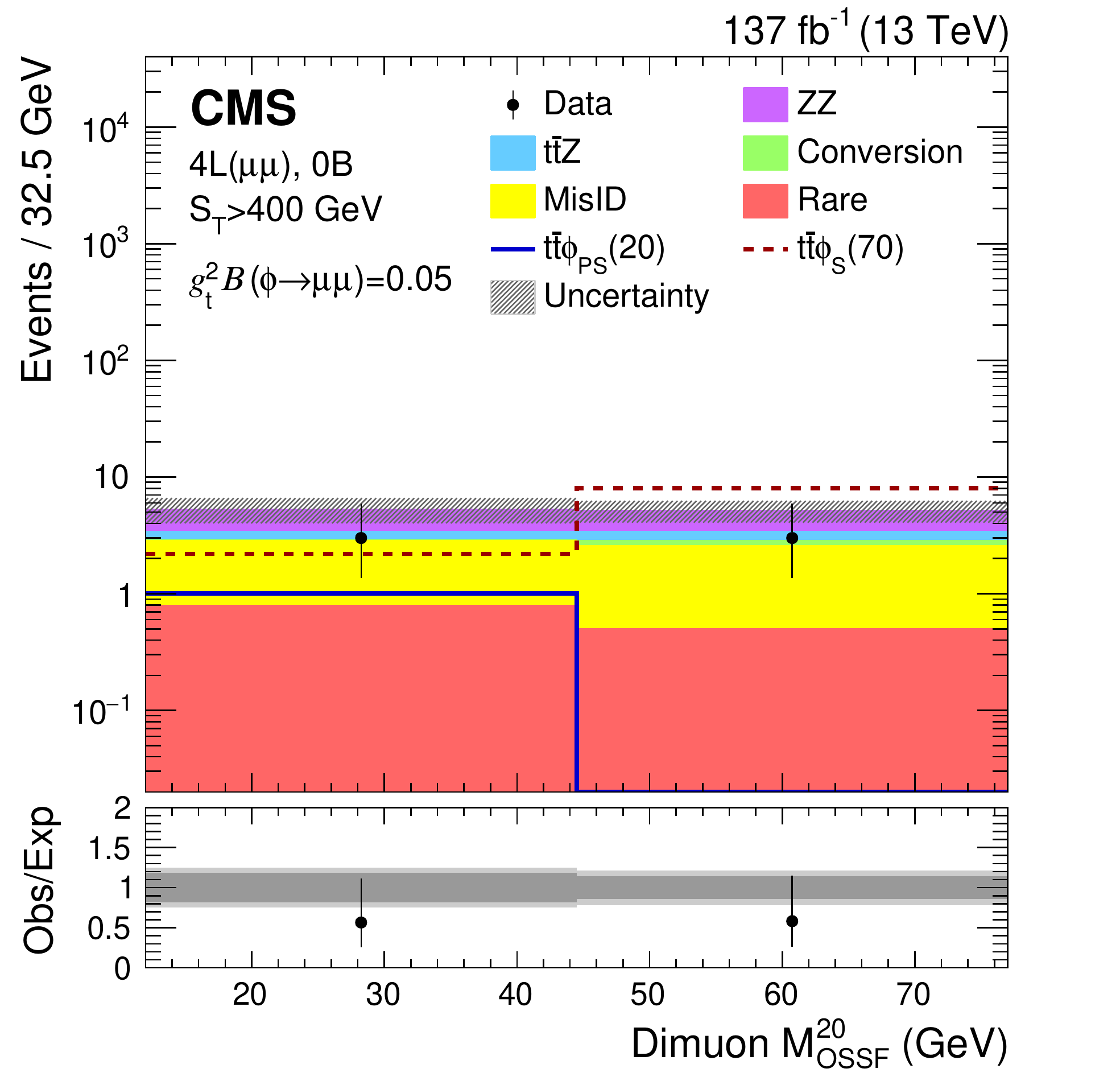} \hspace{.05\textwidth}
\includegraphics[width=.4\textwidth]{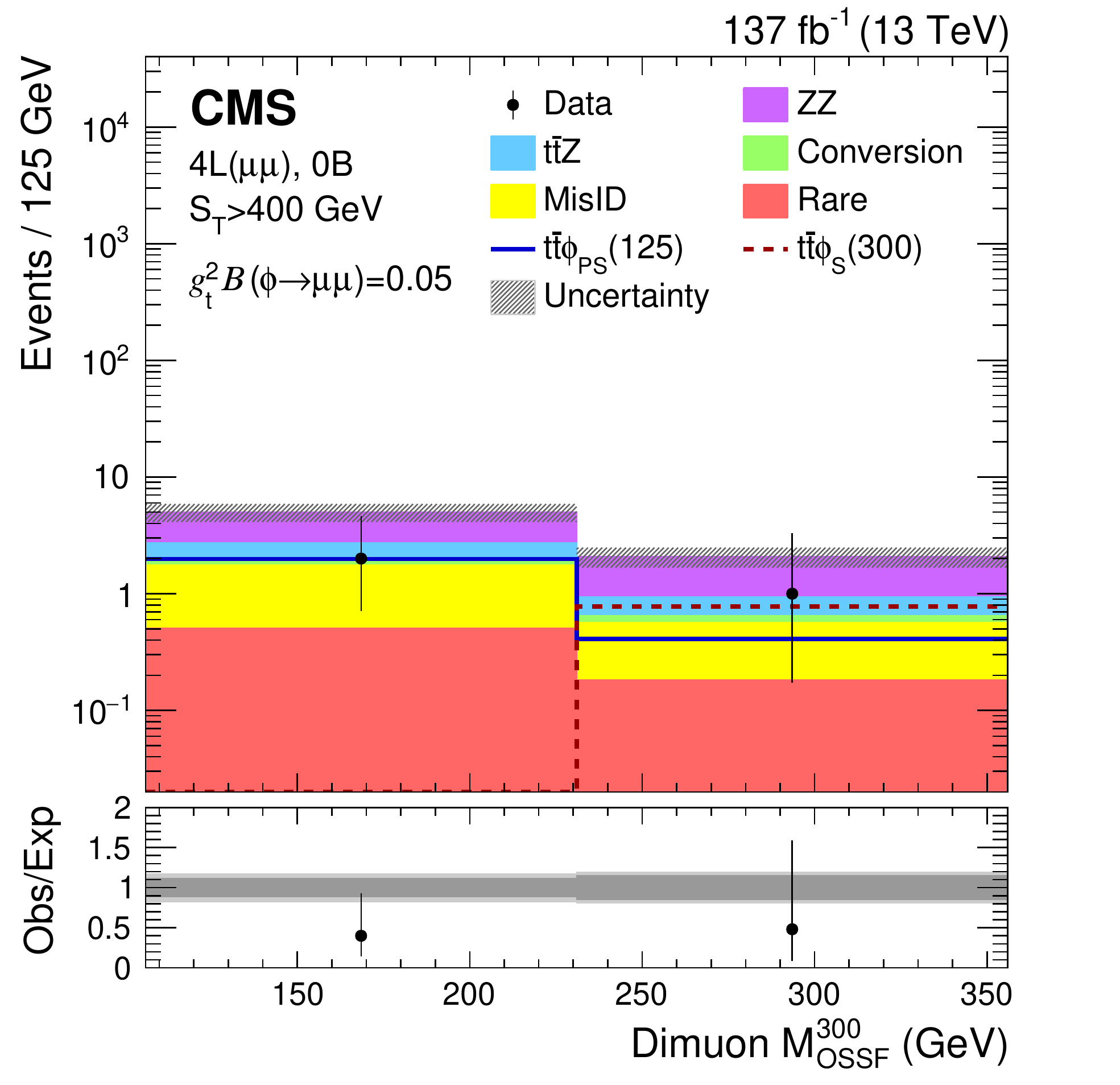}
\includegraphics[width=.4\textwidth]{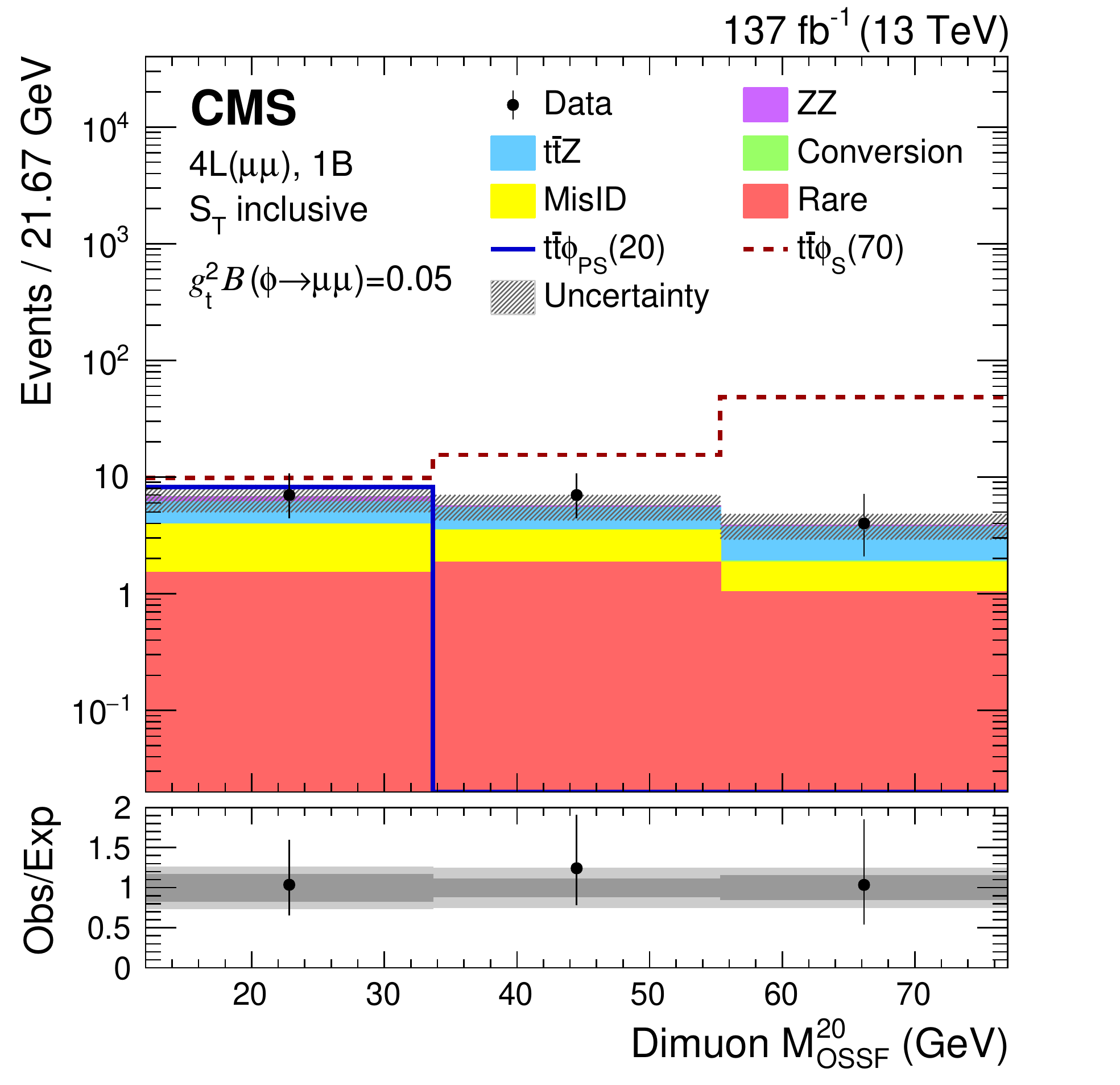} \hspace{.05\textwidth}
\includegraphics[width=.4\textwidth]{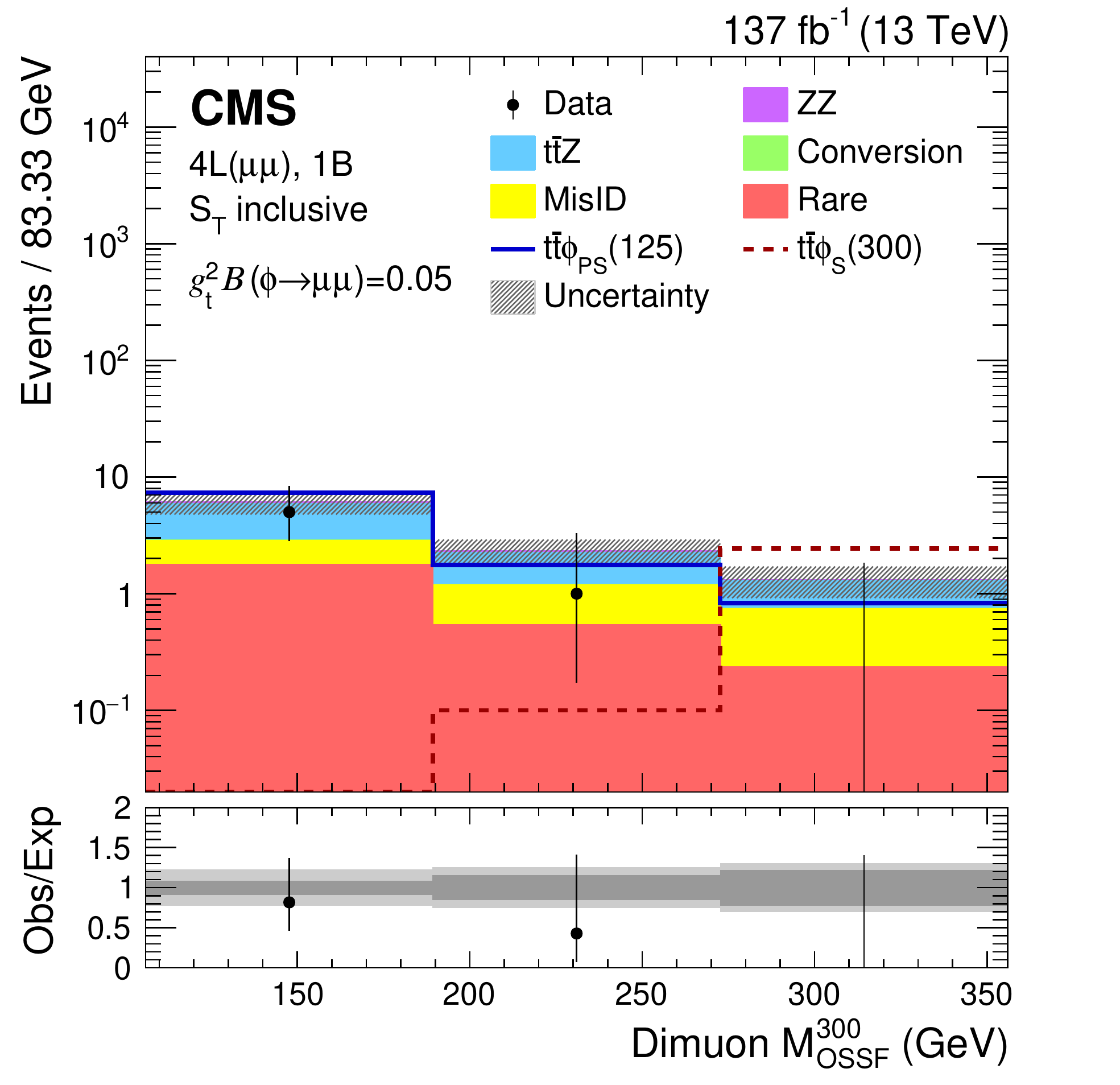}
\caption{
Dimuon $\mossf^{20}$ (left column) and $\mossf^{300}$ (right column) distributions in the 4L($\mu\mu$) $\ttphi$ signal regions.
Upper, center, and lower plots are for 0B $\ST<400\GeV$, 0B $\ST>400\GeV$, and 1B $\ST$-inclusive, respectively.
The total SM background is shown as a stacked histogram of all contributing processes.
The predictions for $\ttphi(\to\mu\mu$) models with a pseudoscalar (scalar) $\phi$ of 20 and 125 (70 and 300)\GeV mass assuming $g_{\PQt}^2\mathcal{B}(\phi\to\mu\mu)=0.05$ are also shown.
The lower panels show the ratio of observed to expected events.
The hatched gray bands in the upper panels and the light gray bands in the lower panels represent the total (systematic and statistical) uncertainty of the backgrounds in each bin, whereas the dark gray bands in the lower panels represent only the statistical uncertainty of the backgrounds.
The rightmost bins do not contain the overflow events as these are outside the probed mass range.
\label{fig:ttPhiMu4LSR}}
\end{figure}

The distributions of expected SM backgrounds and observed event yields in the signal regions as defined in Tables~\ref{tab:seesawSRs} and~\ref{tab:ttphiSRs} are given in Figs.~\ref{fig:Seesaw3LSR}--\ref{fig:Seesaw4LSR} and~\ref{fig:ttPhiEle0B3LSR}--\ref{fig:ttPhiMu4LSR} for the type-III seesaw model and the $\ttphi$ model, respectively.
The figures also show the predicted yields for type-III seesaw models with $\Sigma$ masses of 300 and 700\GeV in the flavor-democratic scenario as well as for $\ttphi$ models with a pseudoscalar (scalar) $\phi$ mass of 20 and 125 (70 and 300)\GeV assuming $g_{\PQt}^2\mathcal{B}(\phi\to{\Pe\Pe}/\mu\mu)=0.05$.

We perform a goodness-of-fit test based on the saturated model method ~\cite{Baker:1984} to quantify the local deviations between the background-only hypothesis and the observed data, without considering the look-elsewhere effect~\cite{Gross:2010qma}. The most significant local deviation from the SM expectation in the signal regions is found in the 3L($\mu\mu$) 1B $\ST<400\GeV$ high mass $\ttphi$ channel (Fig.~\ref{fig:ttPhiMu1B3LSR}) by selecting the bins with $\mossf^{300}>206\GeV$, resulting in a data excess of approximately 3.2 standard deviations.
Similarly, by examining other deviations from the SM, we observe a local data deficit of 2.5 standard deviations in the $10<\mossf^{20}<15\GeV$ bin of the 3L($\Pe\Pe$) 0B $400<\ST<800\GeV$ channel (Fig.~\ref{fig:ttPhiEle0B3LSR}), and a local data excess of 2.5 standard deviations in the $60<\mossf^{20}<65\GeV$ bin of the 3L($\mu\mu$) 1B $400<\ST<800\GeV$ channel (Fig.~\ref{fig:ttPhiMu1B3LSR}).
Other deviations are less significant.
Overall, the observations are found to be globally consistent with the SM predictions within 2.7 standard deviations,
and no statistically significant excess compatible with the signal models probed is observed.

Upper limits at 95\% confidence level (\CL) are set on the product of the signal production cross sections and branching fractions using a modified frequentist approach with the \CLs criterion~\cite{Junk:1999kv,Read:2002hq} and the asymptotic approximation for the test statistic~\cite{Cowan:2010js,ATLAS:2011tau}.
Upper limits at 95\% \CL are also set on the product of the branching fractions and the square of the scalar or pseudoscalar Yukawa coupling in the $\ttphi$ model.
A binned maximum-likelihood fit is performed to discriminate between the potential signal and the SM background processes for both signal models separately.
All of the $\LT$+\ptmiss and $\MTthree$ bins are used for the seesaw signal masses under consideration,
whereas the appropriate subset of the lepton flavor and dilepton mass bins is used for a given $\phi$ mass and branching fraction scenario in the $\ttphi$ signal model,
such that the low (high) dielectron and dimuon mass spectra are considered for a light (heavy) $\ttphi$ signal with the $\phi\to{\Pe\Pe}$ and $\phi\to\mu\mu$ decays, respectively.

The uncertainties in the mean values of both the expected signal and background yields are treated as nuisance parameters modeled by log-normal and gamma distributions for systematic and statistical uncertainties, respectively.
Statistical uncertainties in the signal and background yields in each bin and year are assumed to be fully uncorrelated,
whereas all systematic uncertainties are assumed to be fully correlated among the signal bins in a given year.
The correlation model of all nuisance parameters across the datasets collected in different years is summarized in Table~\ref{tab:systematics}.

\begin{figure}[!htp]
\centering
\includegraphics[width=.6\textwidth]{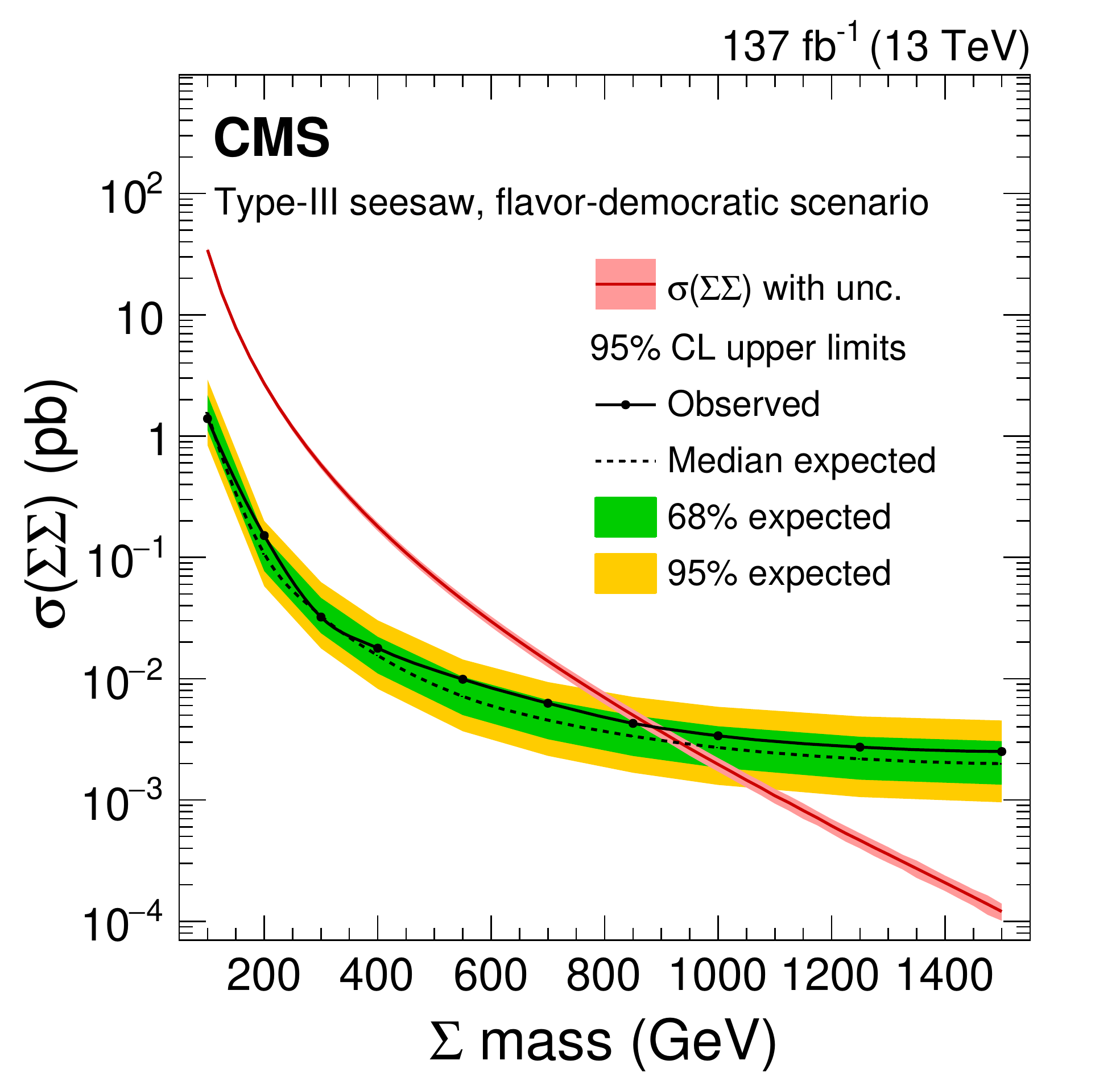}
\caption{
The 95\% confidence level expected and observed upper limits on the total production cross section of heavy fermion pairs.
The inner (green) and the outer (yellow) bands indicate the regions containing 68 and 95\%, respectively, of the distribution of limits expected under the background-only hypothesis.
Also shown are the theoretical prediction for the cross section and the associated uncertainty of the $\Sigma$ pair production via the type-III seesaw mechanism.
Type-III seesaw heavy fermions are excluded for masses below 880\GeV (expected limit 930\GeV) in the flavor-democratic scenario.
\label{fig:seesawLimitComb}}
\end{figure}

\begin{figure}[!htp]
\centering
\includegraphics[width=.4\textwidth]{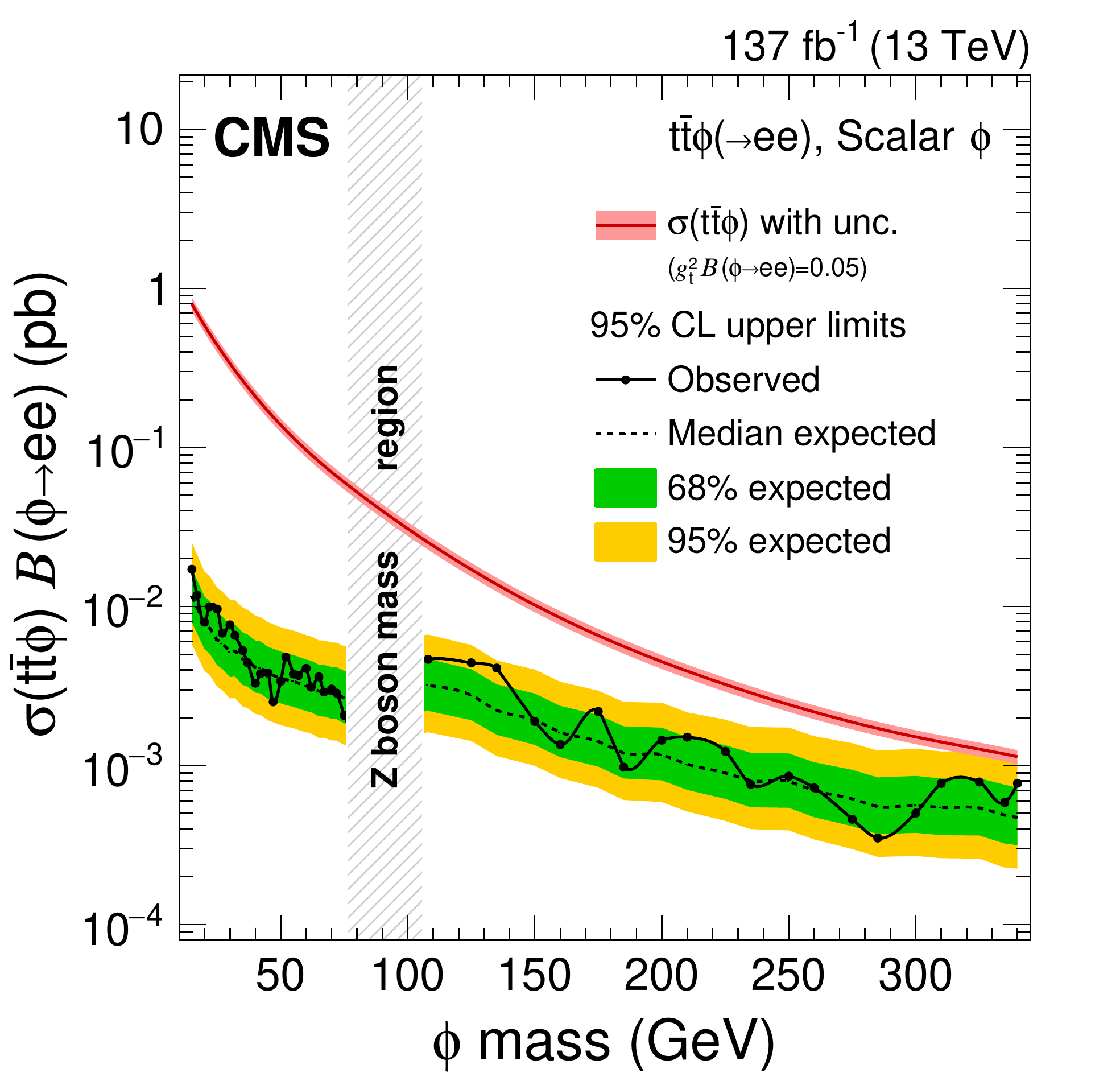} \hspace{.05\textwidth}
\includegraphics[width=.4\textwidth]{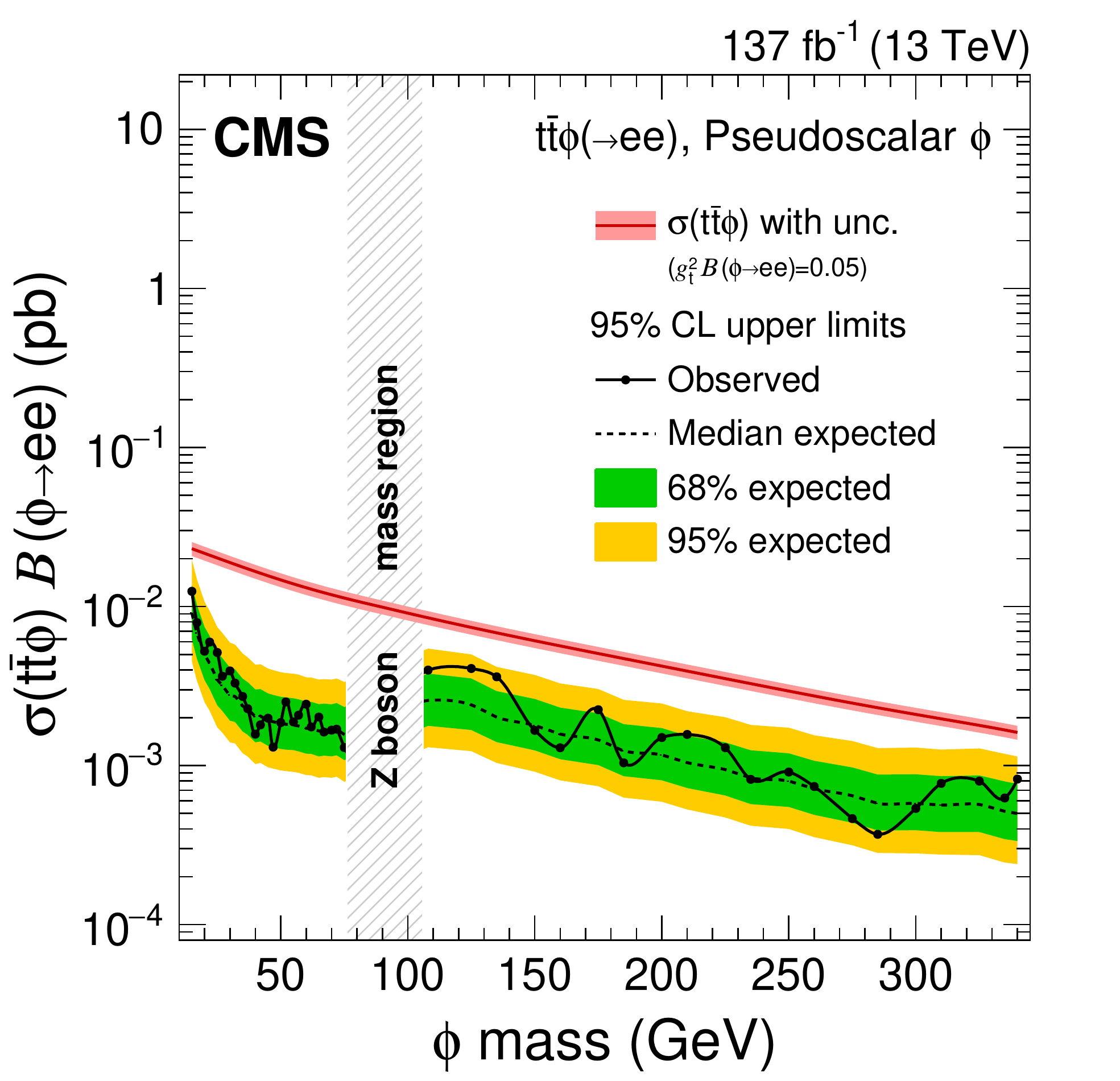}
\includegraphics[width=.4\textwidth]{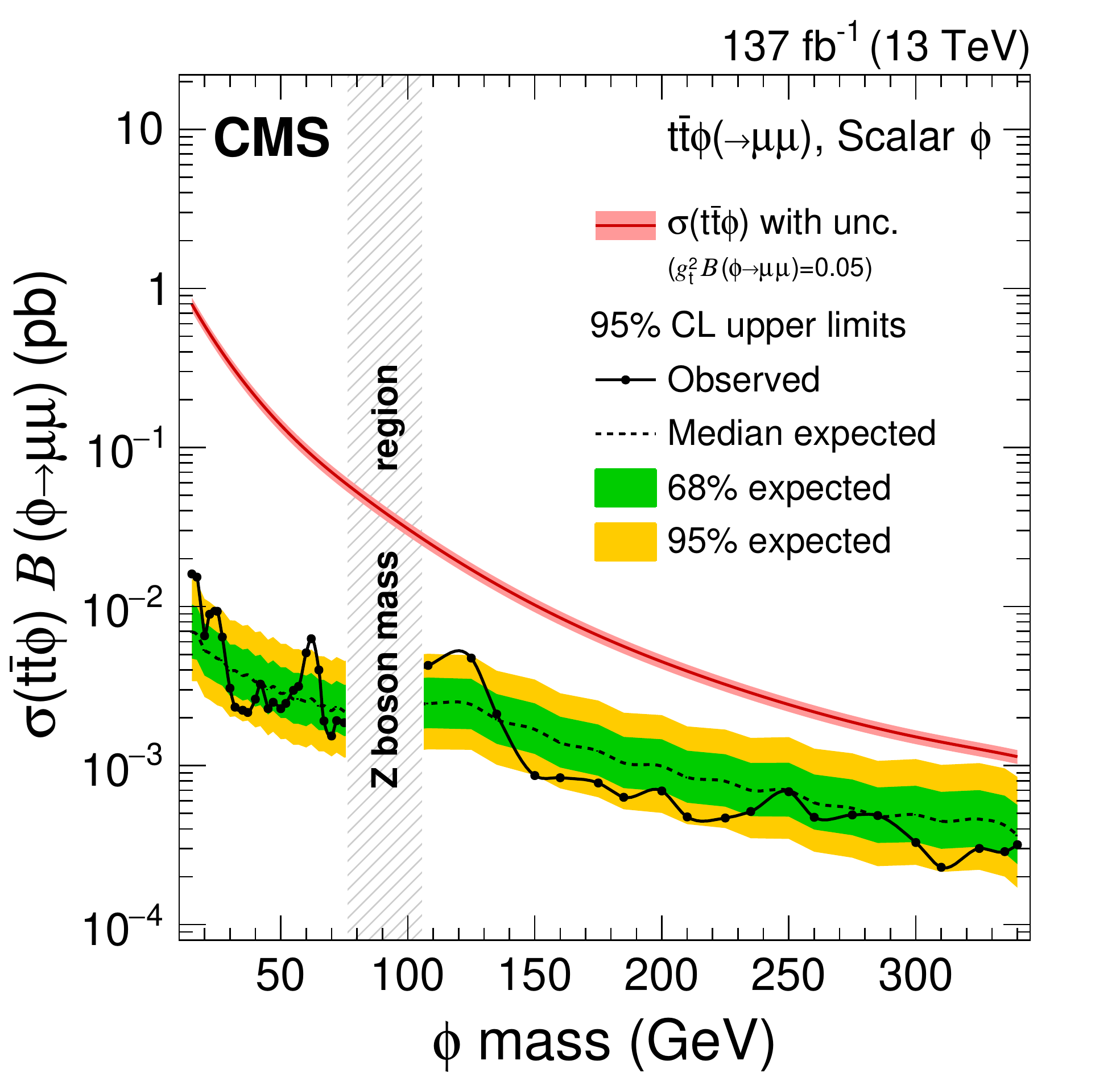} \hspace{.05\textwidth}
\includegraphics[width=.4\textwidth]{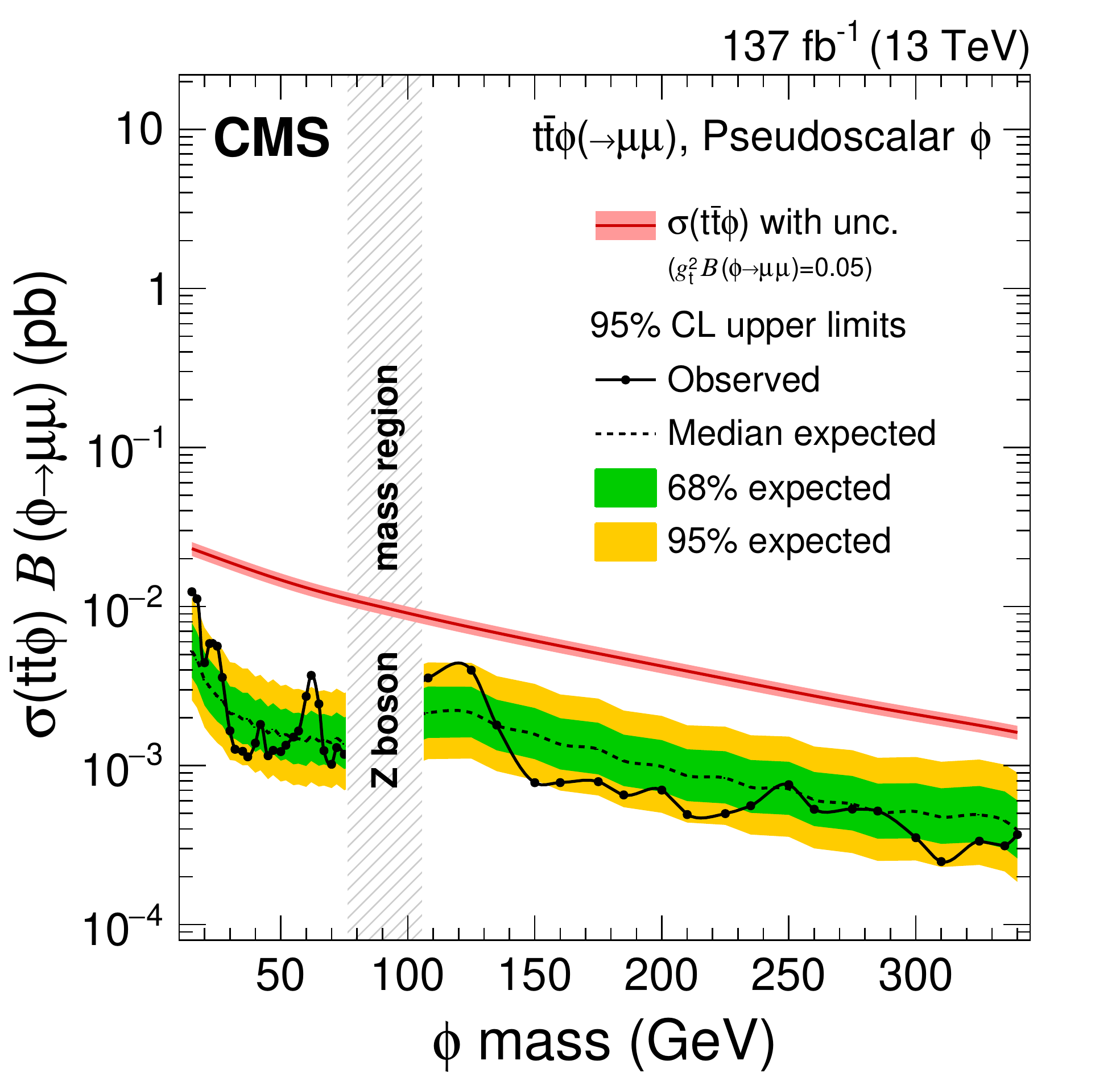}
\caption{
The 95\% confidence level expected and observed upper limits on the product of the signal production cross section and branching fraction of a scalar $\phi$ boson in the dielectron (upper left) and dimuon (lower left) channels, and of a pseudoscalar $\phi$ boson in the dielectron (upper right) and dimuon (lower right) channels, where $\phi$ is produced in association with a top quark pair.
The inner (green) and the outer (yellow) bands indicate the regions containing 68 and 95\%, respectively, of the distribution of limits expected under the background-only hypothesis. The vertical hatched gray band indicates the mass region corresponding to the $\PZ$ boson veto.
Also shown are the theoretical predictions for the product of the production cross section and branching fraction of the $\ttphi$ model, with their uncertainties, and assuming ${g}_{t}^2\mathcal{B}(\phi\to{\Pe\Pe}/\mu\mu)=0.05$. 
All $\ttphi$ signal scenarios are excluded for the product of the production cross section and branching fraction above 1--20\unit{fb} for $\phi$ masses in the range of 15--75\GeV, and above 0.3--5\unit{fb} for $\phi$ masses in the range of 108--340\GeV. 
\label{fig:ttPhiLimitsV1}}
\end{figure}

\begin{figure}[!htp]
\centering
\includegraphics[width=.4\textwidth]{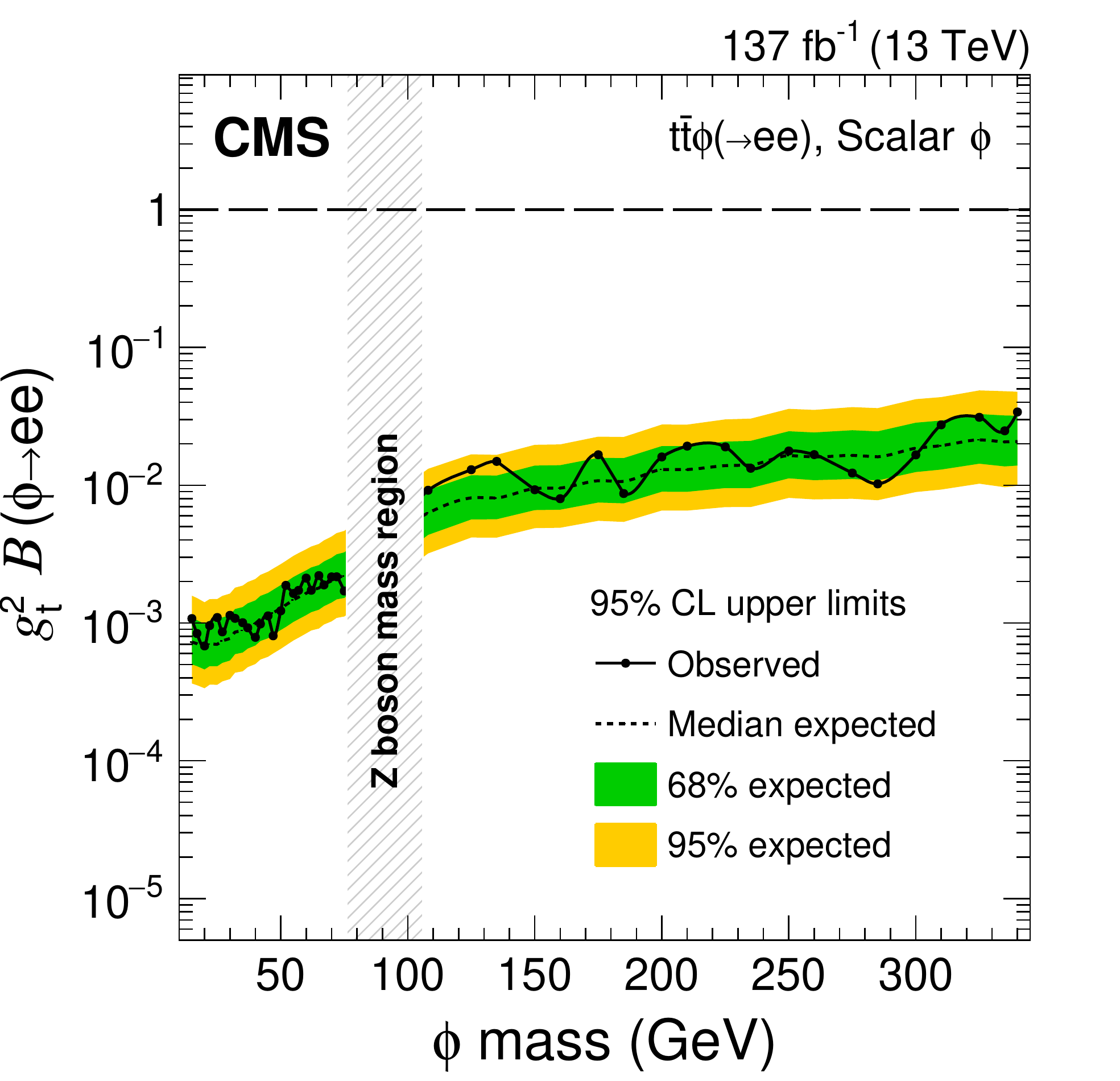} \hspace{.05\textwidth}
\includegraphics[width=.4\textwidth]{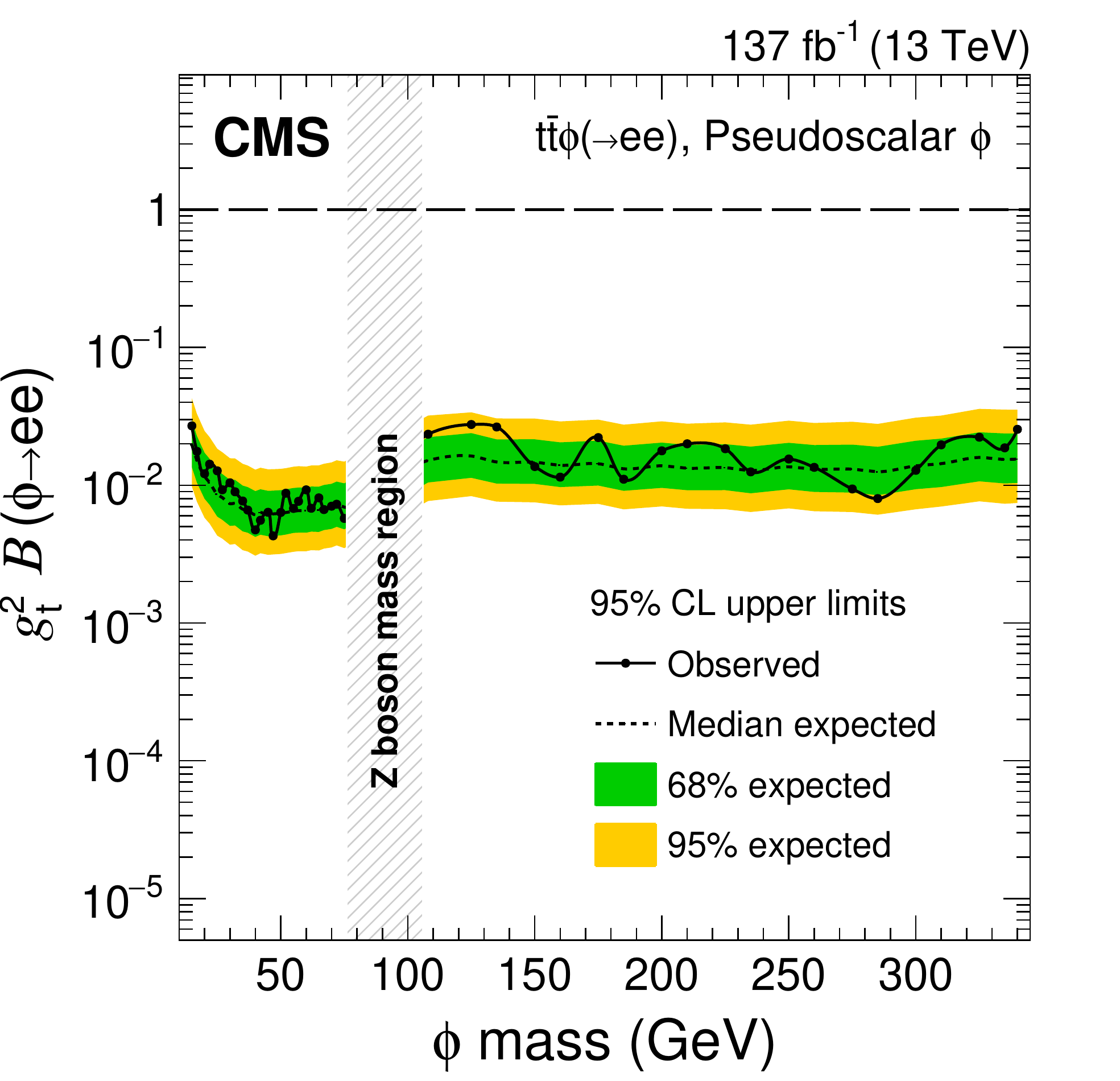}
\includegraphics[width=.4\textwidth]{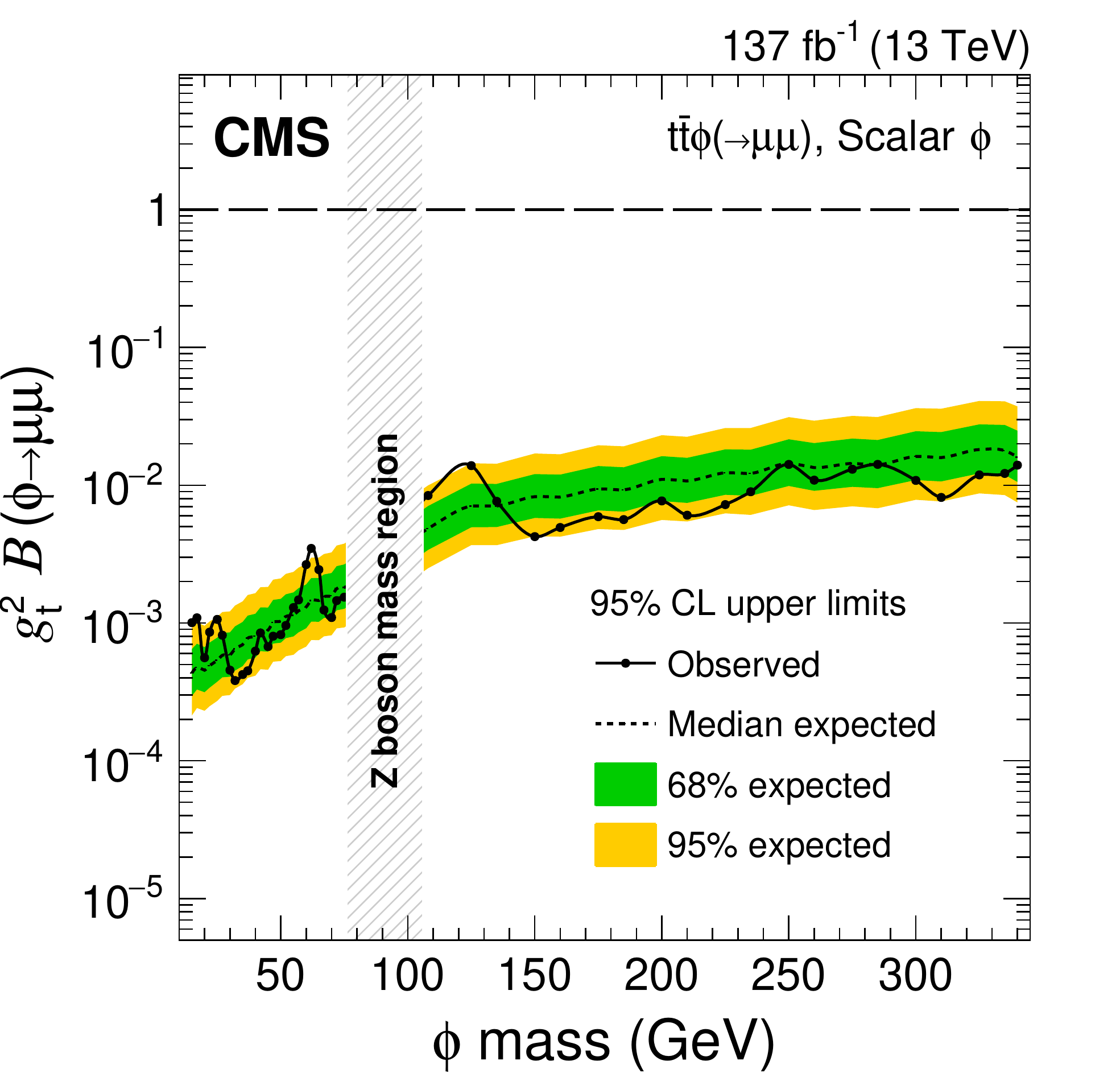} \hspace{.05\textwidth}
\includegraphics[width=.4\textwidth]{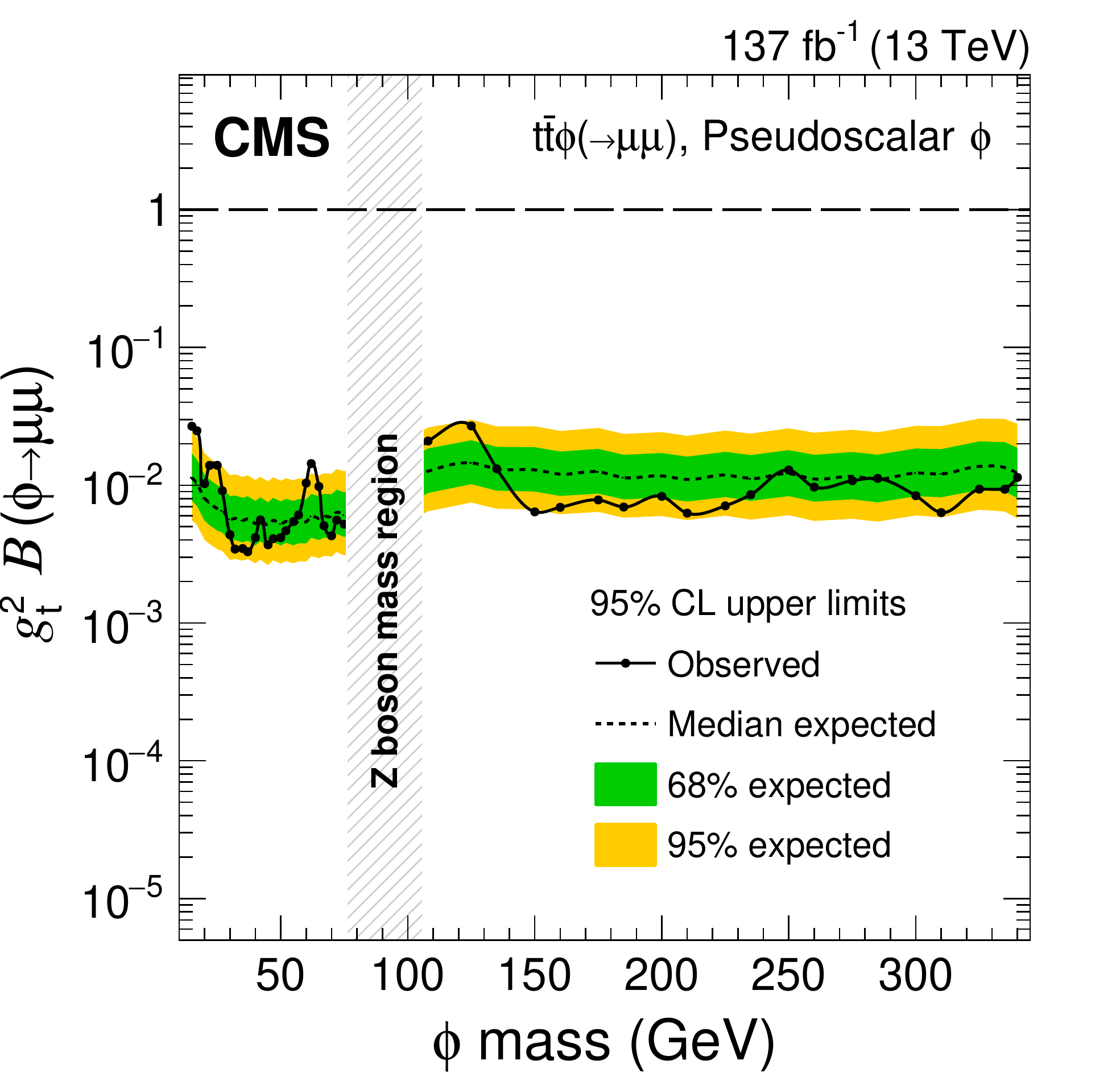}
\caption{
The 95\% confidence level expected and observed upper limits on the product of the square of the Yukawa coupling to top quarks and branching fraction of a scalar $\phi$ boson in the dielectron (upper left) and dimuon (lower left) channels, and of a pseudoscalar $\phi$ boson in the dielectron (upper right) and dimuon (lower right) channels, where $\phi$ is produced in association with a top quark pair.
The inner (green) and the outer (yellow) bands indicate the regions containing 68 and 95\%, respectively, of the distribution of limits expected under the background-only hypothesis.  The vertical hatched gray band indicates the mass region corresponding to the $\PZ$ boson veto.
The dashed horizontal line marks the unity value of the product of the square of the Yukawa coupling to top quarks and the branching fraction. 
Assuming a Yukawa coupling of unit strength to top quarks, the branching fraction of new scalar (pseudoscalar) bosons to dielectrons or dimuons above 0.0004--0.004 (0.004--0.03) are excluded for masses in the range of 15--75\GeV, and above 0.004--0.04 (0.006--0.03) for masses in the range of 108--340\GeV. 
\label{fig:ttPhiLimitsV2}}
\end{figure}

The observed and expected upper limits on the production cross section $\sigma(\Sigma\Sigma)$ in the type-III seesaw signal model are given in Fig.~\ref{fig:seesawLimitComb}.
Type-III seesaw heavy fermions are excluded at 95\% \CL with masses below 880\GeV assuming the flavor-democratic scenario.
Similarly, the upper limits on $\sigma($\ttphi$)\mathcal{B}(\phi\to{\Pe\Pe}/\mu\mu)$ and $g_{\PQt}^2\mathcal{B}(\phi\to{\Pe\Pe}/\mu\mu)$ in the $\ttphi$ signal model are shown in Figs.~\ref{fig:ttPhiLimitsV1} and \ref{fig:ttPhiLimitsV2}, respectively.
In the $\ttphi$ signal model, we exclude cross sections above 1--20\unit{fb} for $\phi$ masses in the range of 15--75\GeV, and above 0.3--5\unit{fb} for $\phi$ masses in the range of 108--340\GeV. 
Furthermore, $g_{\PQt}^2\mathcal{B}(\phi\to{\Pe\Pe}/\mu\mu)$ above (0.4--4)$\ten{-3}$ for the scalar and above (0.4--3)$\ten{-2}$ for the pseudoscalar scenarios are excluded for $\phi$ masses in the 15--75\GeV range, whereas the two models perform similarly for masses 108--340\GeV and are excluded above (0.4--4)$\ten{-2}$ for the scalar and above (0.6--3)$\ten{-2}$ for the pseudoscalar scenarios.
Uncertainties in the production cross sections due to scale and PDF choices are considered for both signal models~\cite{Fuks:2012qx,Fuks:2013vua,deFlorian:2016spz}, and are also shown in Figs.~\ref{fig:seesawLimitComb} and \ref{fig:ttPhiLimitsV1}.

The differences in the low-mass exclusion limits of scalar and pseudoscalar $\ttphi$ models result from the kinematic structure of the couplings, which affect both the production cross section and the signal efficiency of the $\phi$ bosons.
The coupling of a scalar boson to a fermion is momentum independent, whereas that of a pseudoscalar boson is proportional to the momentum in the low momentum limit.
Therefore, the low $\phi$ momentum part of the production cross section is suppressed in the pseudoscalar model in comparison to the scalar model for $\phi$ masses below the top quark mass scale, while both production cross sections are similar for $\phi$ masses at and above the top quark mass scale.
Furthermore, this coupling structure results in more pseudoscalar $\phi$ bosons in the Lorentz-boosted region compared to the scalar $\phi$ bosons,
yielding more energetic leptons with higher selection efficiencies.
The product of the fiducial acceptance and the event selection efficiency for the type-III seesaw and the $\ttphi$ models for various signal mass hypotheses,
calculated after all analysis selection requirements, are given in Table~\ref{tab:signalEffs}.

\begin{table}[!htp]
\centering
\topcaption{Product of the fiducial acceptance and the event selection efficiency for the signal models at various signal mass hypotheses calculated after all analysis selection requirements.
} \label{tab:signalEffs}
\resizebox{\textwidth}{!}{
\begin{tabular}{ l l l l l l l l l l l l l l l l }
\hline
\multicolumn{1}{l}{Signal model} &  \multicolumn{15}{c}{Product of acceptance and efficiency (\%)} \\[0.5ex] \hline
\multicolumn{13}{l}{Type-III seesaw (flavor-democratic scenario)}  & & & \\[0.5ex]
$\Sigma$ mass ({\GeVns})   & 100  & 200  & 300  & 400  & 550  & 700  & 850  & 1000 & 1250 & 1500 & & & & & \\[\cmsTabSkip]
                      & 0.32 & 1.82 & 2.63 & 3.02 & 3.29 & 3.34 & 3.29 & 3.21 & 2.99 & 2.82 & & & & & \\[\cmsTabSkip]
\multicolumn{1}{l}{$\ttphi$}\\[0.5ex]
$\phi$ mass ({\GeVns})             & 15   &  20  & 25  &  30  & 40 & 50 & 60 & 70 & 75 & 108 & 125 & 150 & 200 & 250 & 300 \\[\cmsTabSkip]
Scalar $\phi(\to{\Pe\Pe}$) &  0.85 & 1.29 & 1.67 & 2.02 & 2.74 & 3.44 & 4.25 & 5.16 & 4.95 & 5.53 & 8.32 & 9.00 & 10.3 & 11.1 & 11.5 \\
Scalar $\phi(\to\mu\mu$) & 1.54 & 2.16 & 2.81 & 3.35 & 4.38 & 5.29 & 6.40 & 7.69 & 7.56 & 8.74 & 11.6 & 12.3 & 14.0 & 14.8 & 15.3 \\
Pseudoscalar $\phi(\to{\Pe\Pe}$) & 0.96 & 1.81 & 2.69 & 3.45 & 4.88 & 5.82 & 6.62 & 7.35 & 6.83 & 6.8 & 9.77 & 10.4 & 11.0 & 11.4 & 11.9 \\
Pseudoscalar $\phi(\to\mu\mu$) & 1.69 & 2.95 & 4.24 & 5.38 & 7.14 & 8.46 & 9.73 & 10.4 & 9.93 & 10.3 & 13.4 & 14.0 & 14.9 & 15.2 & 15.9 \\  \hline
\end{tabular}}
\end{table}

\section{Summary}

A search has been performed for physics beyond the standard model, using multilepton events in 137\fbinv of $\Pp\Pp$ collision data at $\sqrt{s} = 13\TeV$, collected with the CMS detector in 2016--2018.
The observations are found to be consistent with the expectations from standard model processes, with no statistically significant signal-like excess in any of the probed channels.
The results are used to constrain the allowed parameter space of the targeted signal models.
At 95\% confidence level, heavy fermions of the type-III seesaw model with masses below 880\GeV are excluded assuming identical $\Sigma$ decay branching fractions across all lepton flavors. This is the most restrictive limit on the flavor-democratic scenario of the type-III seesaw model to date.
Assuming a Yukawa coupling of unit strength to top quarks, branching fractions of new scalar (pseudoscalar) bosons to dielectrons or dimuons above
0.004 (0.03) are excluded at 95\% confidence level for masses in the range 15--75\GeV, and above 0.04 (0.03) for masses in the range 108--340\GeV. These are the first limits in these channels on an extension of the standard model with scalar or pseudoscalar particles.

\begin{acknowledgments}

We thank M. J. Strassler for drawing our attention to the need to include top quark pair production with three-body top quark decays, ${\PQt}\to\cPqb\PW\phi$, in the $\ttphi$  signal model.

We congratulate our colleagues in the CERN accelerator departments for the excellent performance of the LHC and thank the technical and administrative staffs at CERN and at other CMS institutes for their contributions to the success of the CMS effort. In addition, we gratefully acknowledge the computing centers and personnel of the Worldwide LHC Computing Grid for delivering so effectively the computing infrastructure essential to our analyses. Finally, we acknowledge the enduring support for the construction and operation of the LHC and the CMS detector provided by the following funding agencies: BMBWF and FWF (Austria); FNRS and FWO (Belgium); CNPq, CAPES, FAPERJ, FAPERGS, and FAPESP (Brazil); MES (Bulgaria); CERN; CAS, MoST, and NSFC (China); COLCIENCIAS (Colombia); MSES and CSF (Croatia); RPF (Cyprus); SENESCYT (Ecuador); MoER, ERC IUT, PUT and ERDF (Estonia); Academy of Finland, MEC, and HIP (Finland); CEA and CNRS/IN2P3 (France); BMBF, DFG, and HGF (Germany); GSRT (Greece); NKFIA (Hungary); DAE and DST (India); IPM (Iran); SFI (Ireland); INFN (Italy); MSIP and NRF (Republic of Korea); MES (Latvia); LAS (Lithuania); MOE and UM (Malaysia); BUAP, CINVESTAV, CONACYT, LNS, SEP, and UASLP-FAI (Mexico); MOS (Montenegro); MBIE (New Zealand); PAEC (Pakistan); MSHE and NSC (Poland); FCT (Portugal); JINR (Dubna); MON, RosAtom, RAS, RFBR, and NRC KI (Russia); MESTD (Serbia); SEIDI, CPAN, PCTI, and FEDER (Spain); MOSTR (Sri Lanka); Swiss Funding Agencies (Switzerland); MST (Taipei); ThEPCenter, IPST, STAR, and NSTDA (Thailand); TUBITAK and TAEK (Turkey); NASU (Ukraine); STFC (United Kingdom); DOE and NSF (USA).

\hyphenation{Rachada-pisek} Individuals have received support from the Marie-Curie program and the European Research Council and Horizon 2020 Grant, contract Nos.\ 675440, 752730, and 765710 (European Union); the Leventis Foundation; the A.P.\ Sloan Foundation; the Alexander von Humboldt Foundation; the Belgian Federal Science Policy Office; the Fonds pour la Formation \`a la Recherche dans l'Industrie et dans l'Agriculture (FRIA-Belgium); the Agentschap voor Innovatie door Wetenschap en Technologie (IWT-Belgium); the F.R.S.-FNRS and FWO (Belgium) under the ``Excellence of Science -- EOS" -- be.h project n.\ 30820817; the Beijing Municipal Science \& Technology Commission, No. Z181100004218003; the Ministry of Education, Youth and Sports (MEYS) of the Czech Republic; the Lend\"ulet (``Momentum") Program and the J\'anos Bolyai Research Scholarship of the Hungarian Academy of Sciences, the New National Excellence Program \'UNKP, the NKFIA research grants 123842, 123959, 124845, 124850, 125105, 128713, 128786, and 129058 (Hungary); the Council of Science and Industrial Research, India; the HOMING PLUS program of the Foundation for Polish Science, cofinanced from European Union, Regional Development Fund, the Mobility Plus program of the Ministry of Science and Higher Education, the National Science Center (Poland), contracts Harmonia 2014/14/M/ST2/00428, Opus 2014/13/B/ST2/02543, 2014/15/B/ST2/03998, and 2015/19/B/ST2/02861, Sonata-bis 2012/07/E/ST2/01406; the National Priorities Research Program by Qatar National Research Fund; the Ministry of Science and Education, grant no. 3.2989.2017 (Russia); the Programa Estatal de Fomento de la Investigaci{\'o}n Cient{\'i}fica y T{\'e}cnica de Excelencia Mar\'{\i}a de Maeztu, grant MDM-2015-0509 and the Programa Severo Ochoa del Principado de Asturias; the Thalis and Aristeia programs cofinanced by EU-ESF and the Greek NSRF; the Rachadapisek Sompot Fund for Postdoctoral Fellowship, Chulalongkorn University and the Chulalongkorn Academic into Its 2nd Century Project Advancement Project (Thailand); the Nvidia Corporation; the Welch Foundation, contract C-1845; and the Weston Havens Foundation (USA).

\end{acknowledgments}

\bibliography{auto_generated}
\cleardoublepage \appendix\section{The CMS Collaboration \label{app:collab}}\begin{sloppypar}\hyphenpenalty=5000\widowpenalty=500\clubpenalty=5000\input{EXO-19-002-authorlist.tex}\end{sloppypar}
\end{document}

%% file: EXO-19-002-authorlist.tex
\vskip\cmsinstskip
\textbf{Yerevan Physics Institute, Yerevan, Armenia}\\*[0pt]
A.M.~Sirunyan$^{\textrm{\dag}}$, A.~Tumasyan
\vskip\cmsinstskip
\textbf{Institut f\"{u}r Hochenergiephysik, Wien, Austria}\\*[0pt]
W.~Adam, F.~Ambrogi, T.~Bergauer, M.~Dragicevic, J.~Er\"{o}, A.~Escalante~Del~Valle, M.~Flechl, R.~Fr\"{u}hwirth\cmsAuthorMark{1}, M.~Jeitler\cmsAuthorMark{1}, N.~Krammer, I.~Kr\"{a}tschmer, D.~Liko, T.~Madlener, I.~Mikulec, N.~Rad, J.~Schieck\cmsAuthorMark{1}, R.~Sch\"{o}fbeck, M.~Spanring, D.~Spitzbart, W.~Waltenberger, C.-E.~Wulz\cmsAuthorMark{1}, M.~Zarucki
% ...
\vskip\cmsinstskip
\dag: Deceased\\
1:  Also at Vienna University of Technology, Vienna, Austria\\
% ...
